\documentclass[useAMS,amsgen,usenatbib]{mn2e}
\usepackage{psfig,epsfig,times}
\newcounter{fred}
\def\lx{\mathrel{L_X}}
\def\tx{\mathrel{kT_X}}
\def\keV{\mathrel{{\rm keV}}}
\def\txtot{\mathrel{kT_{X{\rm ,tot}}}}
\def\txann{\mathrel{kT_{X{\rm ,ann}}}}
\def\drpeak{\mathrel{\Delta r_{\rm peak}}}
\def\ls{\mathrel{\hbox{\rlap{\hbox{\lower4pt\hbox{$\sim$}}}\hbox{$<$}}}}
\def\gs{\mathrel{\hbox{\rlap{\hbox{\lower4pt\hbox{$\sim$}}}\hbox{$>$}}}}
\def\xc{\mathrel{x_{\rm c}}}
\def\yc{\mathrel{y_{\rm c}}}
\def\rc{\mathrel{r_{\rm core}}}
\def\rt{\mathrel{r_{\rm cut}}}
\def\sigo{\mathrel{\sigma_{\rm o}}}

\def\Mtot{\mathrel{M_{\rm tot}}}
\def\Mcen{\mathrel{M_{\rm cen}}}
\def\texp{\mathrel{T_{\rm exp}}}
\def\R702{\mathrel{R_{702}}}
\def\dos{\mathrel{D_{\rm OS}}}
\def\dls{\mathrel{D_{\rm LS}}}
\def\dol{\mathrel{D_{\rm OL}}}
\def\lobs{\mathrel{\lambda_{\rm obs}}}

\def\ergs{\mathrel{\rm erg\,s^{-1}}}
\def\kms{\mathrel{\rm km\,s^{-1}}}
\def\nH{\mathrel{N_{\rm H}}}
\def\Msol{\mathrel{\rm M_{\odot}}}

\def\zs{\mathrel{z_{\rm S}}}
\def\zl{\mathrel{z_{\rm L}}}
\def\tv{\mathrel{\bf{\vec\tau}}}
\def\gv{\mathrel{\vec g}}
\def\gamv{\mathrel{\bf{\vec\gamma}}}

\title[X--ray Luminous Clusters at z=0.2]{A \emph{Hubble Space
	Telescope} Lensing Survey of X--ray Luminous Galaxy Clusters:
	IV. Mass, Structure and Thermodynamics of Cluster Cores at
	z=0.2}

\author[Smith, G.P., et al.]{Graham P.\ Smith$^{1,2}$,
	 Jean-Paul Kneib$^{1,3}$,
	 Ian Smail$^{4}$,
	 Pasquale Mazzotta$^{4,5}$,\newauthor
	 Harald Ebeling$^{6}$ \&
	 Oliver Czoske$^{3,7}$ \vspace{1mm}\\
	 $^1$ California Institute of Technology, Mail Code 105--24,
	 Pasadena, CA 91125, USA. Email: gps@astro.caltech.edu\\
	 $^2$ Department of Physics, University of Durham, 
	 South Road, Durham DH1 3LE, UK\\
	 $^3$ Observatoire Midi-Pyr\'en\'ees, 14 Avenue E.\,Belin, 
	 31400 Toulouse, France\\
	 $^4$ Institute for Computational Cosmology, University of
	 Durham, South Road, Durham, DH1 3LE, UK\\
	 $^5$ Harvard-Smithsonian Center for Astrophysics, 60 Garden
	 Street, Cambridge, MA 02138, USA\\
	 $^6$ Institute for Astronomy, University of Hawaii, 
	 2680 Woodlawn Drive, Honolulu, HI\,96822, USA\\
	 $^7$ Institut f\"ur Astrophysik und Extraterrestrische
	 Forschung, Universit\"at Bonn, Auf dem H\"ugel 71, 53121
	 Bonn, Germany} 

\date{Draft Version -- March 24, 2004}

\pagerange{000--000}

\begin{document}

\addtolength{\topmargin}{-15mm}

\maketitle

\begin{abstract}
We present a comprehensive space--based study of ten X--ray luminous
galaxy clusters ($L_X{\ge}8{\times}10^{44}{\rm erg
\,s^{-1}}$[0.1--2.4\,keV]) at $z{=}0.2$.  \emph{Hubble Space Telescope
(HST)} observations reveal numerous gravitationally--lensed arcs for
which we present four new spectroscopic redshifts, bringing the total
to thirteen confirmed arcs in this cluster sample.  The confirmed arcs
reside in just half of the clusters; we thus obtain a firm lower limit
on the fraction of clusters with a central projected mass density
exceeding the critical density required for strong--lensing of $50\%$.
We combine the multiple--image systems with the weakly--sheared
background galaxies to model the total mass distribution in the
cluster cores ($R{\le}500{\rm kpc}$).  These models are complemented
by high--resolution X--ray data from \emph{Chandra} and used to
develop quantitative criteria to classify the clusters as relaxed or
unrelaxed.  Formally, $(30{\pm}20)\%$ of the clusters form a
homogeneous sub--sample of relaxed clusters; the remaining
$(70{\pm}20)\%$ are unrelaxed and are a much more diverse population.
Most of the clusters therefore appear to be experiencing a
cluster--cluster merger, or relaxing after such an event.  We also
study the normalization and scatter of scaling relations between
cluster mass, luminosity and temperature.  The scatter in these
relations is dominated by the unrelaxed clusters and is typically
${\sigma}{\simeq}0.4$.  Most notably, we detect 2--3\,times more
scatter in the mass--temperature relation than theoretical simulations
and models predict.  The observed scatter is also asymmetric -- the
unrelaxed systems are systematically 40\% hotter than the relaxed
clusters at 2.5--${\sigma}$ significance.  This structural segregation
should be a major concern for experiments designed to constrain
cosmological parameters using galaxy clusters.  Overall our results
are consistent with a scenario of cluster--cluster merger induced
boosts to cluster X--ray luminosities and temperatures.
\end{abstract}


\section{Introduction}\label{intro}

\setcounter{footnote}{7}

\normalsize

Massive galaxy clusters are the largest collapsed structures in the
Universe ($M_{\rm virial}{\simeq}10^{15}{\rm M_\odot}$), containing
vast quantities of the putative dark matter (DM), hot intracluster gas
(${\tx}{\simeq}7{\rm keV}$), and galaxies ($n_{\rm gal}{\sim}10^3$).
These rare systems stand at the nodes of the ``cosmic web'' as defined
by the large--scale filamentary structure seen in both galaxy redshift
surveys (e.g.\ de~Lapparent, Geller \& Huchra 1986; Shectman et al.\
1996; Vettolani et al.\ 1997; Peacock et al.\ 2001; Zehavi et al.\
2002) and numerical simulations of structure formation (e.g.\ Bond et
al.\ 1996; Yoshida et al.\ 2001; Evrard et al.\ 2002).  Clusters are
inferred to assemble by accreting matter along the filamentary axes,
slowly ($t_{\rm crossing}{\sim}2$--3\,Gyr) ingesting DM, gas and stars
into their deep gravitational potential wells.

Clusters have long been recognized as cosmological probes.  For
example, the evolution of cluster substructure with look--back--time
is in principal a powerful diagnostic of the cosmological parameters
(Gunn \& Gott 1972; Peebles 1980; Richstone, Loeb \& Turner 1992;
Evrard et al.\ 1993).  A complementary probe is to constrain the
matter density of the universe and the normalization of the matter
power spectrum using the cluster mass function.  However, it is
currently not possible to measure the cluster mass function directly.
More easily accessible surrogates such as the X--ray luminosity and
temperature functions are therefore used in combination with scaling
relations between the relevant quantities (e.g.\ Eke et al.\ 1996;
Reiprich \& Boehringer 2002; Viana, Nichol \& Liddle 2002; Allen et
al.\ 2003).  A critical component of such analyses is the precision to
which the scaling relations are known.  Samples of X--ray selected
clusters are now of a sufficient size that systematic uncertainties
may be comparable with the statistical uncertainties, and therefore
deserve careful analysis before robust cosmological conclusions may be
drawn (e.g.\ Smith et al.\ 2003).  Measurements of the
Sunyaev--Zeldovich effect (SZE) are also emerging as a powerful
cosmological tool (Carlstrom, Holder \& Reese 2002).  Cosmological SZE
surveys will rely on the cluster mass--temperature relationship in a
similar manner to cosmological X--ray surveys.  Such experiments may
therefore also be compromised if astrophysical systematics are
identified and carefully eliminated from the analysis (e.g.\ Majumdar
\& Mohr 2003).  Detailed study of the assembly and relaxation
histories of clusters, and their global scaling relations as a
function of redshift are therefore vitally important.

To advance our understanding of the assembly, relaxation and
thermodynamics of massive galaxy clusters requires information about
the spatial distribution of DM, hot gas and galaxies in clusters.
Several baryonic mass tracers are available, for example X--ray
emission from the intracluster medium (hereafter ICM -- e.g.\ Jones \&
Forman 1984; Buote \& Tsai 1996; Schuecker et al.\ 2001) and the
angular and line--of--sight velocity distribution of cluster galaxies
(e.g.\ Geller \& Beers 1982; Dressler \& Shectman 1988; West \& Bothun
1990).  These diagnostics have often been used as surrogates for a
direct tracer of the underlying DM distribution.  The major drawback
of this approach is the requirement to assume a relationship between
the luminous and dark matter distributions (e.g.\ that the ICM is in
hydrostatic equilibrium with the DM potential) -- it is precisely
these assumptions that require detailed testing.  This issue is
further aggravated by the expectation that cluster mass distributions
are DM dominated on all but the smallest scales (e.g.\ Smith et al.\
2001; Sand, Treu \& Ellis 2002; Sand et al.\ 2004).  

Gravitational lensing offers a solution to much of this problem, in
that the lensing signal is sensitive to the total mass distribution in
the lens, regardless of its physical nature and state.  Detailed study
of gravitational lensing by massive clusters is therefore an important
opportunity to gain an empirical understanding of the distribution of
DM in clusters.  Indeed, early comparisons between X--ray and
lensing--based mass measurements revealed a factor 2--3 discrepancy
between X--ray and strong--lensing--based cluster mass estimates
(e.g.\ Miralda--Escud\'e \& Babul 1995; Wu \& Fang 1997), although the
agreement between weak--lensing and X--ray measurements was generally
better, albeit within large uncertainties (e.g.\ Squires et al.\ 1996,
1997; Smail et al.\ 1997).  The simplifying assumptions involved in
the X--ray analysis were soon identified as the likely dominant source
of this discrepancy; this was confirmed by several authors (e.g.\
Allen 1998; Wu et al.\ 1998; Wu 2000).  In summary, X--ray and lensing
mass measurements for the most relaxed clusters agree well if the
multi--phase nature of the ICM in cool cores (e.g.\ Allen et al.\
2001) is incorporated into the X--ray analysis.  The situation is more
complex in more dynamically disturbed clusters, with larger
discrepancies being found at smaller projected radii.  The origin of
the X--ray versus lensing mass discrepancy in clusters that do not
contain a cool core is generally attributed to the simplifying
equilibrium and symmetry assumptions of the X--ray analysis.  For that
reason, most modern X--ray cluster analyses that involve measuring
cluster mass using only X--ray data understandably concentrate on
relaxed, cool core clusters (e.g.\ Allen et al.\ 2001).

An important caveat to adopting lensing as the tool of choice to
measure cluster mass is that lensing actually constrains the projected
mass distribution along the line--of--sight to the cluster.  The
addition of three--dimensional information into lensing studies may
therefore be important before final conclusions are drawn.  For
example Czoske et al.'s (2001; 2002) wide--field redshift survey of
Cl\,0024$+$1654 at $z{=}0.395$ revealed that this previously presumed
relaxed strong--lensing cluster (e.g.\ Smail et al.\ 1996; Tyson et
al.\ 1998) cluster is not relaxed, and appears to have suffered a
recent merger along the line--of--sight (see also Kneib et al.\ 2003).

Early gravitational lensing studies of galaxy clusters concentrated on
individual clusters selected because of their prominent arcs (e.g.\
Mellier et al.\ 1993; Kneib et al.\ 1994, 1995, 1996; Smail et al.\
1995a, 1996; Allen et al.\ 1996; Tyson et al.\ 1998).  This
``prominent arc'' selection function was vital to development of the
techniques required to interpret the gravitational lensing signal
(e.g.\ Kneib 1993).  However it also made it difficult to draw
conclusions about galaxy clusters as a population of astrophysical
systems from these studies.  Smail et al.\ (1997) used the
\emph{Hubble Space Telescope (HST)} to study a larger sample of
optically rich clusters, selected originally for the purpose of
studying the cluster galaxies.  As X--ray selected samples became
available in the late--1990's, Luppino et al.\ (1999) also searched
for gravitational arcs in ground--based imaging of 38 X--ray luminous
clusters.  The broad conclusions to emerge from these pioneering
studies were that to use gravitational lensing to learn about clusters
as a population, a selection function that mimics mass--selection as
closely as possible, and the high spatial resolution available from
\emph{HST} imaging are both key requirements.

We are conducting an \emph{HST} survey of an objectively selected
sample of ten X--ray luminous (and thus massive) clusters at
$z{\simeq}0.2$ (Table~\ref{sample}, \S\ref{design}).  Previous papers
in this series have presented (i) a detailed analysis of the density
profile of A\,383 (Smith et al.\ 2001), (ii) a search for
gravitationally--lensed Extremely Red Objects (EROs -- Smith et al.\
2002a) and (iii) near--infrared (NIR) spectroscopy of ERO\,J003707, a
multiply--imaged ERO at $z{=}1.6$ behind the foreground cluster A\,68
(Smith et al.\ 2002b).  This paper describes the gravitational lensing
analysis of all ten clusters observed with \emph{HST} and uses the
resulting models of the cluster cores to measure the mass and
structure of the clusters on scales of $R{\le}500\,{\rm kpc}$.  We
also exploit archival \emph{Chandra} observations and NIR photometry
of likely cluster galaxies to compare the distribution of total mass
in the clusters with the gaseous and stellar components respectively.
This combination of strong--lensing, X--ray and NIR diagnostics enable
us to quantify the prevalence of dynamical immaturity in the X--ray
luminous population at $z{\simeq}0.2$ and to calibrate the high--mass
end of the cluster mass--temperature relationship.

We outline the organization of the paper.  In \S\ref{design} we
describe the survey design and sample selection.  We then describe the
reduction and analysis of the optical data in \S\ref{optical},
comprising the \emph{HST} imaging data (\S\ref{hst}) and new
spectroscopic redshift measurements for arcs in A\,68 and A\,2219
(\S\ref{spec}).  The end--point of \S\ref{optical} is a definition of
the strong-- and weak--lensing constraints available for all ten
clusters.  We apply these constraints in \S\ref{modelling}, to
construct detailed gravitational lens models of the cluster potential
wells; the details of the modelling techniques are described in the
Appendix, and the process of fitting the constraints in each cluster
are described in \S\ref{modelling}.  We then complement these
gravitational lensing results with observations of the X--ray emission
from the clusters' ICM, drawn from the \emph{Chandra} data archive
(\S\ref{xray}).  The main results of the paper are then presented in
\S\ref{results}, including measurements of the mass and maturity of
the clusters and a detailed study of the cluster scaling relations.
We discuss the interpretation of the results in \S\ref{discussion} and
briefly assess their impact on attempts to use clusters as
cosmological probes.  Finally, we summarize our conclusions in
\S\ref{conclusions}

We assume a spatially flat universe with $H_0{=}50{\rm
km\,s^{-1}Mpc^{{-}1}}$ and $q_0{=}0.5$; in this cosmology
$1''{\equiv}4.2\,{\rm kpc}$ at $z{=}0.2$.  Our main results are
insensitive to this choice of cosmology, for example, the cluster mass
measurements would be modified by ${\ls}10\%$ if we adopted the
currently favored values of $\Omega_{\rm M}{=}0.3$,
$\Omega_\Lambda{=}0.7$, $H_0{=}65{\rm km\,s^{-1}Mpc^{{-}1}}$.  We also
adopt the complex deformation,
$\vec\tau{=}\tau_x{+}i\tau_y{=}|\vec\tau| e^{2i\theta}$, as our
measure of galaxy shape when dealing with the weak lensing aspects of
our analysis, where $\tau{=}(a^2{+}b^2)/2ab$ and $\theta$ is the
position angle of the major axis of the ellipse that describes each
galaxy.  We define the terms ``ellipticity'' to mean $\tau$ and
``orientation'' to mean $\theta$.  All uncertainties are quoted at the
68\% confidence level.

\section{Sample Selection}\label{design}

%
%
\setcounter{table}{0}
\begin{table}
\caption{
Summary of \emph{Hubble Space Telescope} Observations
\label{sample}
}
{\small
\begin{tabular}{p{8mm}p{30mm}p{7mm}p{13mm}p{5mm}}
\hline
Cluster & \centering{Central Galaxy}             & \centering{$z$}        & \centering{$\lx^a$} & \centering{$\texp$} \cr
        & \centering{$\alpha,\delta$\,(J2000)}   &                        &                     & \centering{(ks)}    \cr
\hline
A\,68   & 00 37 06.81 ${+}$09 09 24.0 & 0.255      & \raggedleft{$8.4{\pm}2.3$}    & \raggedleft{$7.5$} \cr
A\,209  & 01 31 52.53 ${-}$13 36 40.5 & 0.209      & \raggedleft{$15.2{\pm}1.0$}   & \raggedleft{$7.8$} \cr
A\,267  & 01 52 41.97 ${+}$01 00 26.2 & 0.230      & \raggedleft{$11.1{\pm}0.9$}   & \raggedleft{$7.5$} \cr
A\,383  & 02 48 03.38 ${-}$03 31 45.7 & 0.187      & \raggedleft{$9.8{\pm}0.3$}    & \raggedleft{$7.5$} \cr
A\,773  & 09 17 53.37 ${+}$51 43 37.2 & 0.217      & \raggedleft{$12.5{\pm}2.1$}   & \raggedleft{$7.2$} \cr
A\,963  & 10 17 03.57 ${+}$39 02 49.2 & 0.206      & \raggedleft{$13.4{\pm}1.0$}   & \raggedleft{$7.8$} \cr
A\,1763 & 13 35 20.10 ${+}$41 00 04.0 & 0.228      & \raggedleft{$14.2{\pm}2.1$}   & \raggedleft{$7.8$} \cr
A\,1835 & 14 01 02.05 ${+}$02 52 42.3 & 0.253      & \raggedleft{$38.3{\pm}0.9$}   & \raggedleft{$7.5$} \cr
A\,2218 & 16 35 49.22 ${+}$66 12 44.8 & 0.171      & \raggedleft{$9.0{\pm}0.8$}    & \raggedleft{$6.5$} \cr
A\,2219 & 16 40 19.82 ${+}$46 42 41.5 & 0.228      & \raggedleft{$19.8{\pm}2.2$}   & \raggedleft{$14.4$} \cr
\hline
\end{tabular}
}
\begin{tabular}{l}
\parbox[t]{80mm}{\footnotesize\addtolength{\baselineskip}{-5pt}
$^a$ ${\lx}$ is given in the [0.2--2.4\,keV] pass--band in units of
$10^{44}\,{\rm erg\,s^{{-}1}}$.  Luminosities are taken from XBACs
catalog (Ebeling et al.\ 1996) unless measurements based on pointed
observations are available: A\,383, Smith et al.\ (2001); A\,209,
A\,267, A\,963, A\,1835, Allen et al.\ (2003).

}\\
\end{tabular}
\end{table}

\normalsize

We aim to study massive galaxy clusters, and so would prefer to select
clusters based on their mass.  Mass--selected cluster catalogs
extracted from ground--based observations are gradually becoming
available (e.g.\ Miyazaki et al.\ 2002; Wittman et al.\ 2003), however
the blurring effect of the atmosphere make the completeness of these
weak--lensing cluster catalogs very difficult to characterize
robustly.  These surveys are also unlikely to achieve the sky coverage
(of order full--sky) required to detect a large sample of the rarest
and most massive systems which are the focus of our program.  In
contrast, X--ray selected cluster catalogs (e.g.\ Gioia et al.\ 1990;
Ebeling et al.\ 1998, 2000; De~Grandi et al.\ 1999) based on the
\emph{ROSAT} All--Sky Survey are already available in the public
domain with well--defined completeness limits.  X--ray selection also
influences the choice of survey epoch because the completeness of the
X--ray cluster catalogs at the time that we applied for \emph{HST}
time in Cycle 8 (GO--8249) fell off rapidly beyond $z{\simeq}0.3$.  We
therefore adopt $z{=}0.2$ as the nominal redshift of our cluster
sample.  This redshift is also well--suited to a lensing survey
because the observer--lens, observer--source and lens--source angular
diameter distances ($\dol, \dos, \dls$) that control the power and
efficiency of gravitational lenses render clusters at $z{\simeq}0.2$
powerful lenses for background galaxy populations at
$z{\sim}0.7$--1.5.  This redshift interval is well--matched to the
current generation of optical spectrographs on 10--m class telescopes.

Accordingly, we select ten of the most X--ray luminous clusters
($L_X{\ge}8{\times}10^{44}\,{\rm ergs\,s^{-1}}$, 0.1--2.4\,keV) in a
narrow redshift slice at $0.17{\le}z{\le}0.25$, with minimal
line--of--sight reddening ($E(B{-}V){\le}0.1$) from the XBACs sample
(X--ray Brightest Abell--type Clusters; Ebeling et al.\ 1996).  These
clusters span the full range of X--ray properties (morphology, central
galaxy line emission, cooling flow rate, core radius) found in larger
X--ray luminous samples (e.g.\ Peres et al.\ 1998; Crawford et al.\
1999).  The median X--ray luminosity of the sample is
$13{\times}10^{44}{\rm erg\,s^{-1}}$.  We list the cluster sample in
Table~\ref{sample}.  As XBACs is restricted to Abell clusters (Abell,
Corwin, \& Olowin 1989), the sample is not strictly X--ray selected.
However, a comparison with the X--ray selected \emph{ROSAT} Brightest
Cluster Sample (BCS; Ebeling et al. 1998, 2000a) shows that 18 of the
19 BCS clusters that satisfy our selection criteria are either Abell
or Zwicky clusters.  This confirms that our sample is
indistinguishable from a genuinely X--ray selected sample.

\section{Optical Data and Analysis}\label{optical}

\subsection{\emph{HST} Observations and Data Reduction}\label{hst}

All ten clusters were observed through the F702W filter using the
WFPC2 camera on board \emph{HST}\footnote{Based on observations with
the NASA/ESA \emph{Hubble Space Telescope} obtained at the Space
Telescope Science Institute, which is operated by the Association of
Universities for Research in Astronomy, Inc., under NASA contract NAS
5--26555.}.  The total exposure time for each cluster is listed in
Table~\ref{sample}.  We adopted a three--point dither pattern for the
eight clusters (A\,68, A\,209, A\,267, A\,383, A\,773, A\,963,
A\,1763, A\,1835) observed in Cycle 8: each exposure was shifted
relative to the previous one by ten WFC pixels (${\sim}1.0''$) in $x$
and $y$.  The archival observations of A\,2218 follow the same dither
pattern, except the offsets were three WFC pixels in $x$ and $y$.
A\,2219 was observed with a six--point dither pattern that comprised
two three--point dithers each of which were identical to that used for
the Cycle~8 observations.  These two dither patterns were offset from
each other by 10 pixels in $x$ and $y$.

\begin{figure*}
\centerline{
\psfig{file=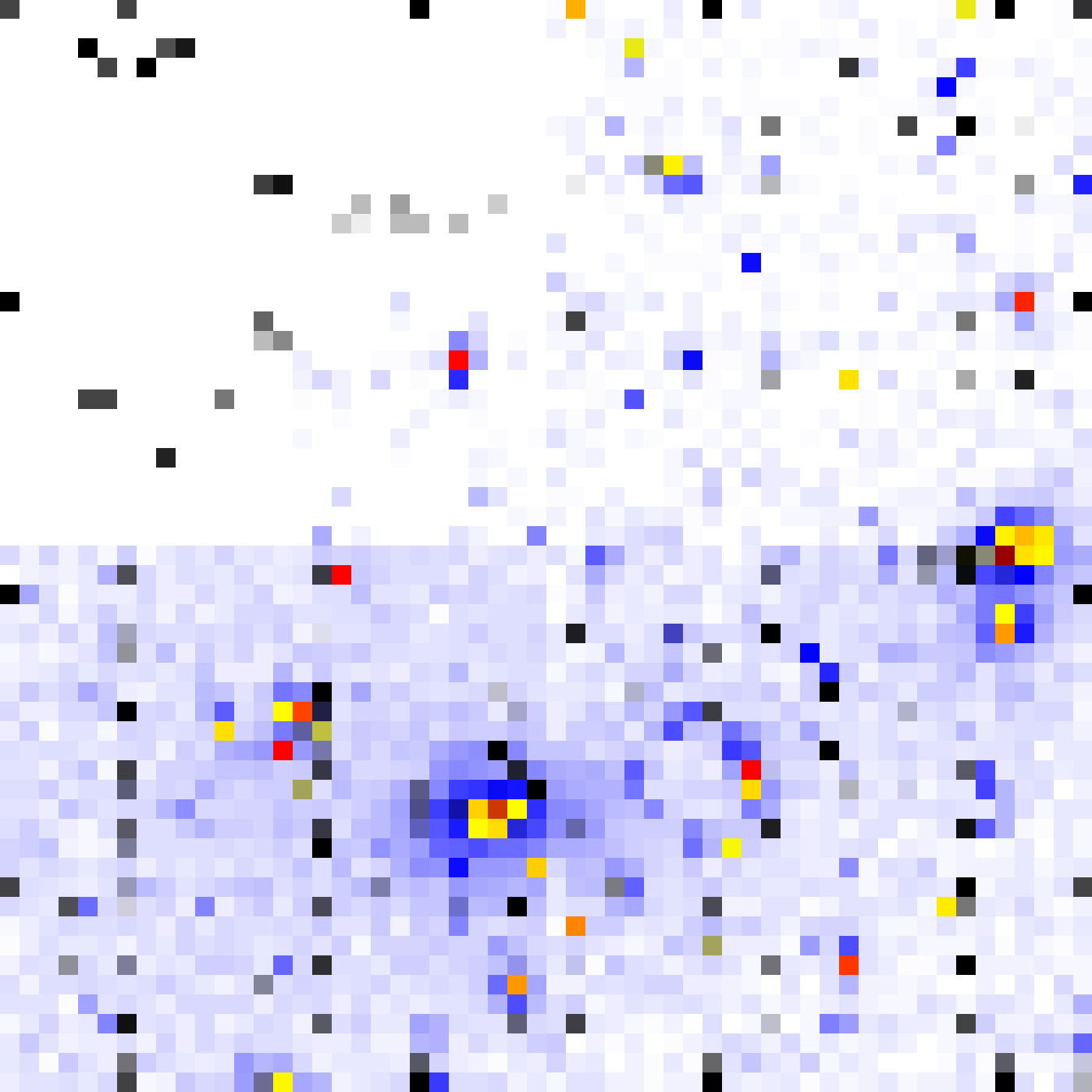,width=55mm}
\hspace{3mm}
\psfig{file=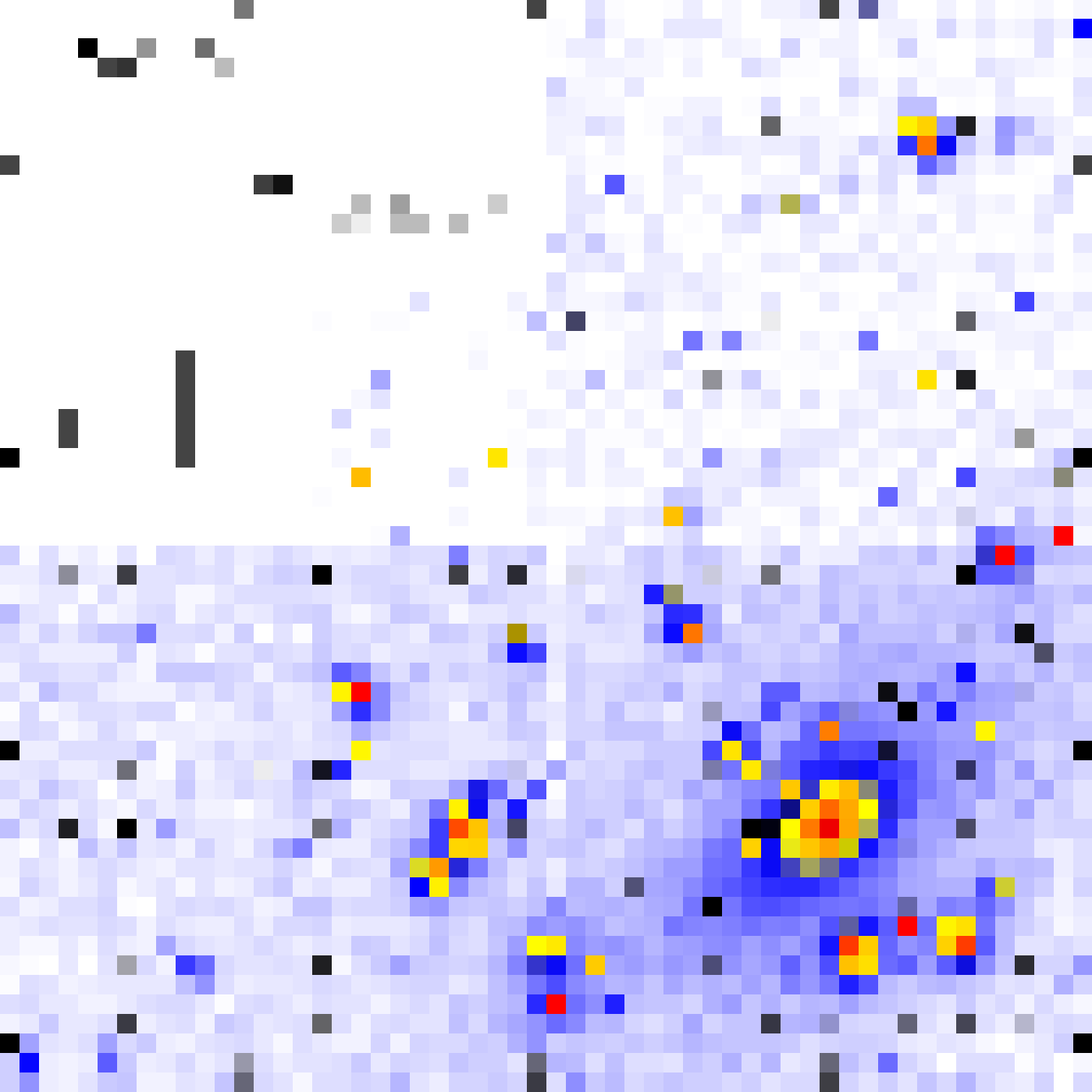,width=55mm}
\hspace{3mm}
\psfig{file=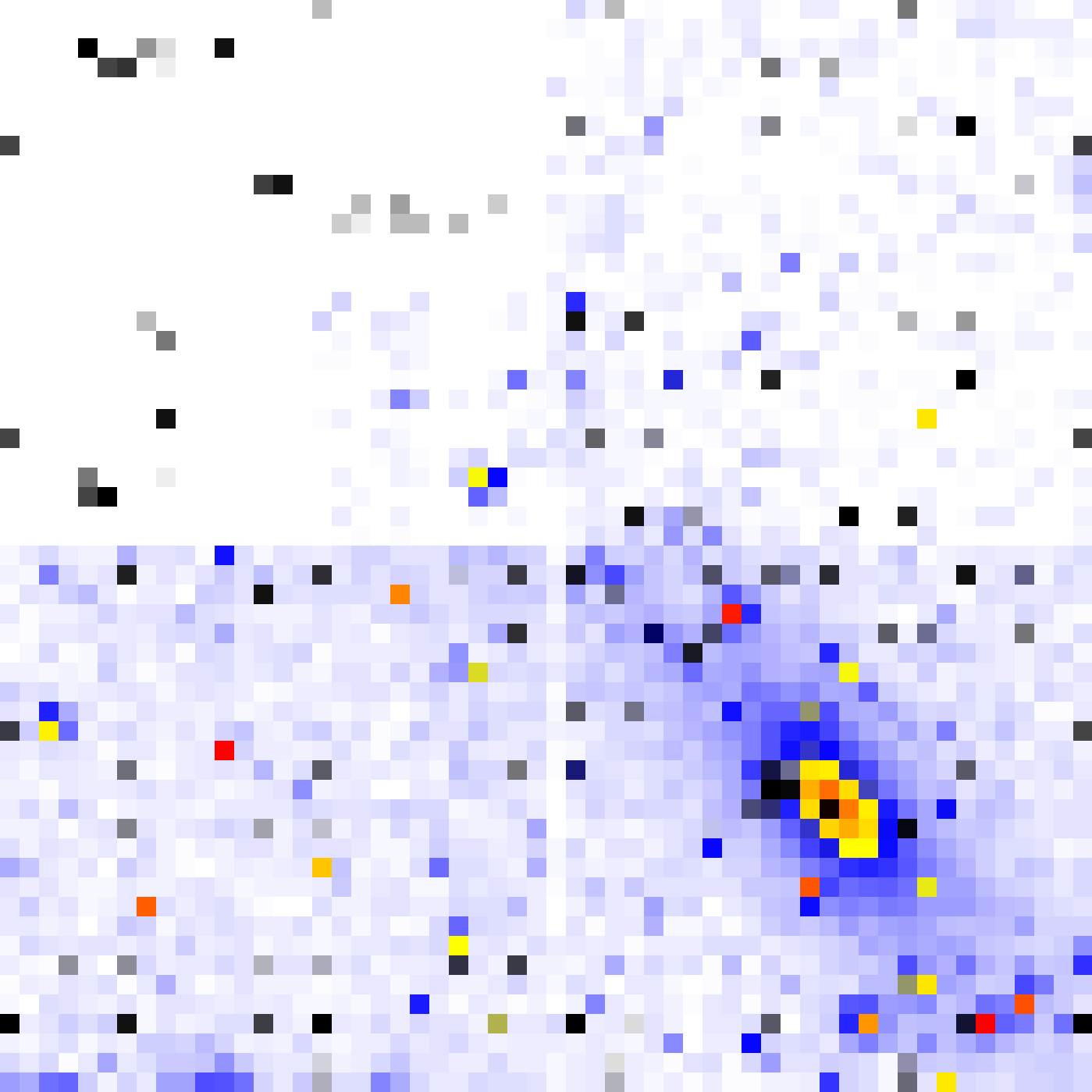,width=55mm}
}
\vspace{3mm}
\centerline{
\psfig{file=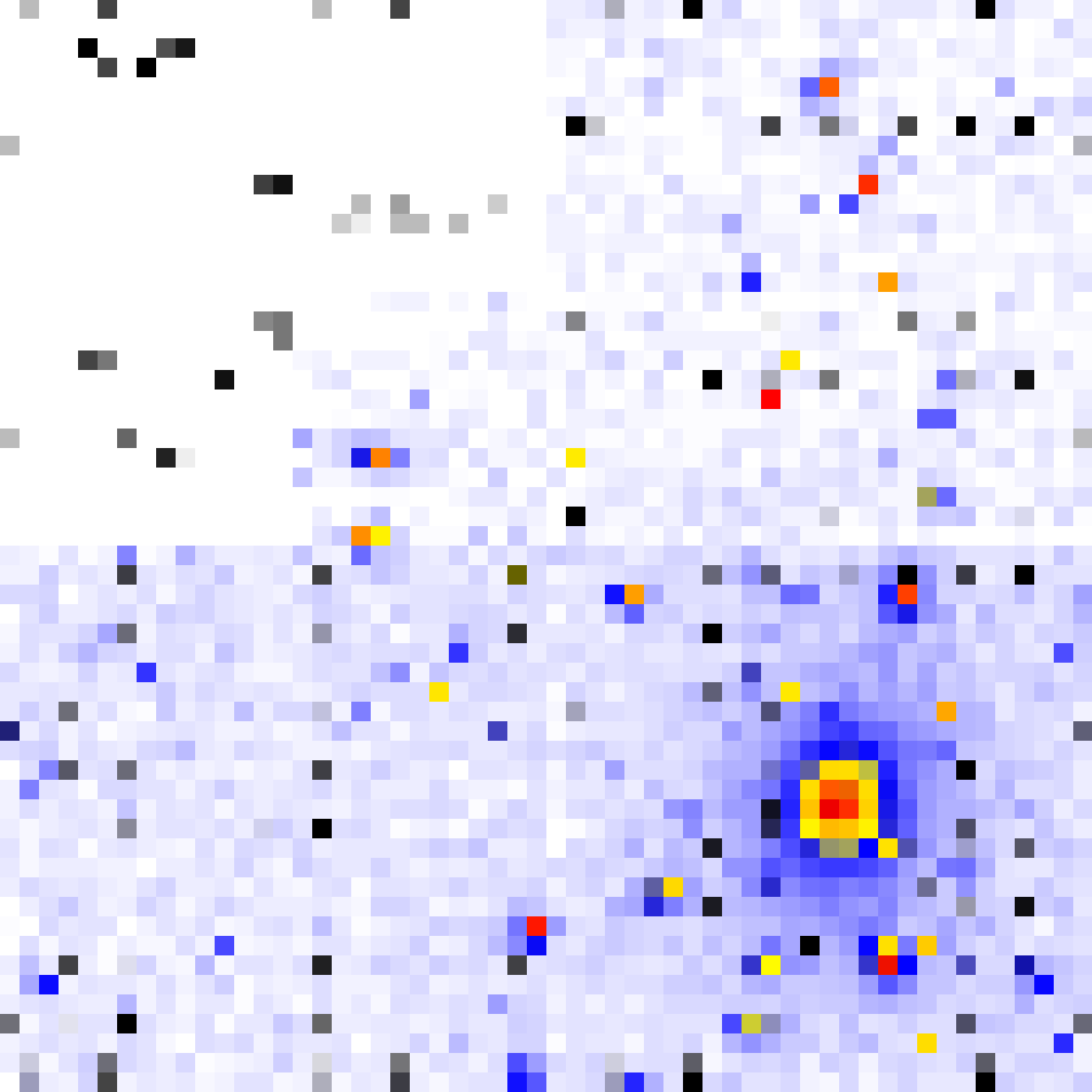,width=55mm}
\hspace{3mm}
\psfig{file=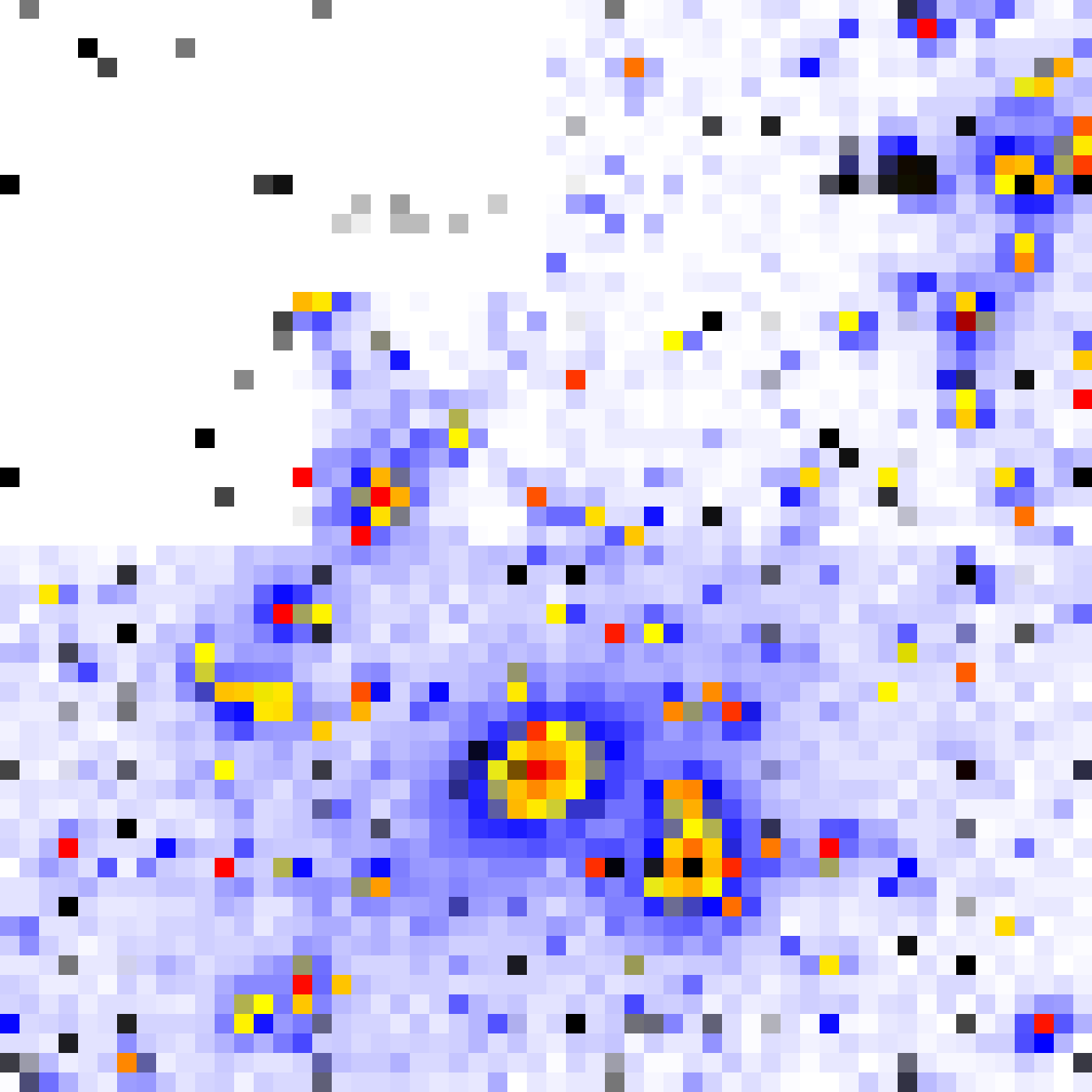,width=55mm}
\hspace{3mm}
\psfig{file=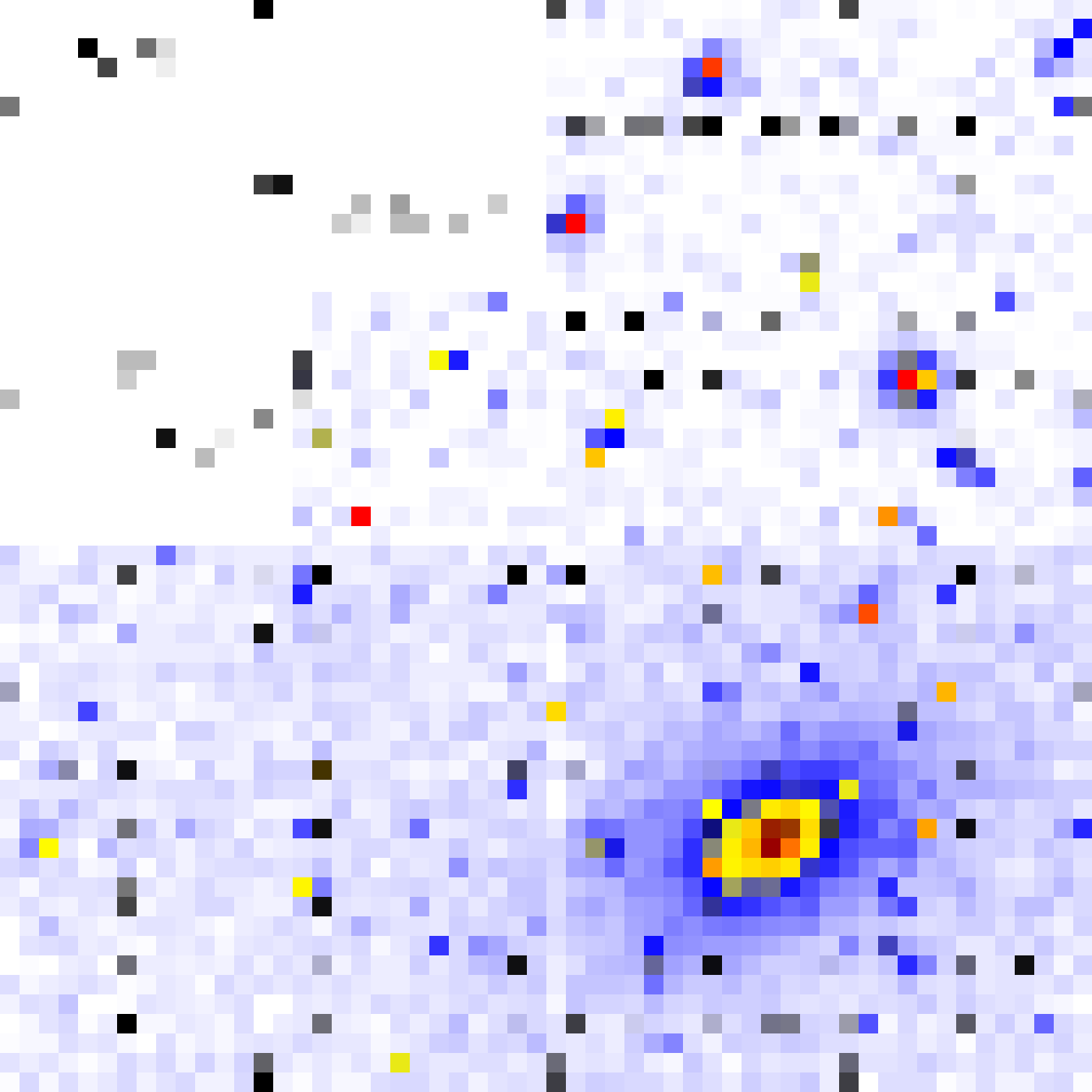,width=55mm}
}
\vspace{3mm}
\centerline{
\psfig{file=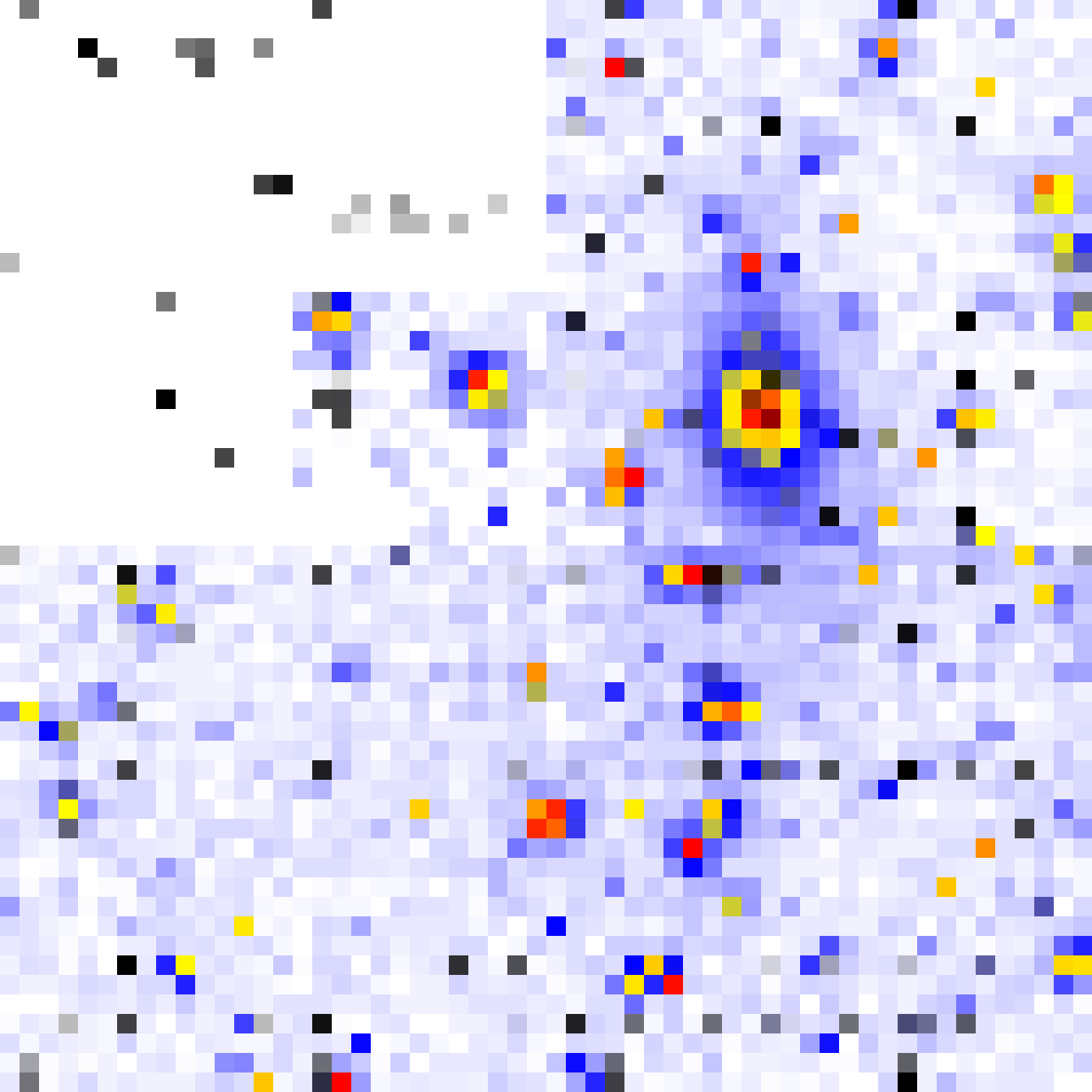,width=55mm}
\hspace{3mm}
\psfig{file=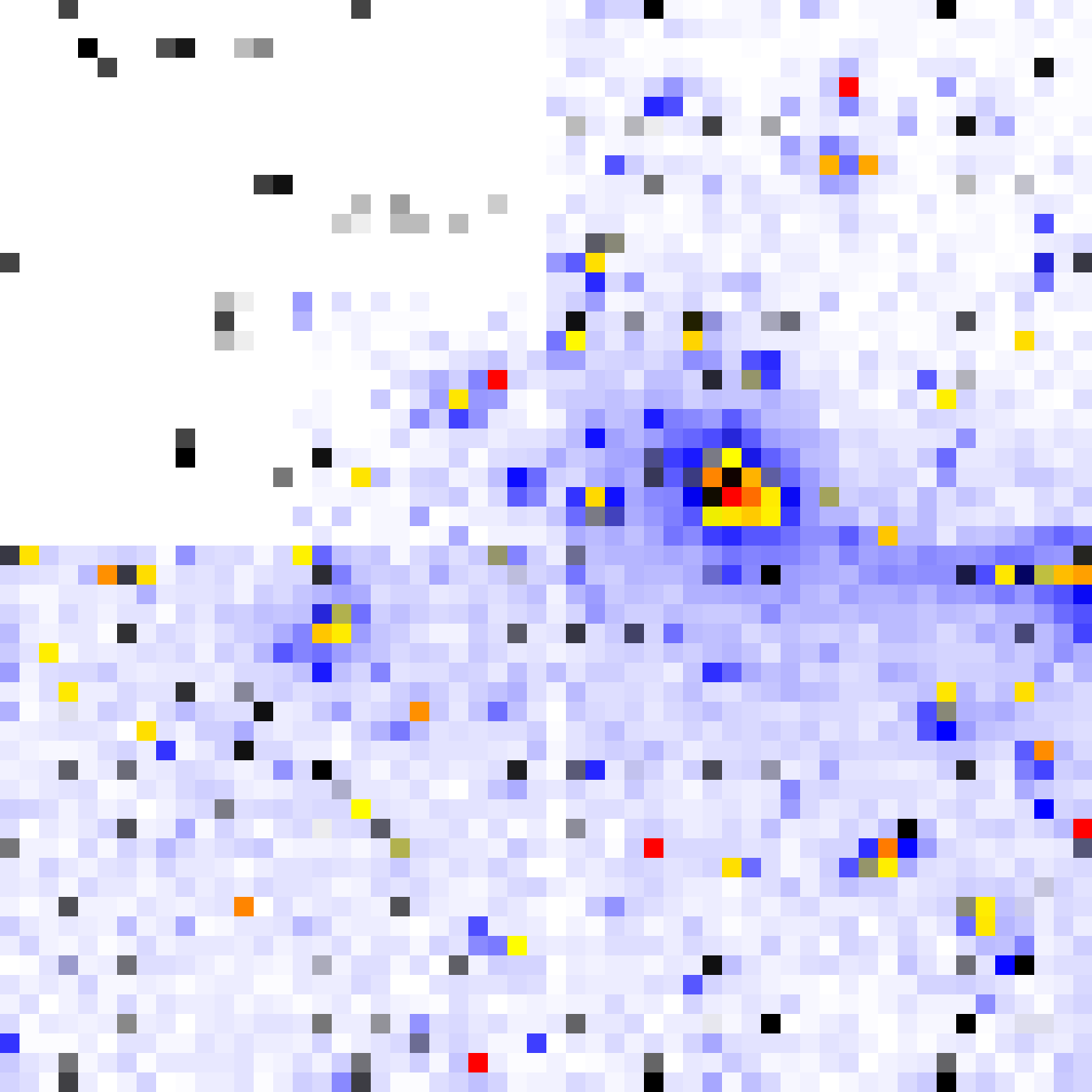,width=55mm}
\hspace{3mm}
\psfig{file=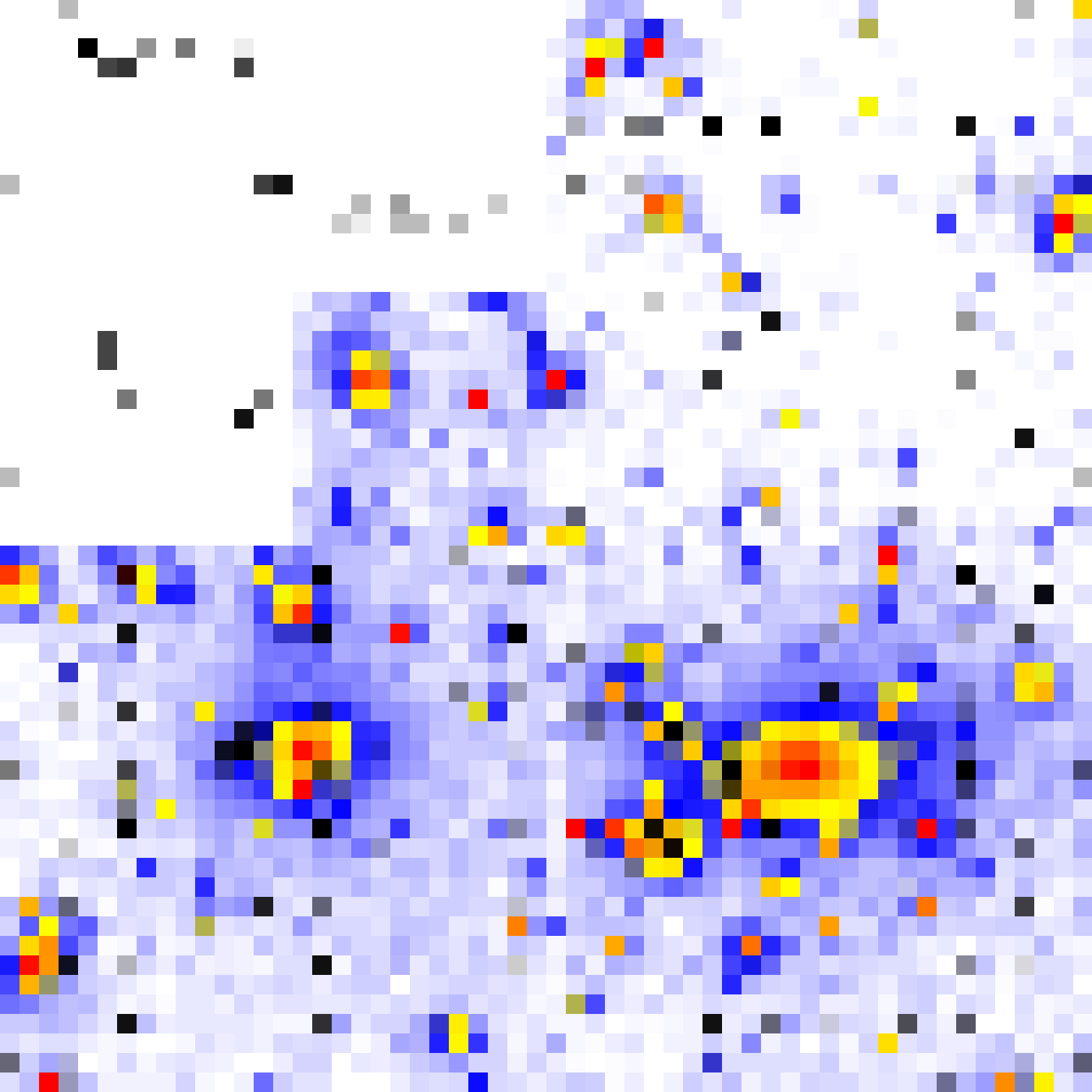,width=55mm}
}
\vspace{3mm}
\centerline{
\psfig{file=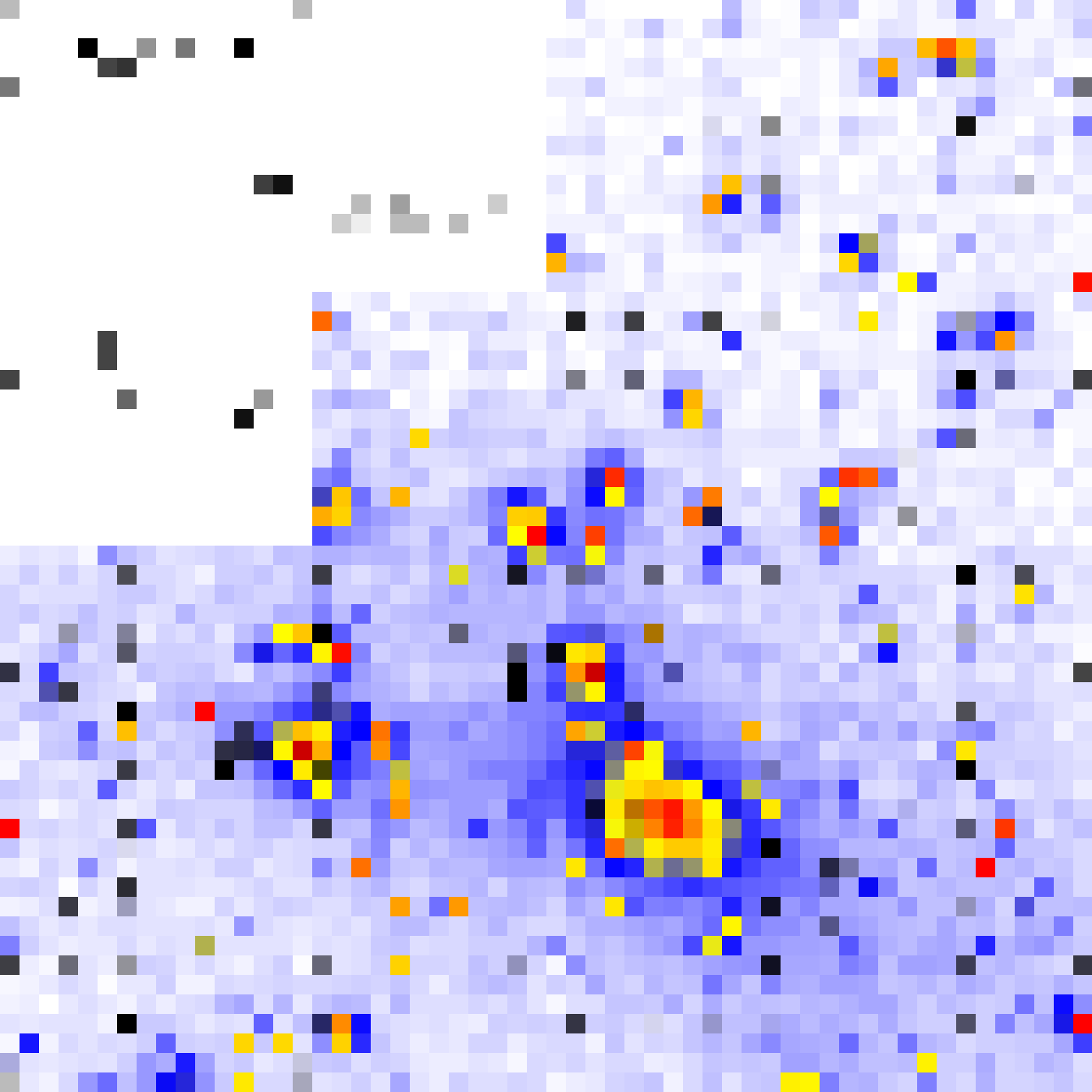,width=55mm}
}
\caption{
The \emph{HST}/WFPC2 frames of all ten clusters on a logarithmic scale
and a false colour table.  We overplot the weak shear field; these
vectors show the mean ellipticity and orientation of faint background
field galaxies on a grid that has been smoothed with a Gaussian of
$\sigma{=}10''$.  The tickmarks are centered on the central galaxy
(Table~\ref{sample}) and are separated by $10''$.
\label{hstdata}
}
\end{figure*}

We measure the actual dither pattern and compare it with the commanded
integer pixel offsets; the median difference between the commanded and
actual offsets is $0.2$\,pixels, and generally lies in the range
${\sim}0$\,--\,$0.4$\,pixels.  The geometrical distortion at the
edge of each chip (Gilmozzi et al.\ 1995; Holtzmann et al.\ 1995;
Trauger et al.\ 1995; Casertano \& Wiggs 2001) translates to an
additional ${\sim}0.2$ pixel shift at the edge of each detector,
falling to zero at the chip centers.  Our observations therefore
sub--sample the $0.1''$ WFC pixels at a level that varies spatially in
the range ${\sim}0$\,--\,$0.5$\,pixels.  We therefore use the {\sc
dither} package (Fruchter \& Hook 2002) to reduce the \emph{HST} data
because this allows us to correct for the geometrical distortion and
to recover a limited amount of spatial information from the
under--sampled WFPC2 point spread function (PSF).  The final reduced
frames (Fig.~\ref{hstdata}) have a pixel scale of $0.05''$ and an
effective resolution of FWHM${=}0.17''$.

\subsection{Identification and Confirmation of Multiple--image
  Candidates}\label{spec} 

The primary reason for observing the cluster cores with \emph{HST} is
to take advantage of the exquisite spatial resolution of these data to
identify multiply--imaged galaxies.  Spectroscopic confirmation of
such systems provides very tight constraints on the absolute mass of
each cluster (roughly an order of magnitude better than is available
from weak--lensing), and the spatial distribution of the cluster mass.

\begin{figure*}
\centerline{
\psfig{file=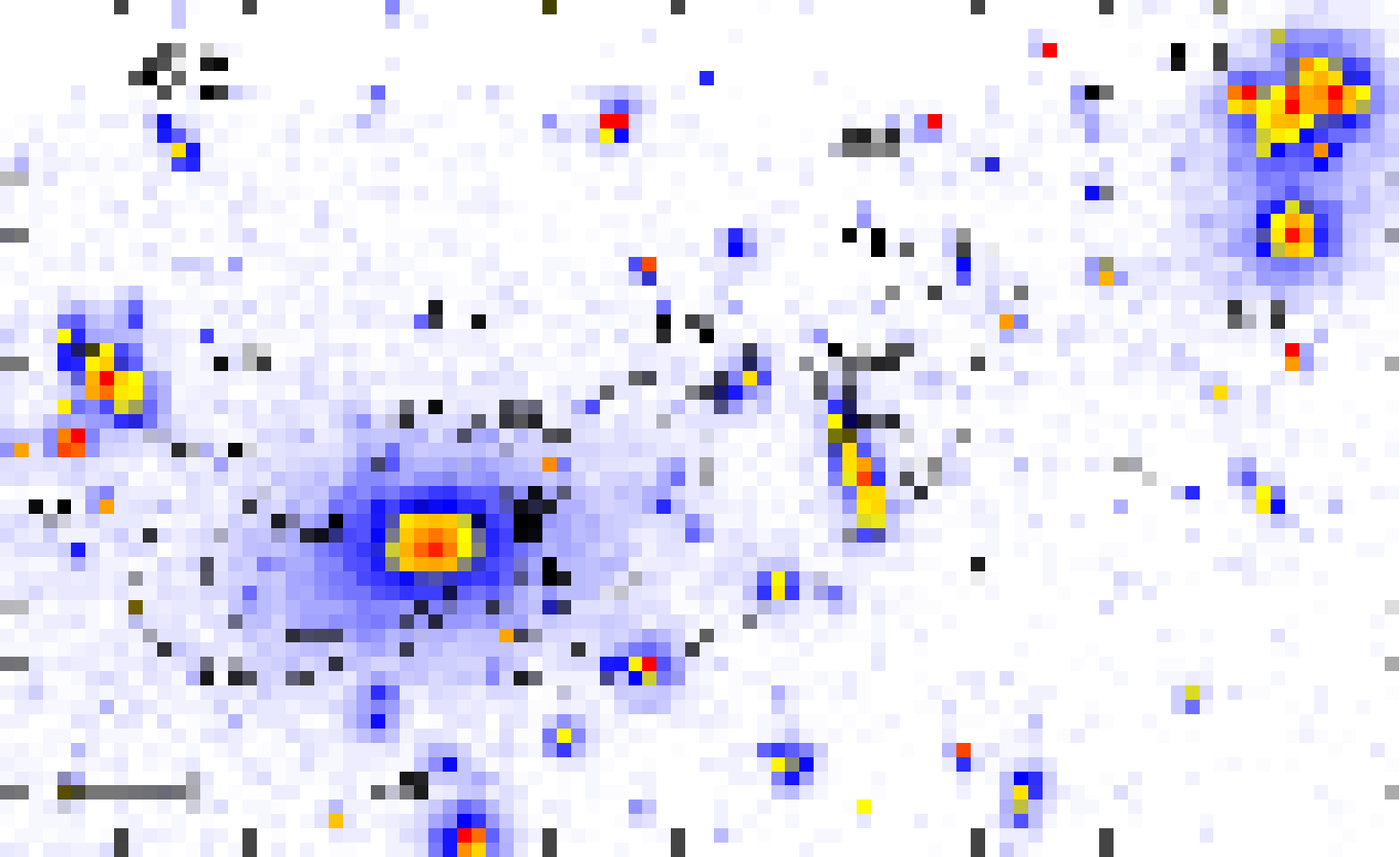,width=140mm}
}
\vspace{3mm}
\centerline{
\psfig{file=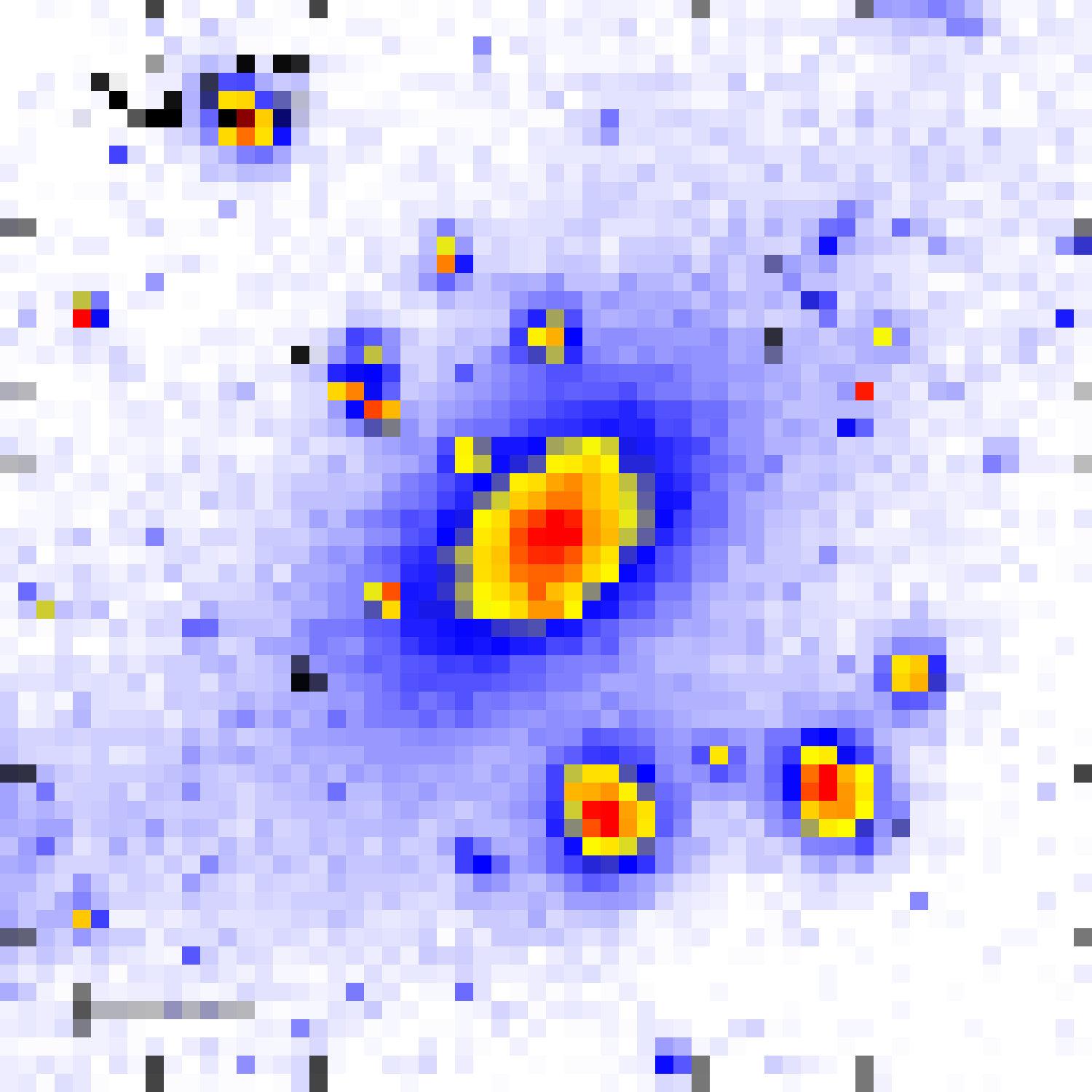,width=57mm}
\hspace{2mm}
\psfig{file=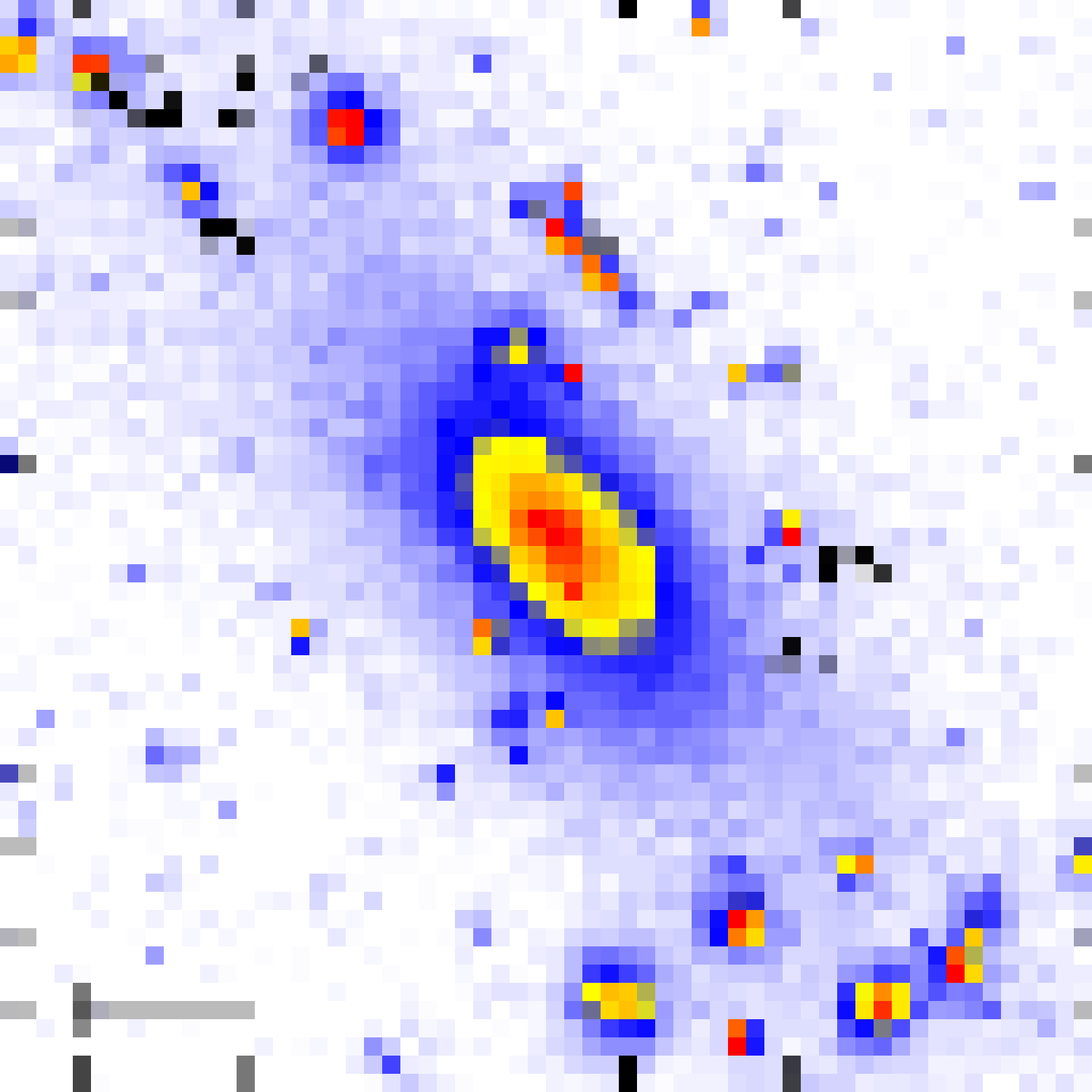,width=57mm}
\hspace{2mm}
\psfig{file=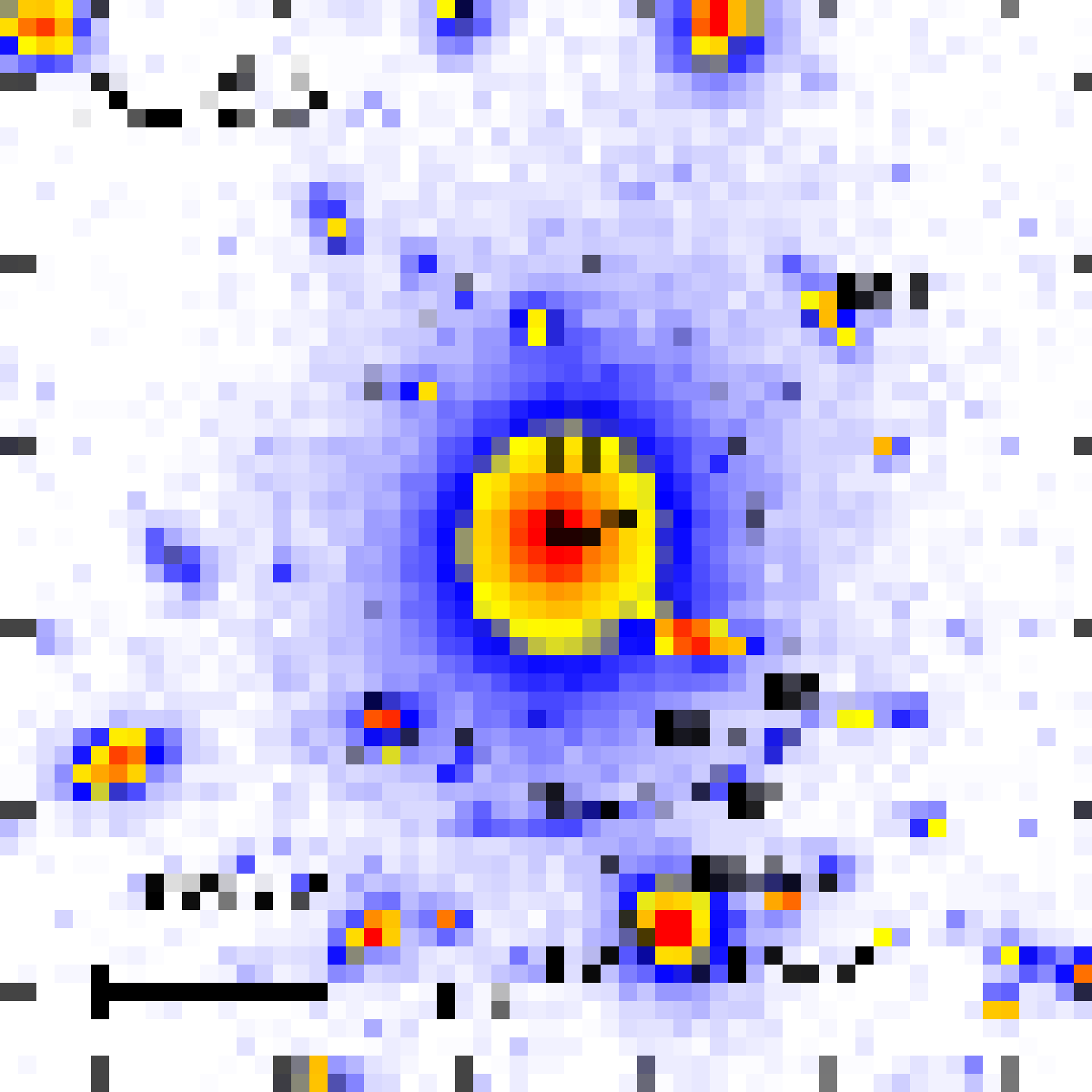,width=57mm}
}
\vspace{3mm}
\centerline{
\psfig{file=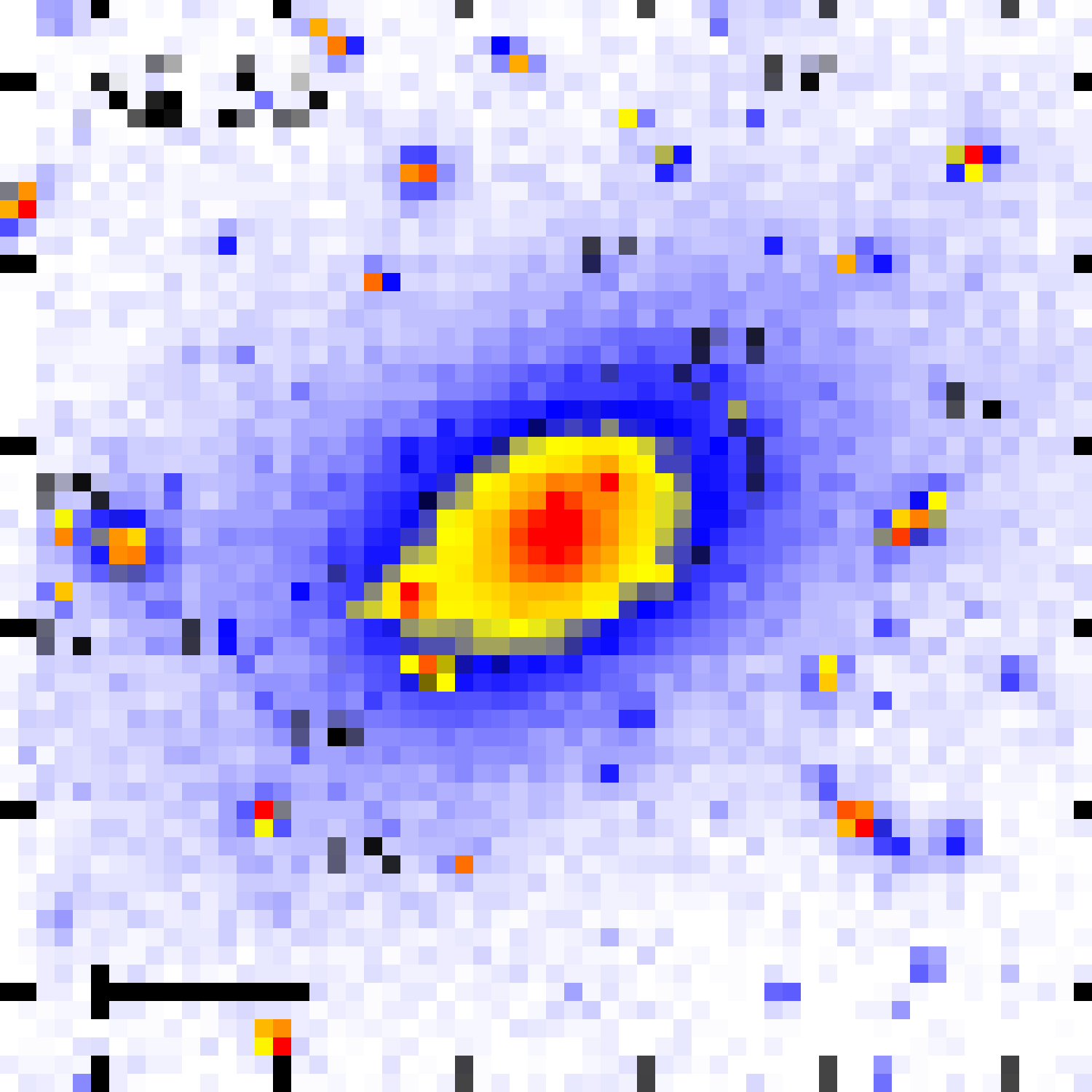,width=57mm}
\hspace{2mm}
\psfig{file=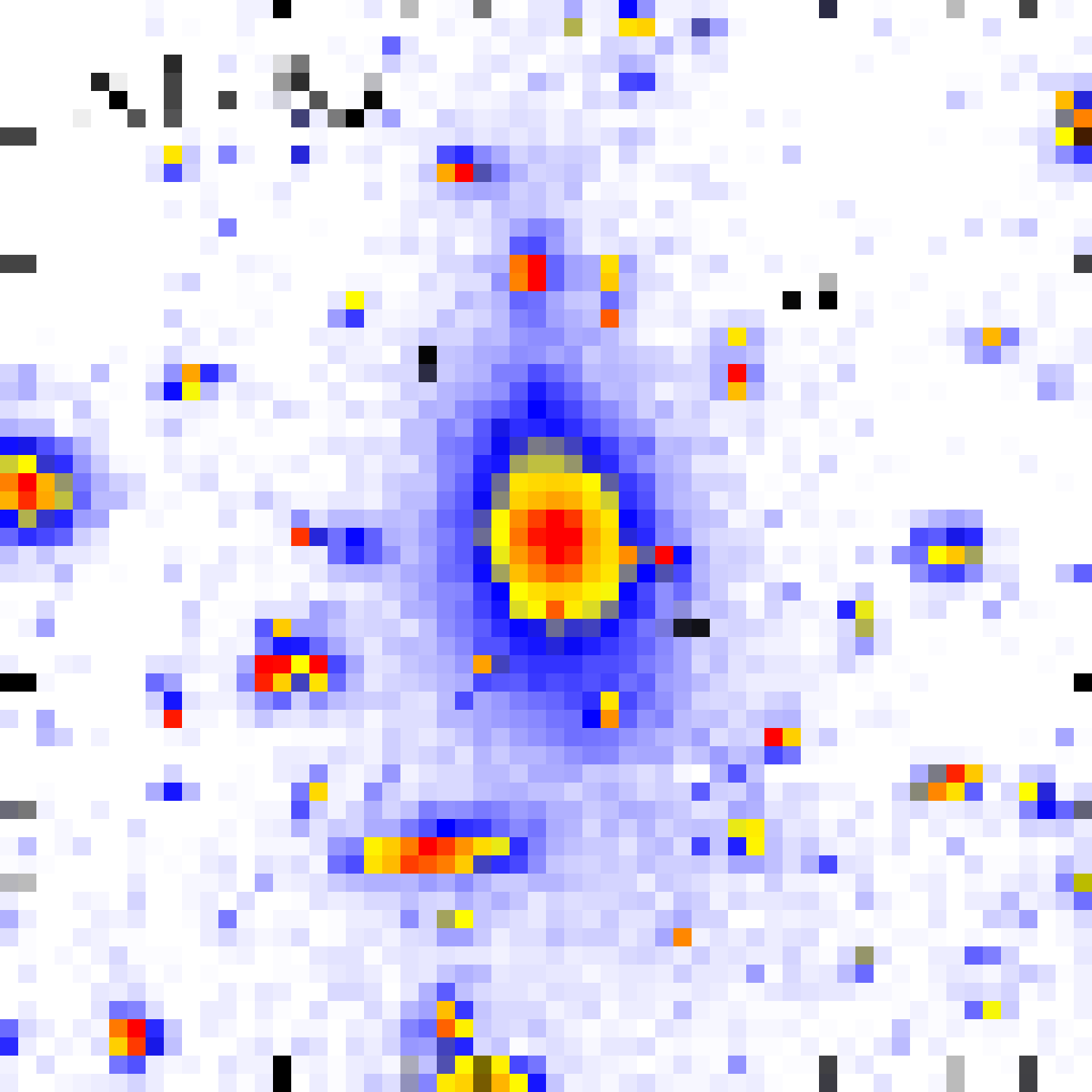,width=57mm}
\hspace{2mm}
\psfig{file=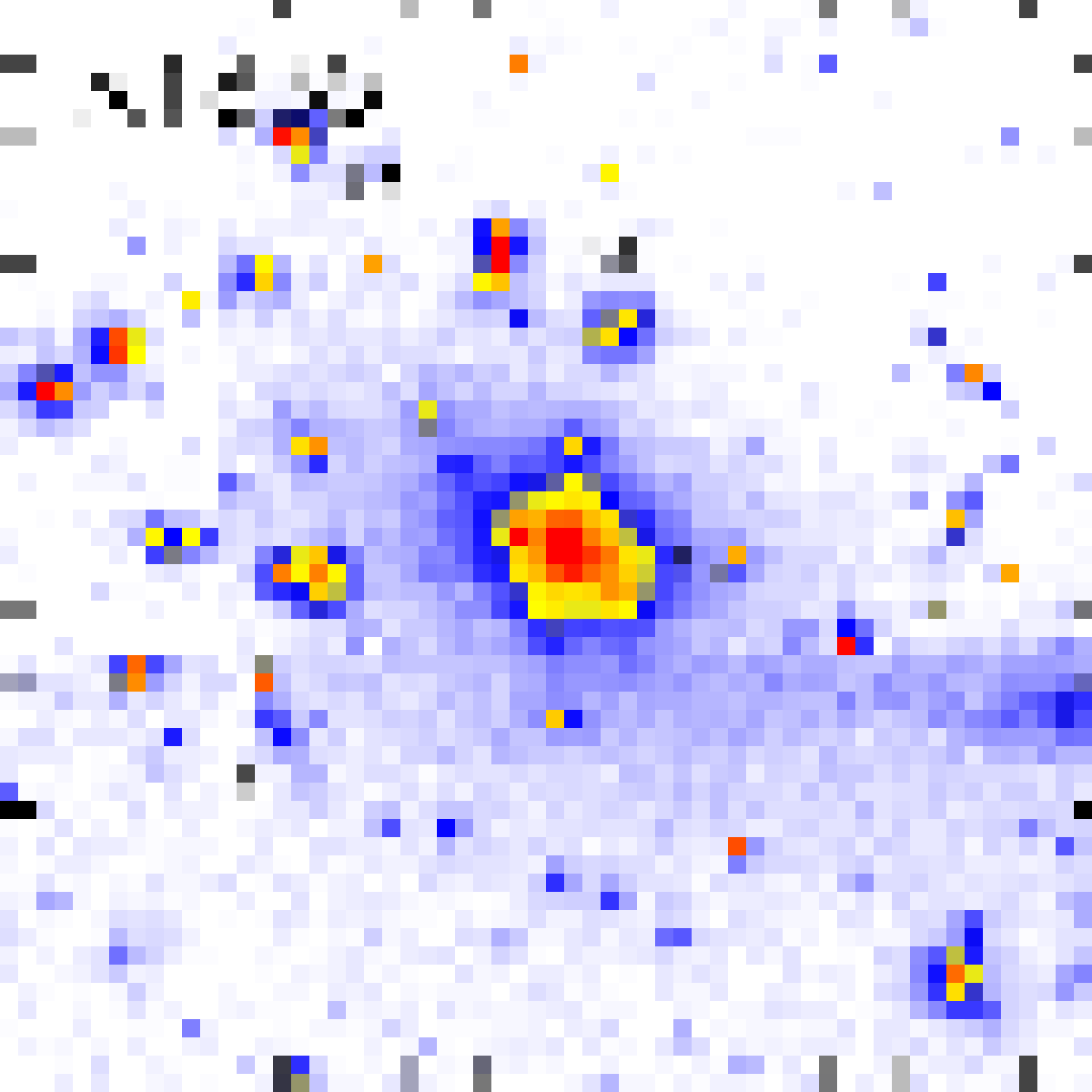,width=57mm}
}
\caption{
Zooms into the central region of seven of the clusters showing the
confirmed and candidate multiple--image systems discussed in the text.
The black curves follow the $z{=}1.60, 5.4$, $z{=}1.01$ and $z{=}0.771$
tangential critical curves for A\,68, A\,383 and A\,963 respectively
as computed from the lens models.  The bar at the bottom left of each
panel shows a physical scale of 50\,kpc.  The orientation of each panel
matches the corresponding panel in Fig.~\ref{hstdata}; tick marks are
spaced at $5''$ intervals.
\label{hstzooms}
}
\end{figure*}

We therefore begin the analysis by searching the \emph{HST} frames for
candidate multiple--image systems.  This search combines visual
inspection of the data (looking for symmetric image pairs and
tangentially or radially distorted arcs) with the SExtractor source
catalogs described in \S\ref{extract}.  The effective surface
brightness limit of the search in regions not affected by bright
cluster galaxies is therefore
$\mu_{702}{\simeq}25$\,mag\,arcsec$^{{-}2}$ (\S\ref{extract}).  To
overcome the influence of bright cluster members, we generated
unsharp--masked versions of the science frames, thus removing most of
the flux from the bright galaxies.  For example this exercise helped
us to identify C19 under the brightest cluster galaxy (BCG) in A\,68
(Fig.~\ref{hstzooms}).  The residual light from the bright galaxies in
these unsharp--masked frames inevitably brightens the
surface--brightness limit of the multiple--image search close to the
cores (central few arcsec) of the subtracted galaxies.  However, the
only images that we expect to find in such locations are strongly
de--amplified counter--images, the brighter images of which should be
easily detectable elsewhere in the frame, if they lie above the
surface brightness detection limit.  We therefore expect the search
for multiple--image systems to be reasonably complete to
$\mu_{702}{\simeq}25$\,mag\,arcsec$^{{-}2}$.

We list the  multiple--image candidates in Table~\ref{mult}, and
mark them in Fig.~\ref{hstzooms}.  Faint sources detected in the
cluster cores are excluded from Table~\ref{mult} if they are not
plausibly multiply--imaged, based on morphological grounds, including
examination of issues relating to the parity of possible counter
images (see Smith 2002 for a more detailed discussion of issues
relating to the identification of multiple--image candidates).  The
clusters are sub--divided in Table~\ref{mult} into those with
spectroscopically confirmed multiple--image systems (top) and those
for which no spectroscopic identifications of genuine multiple--image
systems is yet available (bottom).  Each sub--set of clusters contains
half of the total sample of 10.  Of the five clusters without any
spectroscopically confirmed multiples, two contain convincing
strong--lensing candidates: A\,267 (E2) and A\,1835 (K3).  A firm
lower limit on the fraction of clusters in this sample that contain a
core region with a projected mass density above the critical density
required for strong--lensing is therefore 50\%, although values as
high as ${\sim}$70--80\% are also plausible.

Table~\ref{mult} also lists the spectroscopic redshifts that are
available from other articles in this series (Smith et al.\ 2001,
2002b), and the published literature.  We refer the interested reader
to these articles for the details of the spectroscopy and
multiple--image interpretation.  Note that some of the previously
published multiple--image identifications were based on ground--based
data, and therefore suffered from quite severe uncertainties.  We
discuss in \S\ref{modelling} improvements to the interpretation of the
data that are now possible using the \emph{HST} data presented here.
We also present below new spectroscopic identifications of four
multiple--image candidates, recently obtained with the Keck and Subaru
telescopes.

\begin{table*}
\caption{
Multiple--image Candidates and Spectroscopic Redshifts
\label{mult}
}
{\small
\begin{tabular}{llllll}
\hline
{Cluster} & {Candidate}      & {Redshift$^a$}           & {References}       & {Notes}                                    & {Also known as}               \cr
\hline
\multispan6{\sc\hfil Clusters with Spectroscopically Confirmed Multiple--images [5/10]\hfil}\cr
\hline
A\,68     & C0a/b/c          & $1.60$                   & 1,2                & Triply--imaged ERO                         & EROJ\,003707 \cr
          & C1a/b/c          & $2.6{\pm}0.3$            &                    &                                            & \cr
	  & C2a/b            & $1.5{\pm}0.3$            &                    & \parbox[t]{2.5in}{\addtolength{\baselineskip}{-3pt}Faint image pair; counter--image not detected}                & \cr
          & C4               & $2.625$                  & \S\ref{spec}       & \parbox[t]{2.5in}{\addtolength{\baselineskip}{-3pt}Ly--$\alpha$ in emission.  Singly imaged?}  & \cr
          & C6/C20           & $4{\pm}0.5$              &                    & \parbox[t]{2.5in}{\addtolength{\baselineskip}{-3pt}Pair of images (C6) plus counter image (C20)} &  \cr
          & C8               & $0.861$                  & 3                  & \parbox[t]{2.5in}{\addtolength{\baselineskip}{-3pt}Singly imaged?}                             & \cr
          & C12              & $1.265$                  & 3                  & \parbox[t]{2.5in}{\addtolength{\baselineskip}{-3pt}Singly imaged?}                             & \cr
          & C14              & $0.623$                  & 3                  & \parbox[t]{2.5in}{\addtolength{\baselineskip}{-3pt}Singly imaged?}                             & \cr
          & C15/C16/C17      & $5.4$                    & 4
& \parbox[t]{2.5in}{\addtolength{\baselineskip}{-3pt}Ly--$\alpha$ emitter.} & \cr
          & C19              &                      &                    & \parbox[t]{2.5in}{\addtolength{\baselineskip}{-3pt}Possible radial counter images of part of C0}          & \cr
\hline
A\,383$^b$ & B0/B1/B4        & $1.010$              & 5,6               & \parbox[t]{2.5in}{\addtolength{\baselineskip}{-3pt}Radial and tangential arc system}\cr
           & B2a/b/c/d/e     & $3{\pm}0.5$        & 5                  & \cr
           & B3a/b/c         & $3{\pm}0.5$        & 5                  & \cr
           & B17             & $3{\pm}0.5$        & 5                  & \cr
           & B18             & $0.656$              & 5                  & \cr
\hline
A\,963    & H0               & $0.771$              & 7                  & \parbox[t]{2.5in}{\addtolength{\baselineskip}{-3pt}Three merging images.} & ``Northern'' arc \cr
          & H1/H2/H3         & $1{\pm}0.5$        & 7                    & \parbox[t]{2.5in}{\addtolength{\baselineskip}{-3pt}Group of singly--imaged galaxies?}          & ``Southern'' arc \cr    
          & H6               & $3.269$              & 3                  & \parbox[t]{2.5in}{\addtolength{\baselineskip}{-3pt}Singly--imaged?} \cr
          & H7/H8            & $0.731$              & 3                  & \parbox[t]{2.5in}{\addtolength{\baselineskip}{-3pt}Two singly--imaged galaxies.} \cr
\hline
A\,2218$^c$ & M0a/b/c/d/e    & $0.702$              & 8                  &                                           & \#359/328/337/389 \cr
            & M1a/b/c        & $2.515$              & 9                 &                                           & \#384/468 \cr
            & M2a/b          & $5.576$              & 10                 & \parbox[t]{2.5in}{\addtolength{\baselineskip}{-3pt}Ly--$\alpha$ emitter} \cr
	    & M3a/b/c        & $1.1{\pm}0.1$      & 11                 &                                           & \#444/H6 \cr
            & M4             & $1.034$              & 8,12               &                                           & \#289 \cr
\hline
A\,2219   & P0               & $1.070$              & 13,14,\S\ref{spec} &
          \parbox[t]{2.5in}{\addtolength{\baselineskip}{-3pt}[{\sc oii}] in emission; merging pair of images}         & ${\rm N_{12}}$ \cr
          & P1               &                      &                    & \parbox[t]{2.5in}{\addtolength{\baselineskip}{-3pt}Disk galaxy; edge of disk is counter image of P0}        & ${\rm N_3}$ \cr
          & P2a/b/c          & $2.730$              & 13,14,\S\ref{spec} &                                                         & ${\rm L_{123}}$ \cr
          & P3/P4            & $3.666$               & 14,\S\ref{spec}    &                                                         & A, C\cr
          & P5               &                      & 14,\S\ref{spec}    & \parbox[t]{2.5in}{\addtolength{\baselineskip}{-3pt}Counter image of P3/P4}                                  & B\cr
          & P6/P7/P8         & $2.5{\pm}0.2$      &                    & \parbox[t]{2.5in}{\addtolength{\baselineskip}{-3pt}Faint pair (P6/P7) plus counter--image (P8)}   & \cr
          & P9/P10           & $1.3{\pm}0.2$      &                    & \parbox[t]{2.5in}{\addtolength{\baselineskip}{-3pt}Candidate pair adjacent to P0}                           & \cr
          & P11/P12          & $1.5{\pm}0.3$      &                    & \parbox[t]{2.5in}{\addtolength{\baselineskip}{-3pt}Faint pair}                                              & \cr
\hline
\multispan6{\sc\hfil Clusters with \underline{only} Candidate Multiple--images [5/10]\hfil}\cr
\hline
A\,209    & D0               &                      &                    & \parbox[t]{2.5in}{\addtolength{\baselineskip}{-3pt}Faint arclet -- singly imaged?} \cr
          & D1               &                      &                    & \parbox[t]{2.5in}{\addtolength{\baselineskip}{-3pt}Asymmetric morphology -- singly imaged?} \cr
          & D2               &                      &                    & \parbox[t]{2.5in}{\addtolength{\baselineskip}{-3pt}Disturbed morphology -- singly imaged?} \cr
\hline
A\,267    & E1               & $0.23$               & 15                  & \parbox[t]{2.5in}{\addtolength{\baselineskip}{-3pt}Cluster member} \cr
          & E2a/b            &                      &                    & \parbox[t]{2.5in}{\addtolength{\baselineskip}{-3pt}Faint image pair; counter--image not detected.}\cr
\hline
A\,773    & F0               & $0.650$              & 3                  & \parbox[t]{2.5in}{\addtolength{\baselineskip}{-3pt}Singly--imaged?} \cr
          & F13              & $0.398$              & 3                  & \parbox[t]{2.5in}{\addtolength{\baselineskip}{-3pt}Singly--imaged?} \cr
          & F18              & $0.487$              & 3                  & \parbox[t]{2.5in}{\addtolength{\baselineskip}{-3pt}Singly--imaged?} \cr
          & F19              & $0.425$              & 3                  & \parbox[t]{2.5in}{\addtolength{\baselineskip}{-3pt}Singly--imaged?} \cr
\hline
A\,1763   & J4               &                      & 1                  & \parbox[t]{2.5in}{\addtolength{\baselineskip}{-3pt}Singly imaged or merging pair?}            &  EROJ\,133521{+}4100.4 \cr
\hline
A\,1835   & K0               &                      & 16                  & \parbox[t]{2.5in}{\addtolength{\baselineskip}{-3pt}Radial feature -- associated with BCG?} & A\,1835--B$'$\cr
          & K1               &                      & 16
          & \parbox[t]{2.5in}{\addtolength{\baselineskip}{-3pt}High
          surface brightness arclet -- singly--imaged?}                          & A\,1835--B \cr
          & K2               &                      &                    & \parbox[t]{2.5in}{\addtolength{\baselineskip}{-3pt}Faint arclet -- singly imaged?}                          &  \cr
          & K3               &                      & 16                  & \parbox[t]{2.5in}{\addtolength{\baselineskip}{-3pt}Low surface brightness blue arcs}                        & A\,1835--A \cr
\hline
\end{tabular}
}
\begin{tabular}{l}
\parbox[b]{6.5in}{\footnotesize\addtolength{\baselineskip}{-5pt} 
$^a$ Redshift stated with an error bar are inferred from the lens model of the relevant cluster.

}\\
\parbox[b]{6.5in}{\footnotesize\addtolength{\baselineskip}{-5pt}
$^b$ See Smith et al.\ (2001) for a full list of candidate multiples in A\,383.

}\\
\parbox[b]{6.5in}{\footnotesize\addtolength{\baselineskip}{-5pt}
$^c$ See Kneib et al.\ (1996; 2004, in prep.) for a full list of candidate multiples in A\,2218.

}\\
\parbox[b]{6.5in}{\footnotesize\addtolength{\baselineskip}{-5pt}
\underline{References}

[1] Smith et al.\ (2002a), [2] Smith et al.\ (2002b), [3] Richard et
al.\ (2003, in prep.), [4] Kneib et al.\ (2004a, in prep.), [5] Smith
et al.\ (2001), [6] Sand et al.\ (2004), [7] Ellis et al.\ (1991), [8]
Pell\'o et al.\ (1992), [9] Ebbels et al.\ (1998), [10] Ellis et al.\
(2001), [11] Kneib et al.\ (1996), [12] Swinbank et al.\ (2003), [13]
Smail et al.\ (1995), [14] B\'ezecourt et al.\ (2000), [15] Kneib et
al.\ (2004b, in prep.), [16] Schmidt, Allen \& Fabian (2001),

}\\
\end{tabular}
\end{table*}

\addtocounter{figure}{-1}

\begin{figure*}
\centerline{
\psfig{file=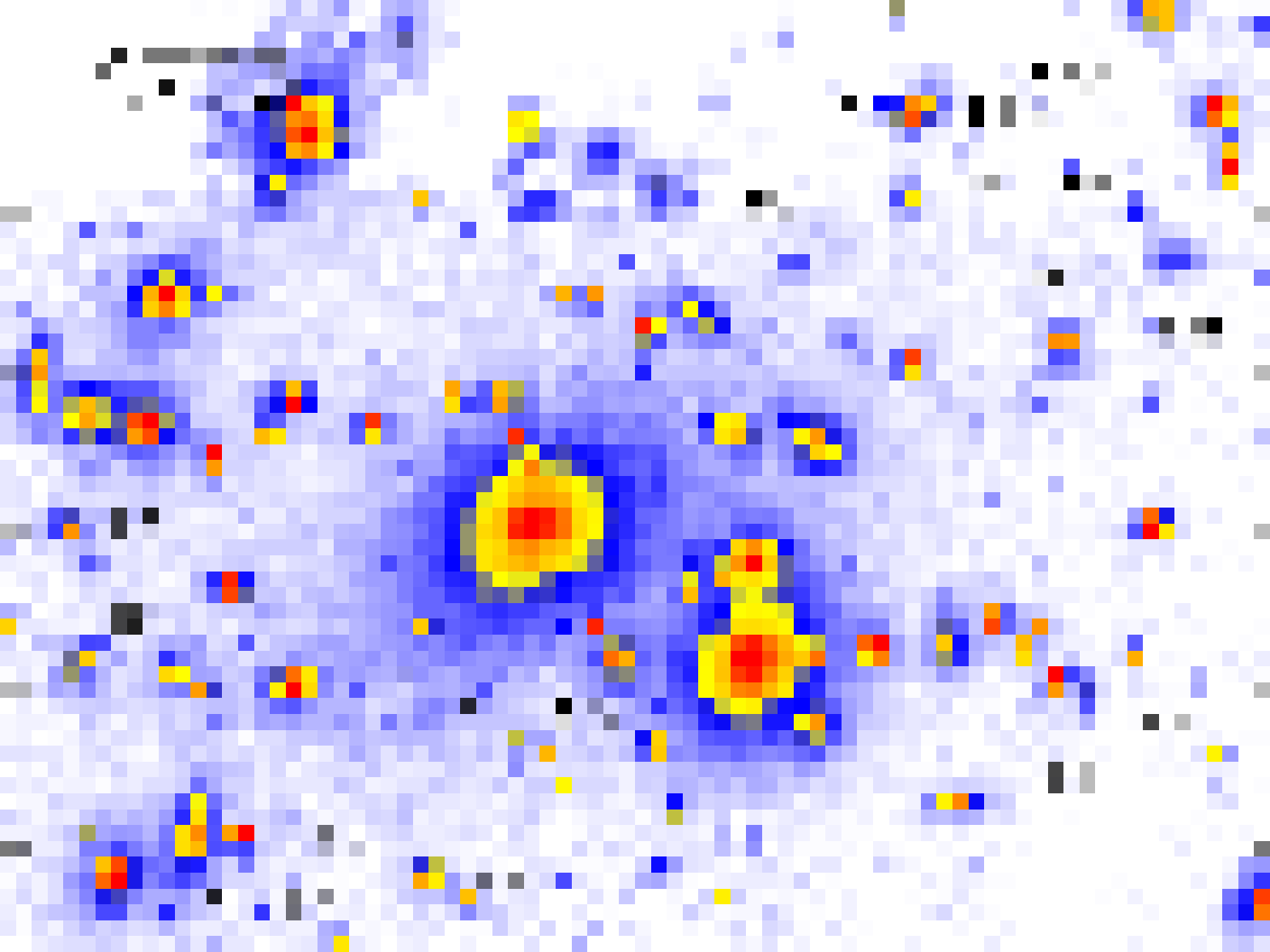,width=110mm}
}
\vspace{2mm}
\centerline{
\psfig{file=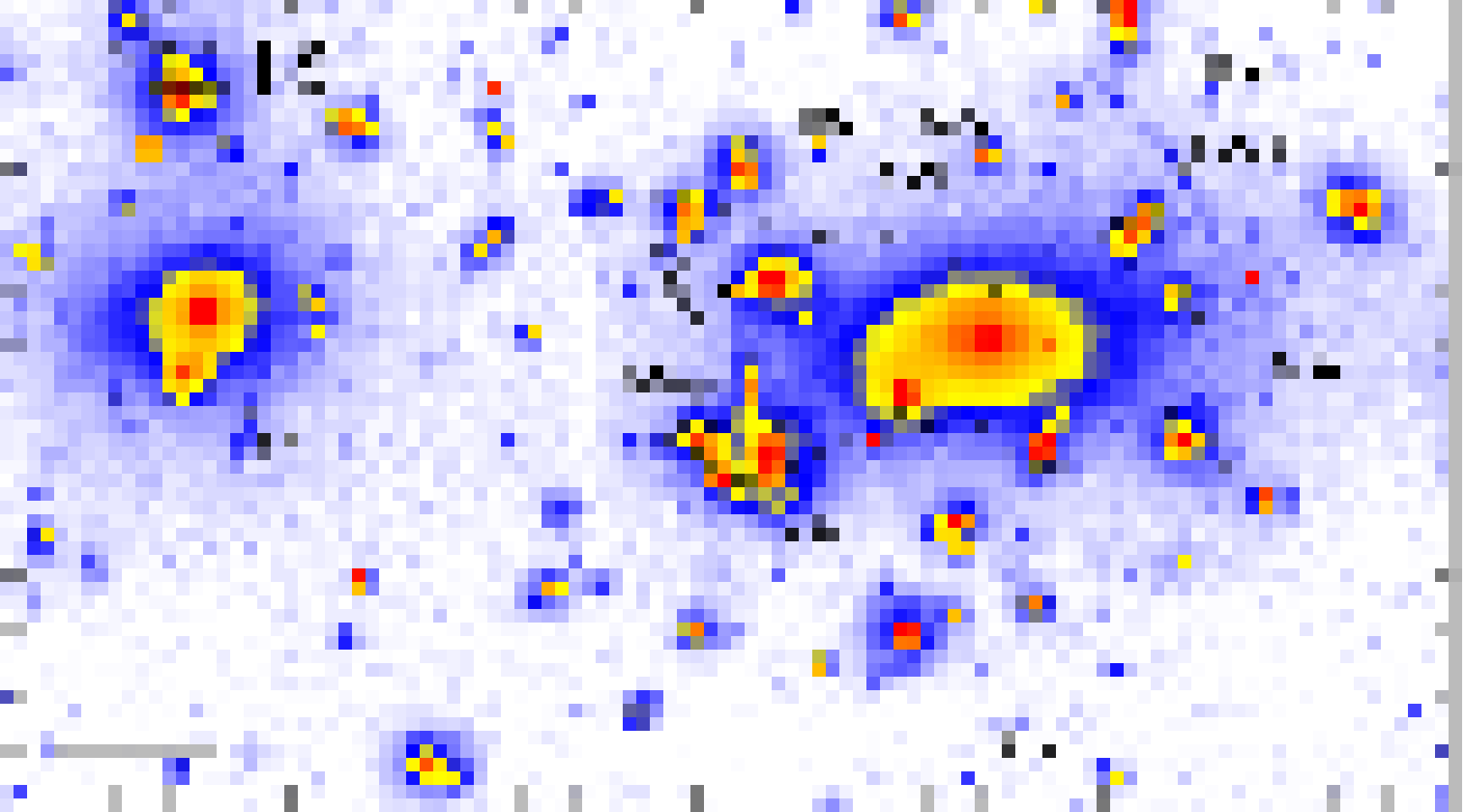,width=110mm}
}
\vspace{2mm}
\centerline{
\psfig{file=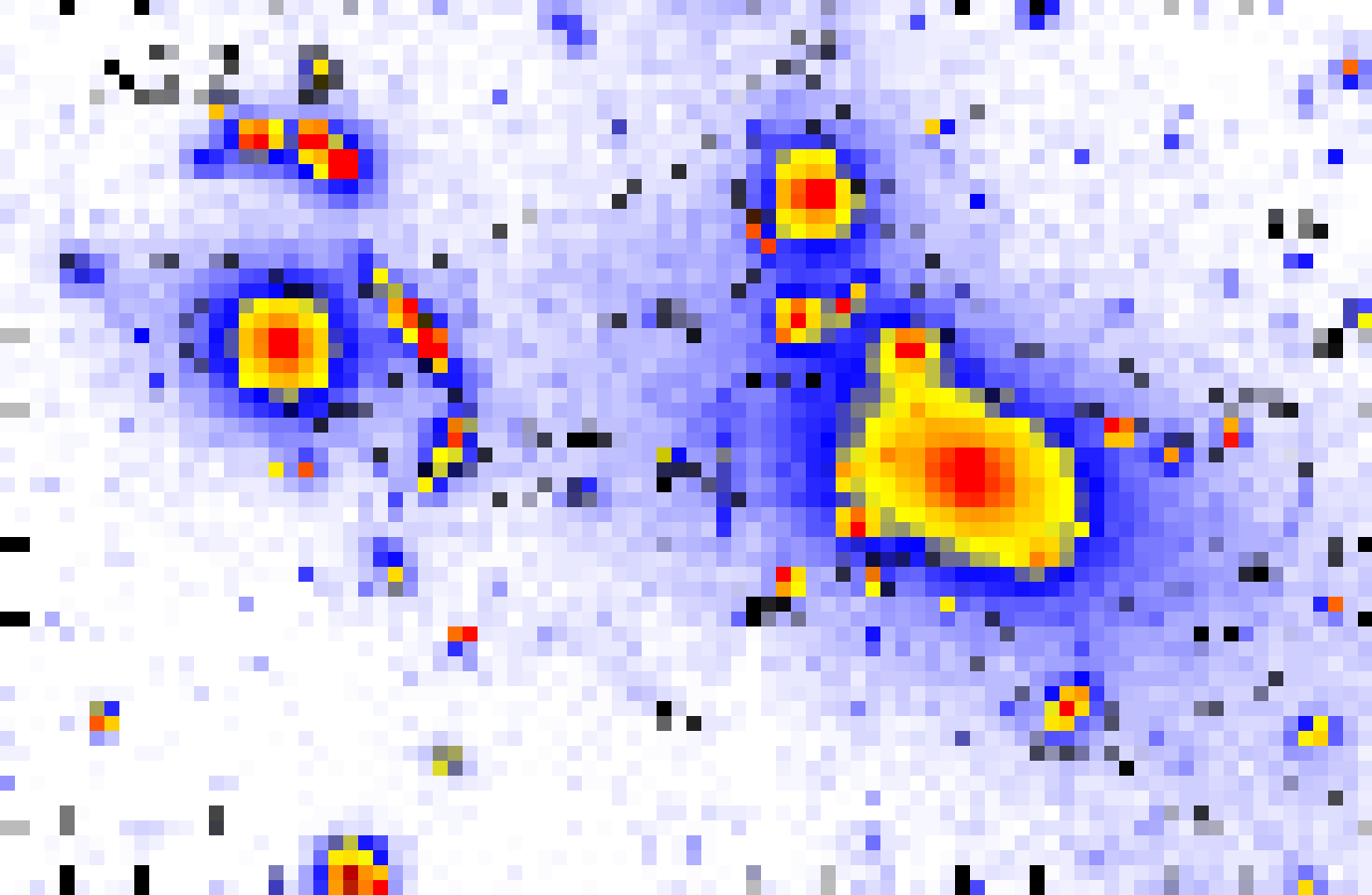,width=110mm}
}
\caption{
[continued]~Zooms into the central region of A\,773, A\,2218 and
A\,2219, showing the 
confirmed and candidate multiple--image systems discussed in the text.
The curves follow the $z{=}0.702$  and $z{=}1.069, 3.666$ tangential
critical curves for A\,2218 and A\,2219 respectively, as computed from
the lens models.  The bar at the bottom left of each
panel shows a physical scale of 50\,kpc.  The
orientation of each panel matches the corresponding panel in
Fig.~\ref{hstdata}; tick marks are spaced at $5''$ intervals.
}
\end{figure*}


\subsubsection{Keck--I/LRIS Observations of A\,68}

\begin{figure}
\centerline{
\psfig{file=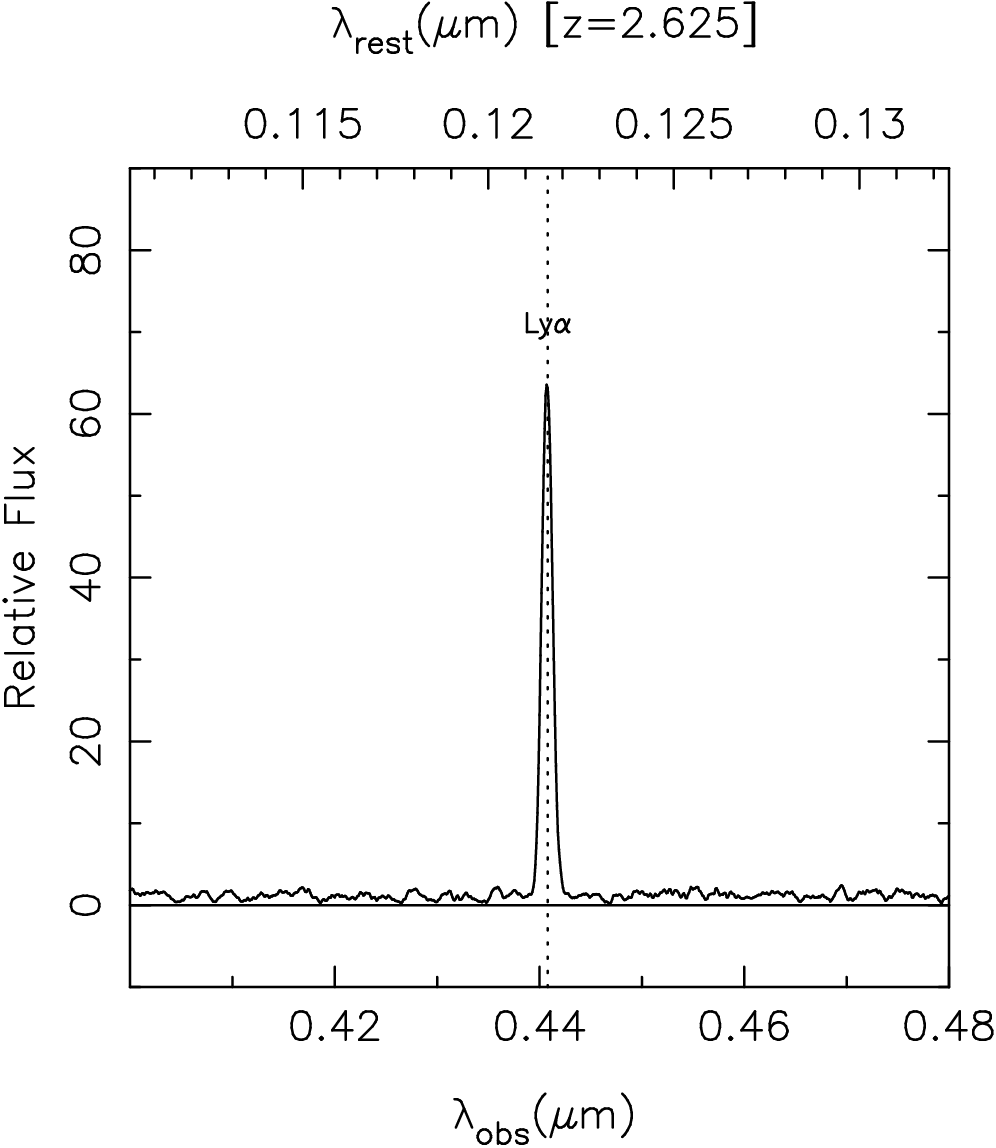,width=78mm,angle=0}
}
\caption{
One--dimensional spectrum of C4 in A\,68, obtained with LRIS.  We
interpret the single strong emission line as Ly--$\alpha$ at
$z{=}2.625$ (see \S\ref{spec} for more details).
\label{a68c4}}
\end{figure}

On November 30, 2002, we conducted deep multi--slit spectroscopy with
the Low Resolution Imager Spectrograph (LRIS; Oke et al 1995) on the
Keck--I telescope\footnote{The W.M.\ Keck Observatory is operated as a
scientific partnership among the California Institute of Technology,
the University of California, and NASA.}, on the cluster A\,68.  The
night had reasonable seeing, ${\sim}0.8''$, but was not photometric
(with some cirrus), thus no spectrophotometric standard stars were
observed.  A\,68 was observed for a total of 7.2\,ks using the D680
dichroic with the 600/7500 grating on the red side and the 400/3400
grism on the blue side.  On the red side the spectral dispersion was
1.28\AA\ pixel$^{-1}$ with a spatial resolution of 0.214\arcsec\
pixel$^{-1}$, and on the blue side, the spectral dispersion was
1.09\AA\ pixel$^{-1}$ with a spatial resolution of 0.135\arcsec\
pixel$^{-1}$ using the blue sensitive 2k${\times}$4k Marconi CCDs.

\begin{figure*}
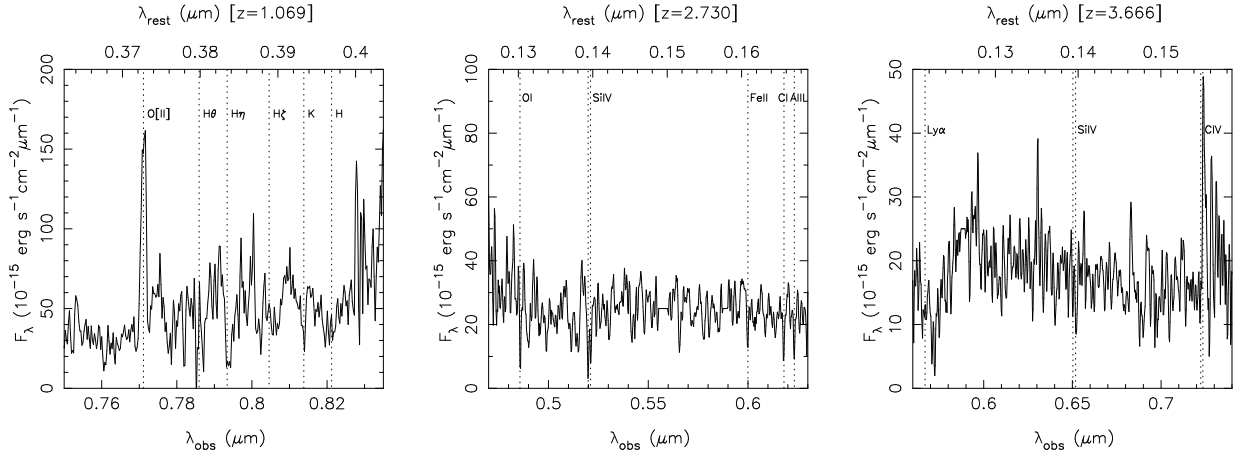

\centerline{
\psfig{file=a2219_zeq1.07b.ps,width=50mm,angle=0}
\hspace{5mm}
\psfig{file=a2219_zeq2.73.ps,width=50mm,angle=0}
\hspace{5mm}
\psfig{file=a2219_zeq3.67.ps,width=50mm,angle=0}
}
\caption{
One--dimensional spectra of P0 (left), P2c (center) and P3 (right)
multiple--image systems in A\,2219, obtained with the FOCAS
spectrograph on Subaru.  The vertical dotted lines in each panel mark
the features used to identify the redshift of each system; the details
of these identifications is discussed in \S\ref{spec}.
\label{a2219spec}}
\end{figure*}

Three multiple--image candidates were targetted in the mask: C0ab, C1c
and C4 (Table~\ref{hstzooms}).  Only C4 has a strong spectral feature --
a single strong emission line at ${\lobs}{=}4404.7$\,\AA\
(Fig.~\ref{a68c4}).  We interpret this line as Ly--$\alpha$
${\lambda}$1216\AA\ at $z{=}2.625$.  The only other possibility would be
[{\sc oii}] at $z{=}0.18$, i.e.\ in front of the cluster.  We consider
this the less likely option given the apparent tangential distortion
of the arclet with respect to the cluster center, and the presence of
C3 and C20 which appear to be lensed galaxies at a similar redshift to
C4 (Fig.~\ref{hstzooms}).  We note however, that it
appears these three arclets are each single images of different
background galaxies.

\subsubsection{Subaru/FOCAS Observations of A\,2219}

On May 29--30, 2001, we conducted deep multi--slit spectroscopy of
A\,2219 with the Faint Object Camera And Spectrograph (FOCAS;
Kashikawa et al 2002) on the Subaru 8.3--m telescope\footnote{Based
on data collected at the Subaru Telescope, which is operated by the
National Astronomical Observatory of Japan.}.  The two nights had
reasonable seeing, ${\rm FWHM}{\simeq}0.8''$, but were not fully
photometric (with some cirrus), nevertheless we obtained a crude flux
calibration using the spectrophotometric standard star ``Wolf1346''.

We observed A\,2219 for a total of 12.6\,ks using the Medium Blue
(300B/mm) grism and the order sorting filter Y47.  We used the MIT
2k$\times$4k CCD detector with a binning factor of 3 in $x$ and 2 in $y$,
this gives us a spectral dispersion of 2.8\AA\ pixel$^{-1}$ and a
spatial resolution of 0.3\arcsec\ pixel$^{-1}$.  Four multiple image
candidates were targetted in the mask: P0 and P2c had previously been
identified by Smail et al.\ (1995) as gravitational arcs, and P3/P4
had been identified by B\'ezecourt et al.\ (2000) as lying at
$z{=}3.6{\pm}0.4$ using photometric redshift techniques.  We list the
results of our observations (see also Figure~\ref{a2219spec}):

\begin{list}{(\roman{fred})}{\usecounter{fred}\setlength{\labelwidth}{5mm}\setlength{\itemindent}{0mm}\setlength{\labelsep}{1.5mm}\setlength{\leftmargin}{6.5mm}\setlength{\itemsep}{2mm}}

\item P0 is identified as a $z{=}1.069{\pm}0.001$ star--forming galaxy
showing a strong [{\sc oii}] $\lambda$3727\AA\ emission plus weak Balmer
and Calcium lines.
\item P2c is identified as a $z{=}2.730{\pm}0.001$ 
galaxy using the following interstellar metal absorption lines: OI
$\lambda$1302.17\AA, SiIV $\lambda$1393.7, 1397.0\AA, FeII
$\lambda$1608.45\AA, CI $\lambda$1656.93\AA, and AlII
$\lambda$1670.79\AA.
\item P3 is identified as a $z{=}3.666{\pm}0.001$ galaxy
using a broad Ly--$\alpha$ absorption feature and the metal absorption
lines OI $\lambda$1302.17\AA, SiIV $\lambda$1393.7, 1397.0\AA, plus
CIV $\lambda$1548.2, 1550.77\AA\ in emission with a broad absorption
feature on the blue side.
\item P4 was also observed, although, it appears that the slit was not
well aligned with the target galaxy, possibly due to a mask--milling
problem.  We do however detect an absorption feature in these data at
the same wavelength as the Ly--$\alpha$ absorption feature in P3.  It
therefore appears that P4 is also at $z{=}3.666$.
\end{list}

We interpret P0 as a pair of merging images straddling the $z{=}1.07$
critical line.  Smail et al.\ (1995) proposed that P1 is the
counter--image of this pair, however B\'ezecourt et al.\ (2000) argued
against P1 because its optical/near--infrared colours are redder than
those of P0.  When constraining the lens model of this cluster with
just this multiple--image system, several alternative counter--images
provided plausible fits.  However when this constraint was combined
with other multiple--image systems, especially P3/P4/P5 which also
lies in the saddle region between the BCG and the group of galaxies to
the South--West, an acceptable model was only possible if P1 is
identified as the counter--image of P0.  The contradiction between
this result and B\'ezecourt et al.'s (2000) photometry is eliminated
by the superb spatial resolution of the \emph{HST} data, because it
reveals that P1 is a disk galaxy, the Southern portion (presumably
part of the disk) of which has a surface--brightness consistent with
that of P0.  This is confirmed by inspection of a colour image of this
field based on Czoske's (2002) $BRI$--band CFH12k imaging of this
cluster, which reveals that the Southern envelope of P1 is also bluer
than the central region, and is consistent with this interpretation.

The spectroscopic identifications of P3 and P4 confirm B\'ezecourt et
al.'s (2000) results.  We identify P5 as the third image of this galaxy.

\subsection{Source Extraction and Analysis}\label{extract}

In addition to the multiple--image constraints described in the
previous section, we also need to construct catalogs of cluster
galaxies and faint, weakly lensed galaxies.  The former are
incorporated into the gravitational lens models (\S\ref{modelling}) as
galaxy--scale perturbations to the overall cluster potential.  The
latter supplement the multiple--image systems to further constrain the
parameters of the lens models.  

\begin{figure*}
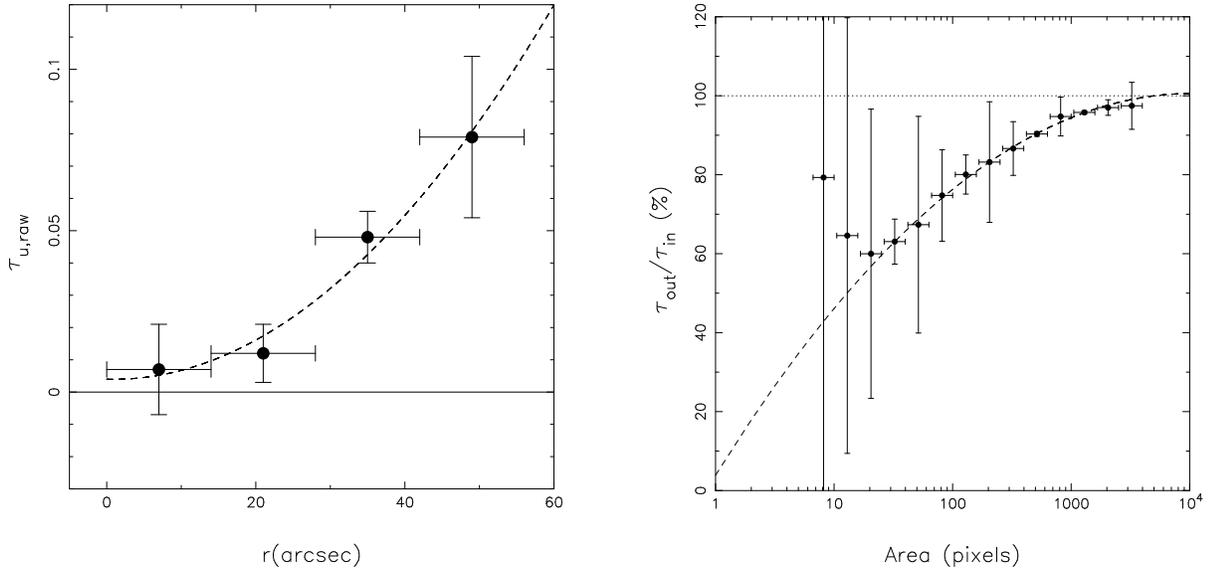

\centerline{
\psfig{file=tautanuncor.ps,width=75mm,angle=-90}
\hspace{10mm}
\psfig{file=tauarea.ps,width=75mm,angle=-90}
}
\caption{
{\sc left} --- Mean tangential shear, $\tau_{\rm u,raw}$, of the
high signal--to--noise star sample as a function of radial distance from
the stacked chip center.  The dashed line shows the polynomial fit to
these data that we used to remove the PSF anisotropy.  {\sc right} ---
The results of Monte Carlo simulations to test how accurately we can
measure the ellipticity of faint galaxies.  The measured ellipticity
as a fraction of the true ellipticity is well--fitted by a second--order
polynomial function (dashed curve).  When the galaxy image is smaller
than ${\sim}30$ pixels, the shape measurements are overwhelmed by the
PSF and the effects of pixelization.  We therefore cut our faint
galaxy catalogs at a ``faint'' limit of area${\ge}30$\,pixels.  We
also use the polynomial function plotted in this figure to correct the
observed ellipticities to intrinsic ellipticities.
}
\label{psfcorr}
\end{figure*}

The first step toward the cluster and background galaxy catalogs is to
analyze the \emph{HST} frames using SExtractor (Bertin \& Arnouts
1996).  We selected all objects with isophotal areas in excess of $7$
pixels ($0.017$\,arcsec$^{2}$) at the
$\mu_{702}{=}25.2$\,mag\,arcsec$^{{-}2}$ isophote
(1.5--$\sigma$/pixel).  All detections centroided within $3''$ of the
edge of the field of view, and within regions affected by diffraction
spikes associated with bright stars are discarded, leaving a total of
8,730 ``good'' detections, of which 193 are classified as stars.  We
estimate from the roll--over in the number counts at faint limits, and
Monte Carlo simulations of our ability to recover artificial faint
test sources with SExtractor, that the 80\% completeness limit of the
\emph{HST} frames is $R_{702}{\simeq}26$.  We also use a simple model
that combines the behaviour of deep $R$--band field galaxy counts
(e.g.\ Smail et al.\ 1995b; Hogg et al.\ 1997) with a composite
cluster luminosity function with a faint end slope of ${\alpha}{=}1$
(e.g.\ Adami et al.\ 2000; Goto et al.\ 2002; De~Propris et al.\ 2003)
to determine at what magnitude limit to divide the source catalogs
into bright and faint sub--samples.  We adopt ${\R702}{=}22$, which
corresponds to 3.5\,mags fainter than an $L^\star$ galaxy at the
cluster redshift.  We estimate conservatively that 20\% of the sources
fainter than this limit may be cluster galaxies, thus contaminating
the sample used for the weak--lensing constraints.  In \S\ref{tests}
we verify that this contamination has a negligible effect on our
results.

\subsubsection{Cluster Galaxies}
\label{clus}

The mass of the galaxy--scale mass components in the lens models
generally scale with their luminosity (see Appendix for details).  We
therefore apply two corrections to the $R_{702}$--band {\sc mag\_best}
magnitudes of the bright galaxies (\S\ref{extract}) to obtain robust
measurements of the luminosities of the cluster galaxies.

Balogh et al.\ (2002) fitted parametrized bulge and disk surface
brightness profiles using {\sc gim2d} (Simard 1998) to the cluster
galaxies in this sample.  We compare the SExtractor {\sc mag\_best}
values in our bright galaxy catalogs (\S\ref{extract}) with Balogh et
al.'s surface photometry of the same galaxies, finding that {\sc
mag\_best} is fainter than the corresponding {\sc gim2d} magnitude.
Typically ${\Delta}{\R702}{\sim}0.1$--0.2, increasing to
${\Delta}{\R702}{\sim}0.5$--1.5 for the brightest cluster members
including the BCGs.  These differences arise because SExtractor
over--estimates the sky background for the brighter and thus larger
cluster galaxies, because as the size of these galaxies approaches the
mesh size used for constructing the local background map, source flux
is absorbed into the background.  A second problem occurs in crowded
cluster cores.  When a smaller galaxy is de--blended from a brighter
galaxy, SExtractor often incorrectly associates pixels from the
brighter galaxy with the fainter, thus over--estimating the flux from
fainter and under--estimating the flux from brighter galaxies.  We
therefore adopt Balogh et al.'s surface photometry as the total
$R_{702}$--band magnitudes of the cluster galaxies.

Optical photometry is more sensitive to ongoing star formation in
cluster galaxies than NIR photometry.  To gain a more reliable measure
of stellar mass in the cluster galaxies we therefore exploit $K$--band
imaging of the cluster fields, obtained as part of our search for
gravitationally--lensed EROs (Smith et al.\ 2002a), to convert the
total $R_{702}$--band magnitudes to total $K$--band magnitudes.  We
subtract the $2''$ aperture $({\R702}{-}K)$ colours of the cluster
galaxies measured by Smith et al.\ (2002a) from the total
$R_{702}$--band magnitudes to obtain total $K$--band magnitudes.
Finally, we convert these magnitudes to rest--frame $K$--band
luminosities, adopting $M_K^\star{=}({-}23.38{\pm}0.03){+}5{\rm
log}\,h$ (Cole et al.\ 2001) and $M_{K\odot}{=}3.39$ (Johnson 1966;
Allen 1973) and estimating $K$--corrections from Mannucci et al.\
(2001).  Summing in quadrature all of the uncertainties arising from
these conversions, we estimate that the $K$--band luminosity of an
$L^\star$ galaxy is good to $10\%$.

\subsubsection{Faint Galaxies}
\label{faint}

In this section we develop and apply several corrections to recover
robust shape measurements of the faint galaxies for use as
weak--lensing constraints on the cluster mass distributions.  The goal
of these corrections is to remove any artificial enhancement or
suppression of image ellipticities from the faint galaxy catalogs.
Such effects arise from the geometry of the focal plane, isotropic and
anisotropic components of the PSF and pixelization of faint galaxy
images.  A correction for the geometric distortion of the focal plane
was applied in the data reduction pipeline using Trauger et al.'s
(1995) polynomial solution (\S\ref{hst}).  We deal with the 
remaining issues in turn below.

The \emph{HST}/WFPC2 PSF varies spatially and temporally at the
${\sim}10\%$ and ${\sim}2\%$ levels respectively (Hoekstra et al.\
1998; Rhodes et al.\ 2000).  We ignore the temporal component because,
as we demonstrate below using simulations, point source ellipticities
of ${\sim}2\%$ are comparable with the noise on the shape
measurements.  We examine the spatial variation of the PSF in the ten
WFPC2 frames, however each field contains just ${\sim}6$ suitable
isolated, high signal--to--noise, unsaturated stars.  We therefore
exploit archival \emph{HST}/WFPC2 observations of a further eight low
luminosity clusters (${\lx}{\le}10^{44}[0.1$--$2.4\,{\rm
keV}]{\ergs}$) at $z{\simeq}0.25$ (Cl\,0818${+}$56, Cl\,0819${+}$70,
Cl\,0841${+}$70, Cl\,0849${+}$37, Cl\,1309${+}$32, Cl\,1444${+}$63,
Cl\,1701${+}$64 and Cl\,1702${+}$64) that were observed in an
identical manner to our Cycle 8 observations (Balogh et al.\ 2002).
These data were processed identically to the primary science data and
stringent selection criteria applied to the combined dataset to
construct a sample of 103 stars from which to derive a PSF correction.

The PSFs of these stars are tangentially distorted with respect to the
center of each WFC chip with the magnitude of the distortion
increasing with distance from each chip center.  The tangential shear
at the edge of each chip is ${\sim}5$--10\%, falling to ${\ls}1$--2\%
at each chip centre, and the variation in distortion pattern between
the chips is negligible, in agreement with Hoekstra et al.\ (1998) and
Rhodes et al.\ (2000).  We therefore stack the three chips to derive a
global solution by fitting a second order polynomial to the tangential
shear as a function of distance from the chip center
(Fig.~\ref{psfcorr}).  After applying this correction to the entire
star sample from all 18 cluster fields, the median tangential shear of
the stars is reduced to the same level as the radial stellar shear
i.e.\ ${\ls}1$--2\%.  We interpret the residuals as random noise, and
test this hypothesis using Monte Carlo simulations.  We insert $10^4$
stellar profiles that are sheared in small increments between zero and
10\% into random blank--sky positions in our science frames.  We then
run the same SExtractor detection algorithm as described in
\S\ref{extract} on these frames.  These simulations reveal that the
position angle of a nearly circular stellar source can only be
measured to ${\ls}10$\% accuracy if the ellipticity of the source is
${\gs}2.5\%$, thus confirming our interpretation of the residuals.  We
use the results of this analysis to correct the shape measurements in
the faint galaxy catalogs.  The corrections all result in a change of
${\ls}0.02$ in the final weak shear constraints listed in
Table~\ref{constr} -- i.e.\ smaller than or comparable with the
statistical uncertainties.

We also use Monte Carlo simulations to estimate the minimum number of
contiguous pixels required for a reliable shape measurement
(${\ls}10$\% uncertainty).  We first select a sample of relatively
bright ($R_{702}{\sim}19$--21) background galaxies to use as test
objects, ensuring that these galaxies cover the observed range of
ellipticities in deep field galaxy surveys (e.g.\ Ebbels 1998).  We
scale and insert these test objects into random blank--sky positions
in the science frames and attempt to detect them by running SExtractor
in the same configuration as used in \S\ref{extract}.  We perform
${\sim}10^6$ realizations spanning the full range of expected apparent
magnitudes and scale sizes of faint galaxies (e.g.\ Smail et al.\
1995b).  The measured ellipticity declines markedly as a fraction of
the input ellipticity for sources with areas smaller than
${\sim}1,000$ pixels.  Also, the smallest galaxy area for which the
uncertainty in its shape measurement is ${\ls}10$\% is ${\sim}30$
pixels.  This limit represents, for the expected ellipticity
distribution ($0{\le}{\tau}{\ls}1.5$ -- Ebbels 1998), the minimum
galaxy size for which both the minor and major axes are resolved by
\emph{HST}/WFPC2.  We therefore adopt 30 contiguous pixels as the
``faint'' limit of our background galaxy catalogs.  We also fit a
second order polynomial to the simulation results in the range
$30{\le}{\rm area}{\le}10^3$ pixels (Fig.~\ref{psfcorr}), and use this
recovery function to correct the measured ellipticity of each source
in the faint galaxy catalogues.

\subsection{Summary of Lens Model Constraints}
\label{constraints}

\begin{table}
\caption{Summary of Model Constraints
\label{constr}
}
{\small
\begin{tabular}{llccc}
\hline
{Cluster} & Multiple--image & \multispan3{\hfil Weak--shear Measurements \hfil} \cr
          & Systems$^a$        & $R_{\rm min}$ & $N_{\rm fgal}$ & $\langle\tau_{\rm u}\rangle^b$ \cr
  & & (kpc) \cr
\hline
A\,68     & \parbox[t]{1.1in}{\raggedright\addtolength{\baselineskip}{-3pt}C0/[C19], [C1],
    [C2], [C6/C20], C15/C16/C17} & $200$ & $343$ & $0.18{\pm}0.03$ \cr
A\,209    & ...                    & $100$ & $431$ & $0.10{\pm}0.02$ \cr
A\,267    & [E2]                   & $200$ & $323$ & $0.06{\pm}0.02$ \cr
A\,383    & \parbox[t]{1.1in}{\raggedright\addtolength{\baselineskip}{-3pt}B0/B1/B4, [B2/B3/B17]}  & $160$ & $357$ & $0.12{\pm}0.02$ \cr
A\,773    & ...                    & $200$ & $297$ & $0.19{\pm}0.03$ \cr
A\,963    & H0                     & $120$ & $455$ & $0.13{\pm}0.02$ \cr
A\,1763   & ...                    & $150$ & $399$ & $0.09{\pm}0.02$ \cr
A\,1835   & ...                    & $300$ & $190$ & $0.20{\pm}0.03$ \cr
A\,2218   & M0, M1, M2, [M3]       & $250$ & $187$ & $0.16{\pm}0.03$ \cr
A\,2219   &
    \parbox[t]{1.1in}{\raggedright\addtolength{\baselineskip}{-3pt}P0/P1,
    P2, P3/P4/P5, [P6/P7/P8], [P9/P10], [P11/P12]} & $200$ & $246$ & $0.15{\pm}0.03$ \cr
\hline
\end{tabular}
}
{\footnotesize\addtolength{\baselineskip}{-5pt}
\begin{tabular}{l}
\parbox[b]{80mm}{$^a$~~ Unconfirmed systems are listed in parenthesis.}\\
\parbox[b]{80mm}{$^b$~~$\langle\tau_{\rm u}\rangle$ is the mean
tangential shear of the faint background galaxies selected for
inclusion in the weak--shear constraints.  We use the shape and
orientation of each galaxy as an individual constraint on the lens
model, and here summarize the strength of these constraints by listing
$\langle\tau_{\rm u}\rangle$ for each cluster.}\\
\end{tabular}
}
\end{table}

In this section we briefly re--cap the strong--lensing constraints
and then describe how the faint galaxy catalogs constructed in
\S\ref{faint} are converted into constraints on the cluster mass
distributions.

The multiple--image systems (Table~\ref{mult}) comprise two
categories: confirmed and unconfirmed.  Confirmed multiples have a
spectroscopic redshift and all the counter--images are either
identified or lie fainter than the detection threshold of the
observations.  The morphology of candidate multiples strongly suggests
that they are multiply--imaged, but the redshift of these systems is
less well defined ($\Delta z{\gs}0.1$) and not all counter--images may
be identified.  The confirmed systems provide constraints on both the
absolute mass of the clusters and the shape of the underlying mass
distribution; in contrast, the unconfirmed systems place additional
constraints on the shape of the cluster potential.  Both categories of
multiple--image constraints probe only the central
$R{\ls}50$--$100{\rm kpc}$ of each cluster.  Therefore, to extend the
constraints to larger radii, we supplement the strong--lensing
constraints with the weakly--sheared background galaxies.

To use the weakly--sheared galaxies as model constraints, we need to
estimate their redshifts.  For that purpose, we use the Hubble Deep
Field North (HDF--N) photometric redshift catalog of Fern\'andez--Soto
et al.\ (1999).  The $N_{\rm pixels}{\ge}30$ limit developed in
\S\ref{faint} is equivalent to a magnitude limit of ${\R702}{\ls}26$.
Using a simple no--evolution model (King \& Ellis 1985) we estimate
that a typical galaxy at $z{\sim}0.5$--1.5 has a colour of
$({\R702}{-}I_{814}){\simeq}0.7$ in the Vega system; converting to AB
magnitudes, this translates into a faint limit of $I_{814,
AB}{\le}25.8$ in the HDF--N catalog.  The median redshift to this limit
is $z_{\rm median}{=}0.9$, with an uncertainty of ${\sim}0.2$,
stemming from the dispersion in galaxy colours at $z{\sim}0.5$--1.5
and uncertainty in the conversion between the $N_{\rm pixels}{\ge}30$
and ${\R702}{\le}26$.  We therefore adopt $z{=}0.9{\pm}0.2$ as the
fiducial redshift of the faint galaxy catalogs.  Note that the
uncertainty in the median redshift contributes just 10--20\% of the
total error budget, which is dominated by the statistical uncertainty
in the shear measurements.  For each cluster we also examine the
region occupied by multiple--image systems in the \emph{HST} frames
and choose a minimum cluster--centric radius ($R_{\rm min}$) for the
inclusion of faint galaxies in the model constraints.  We test the
robustness of these choices by perturbing the $R_{\rm min}$ values by
${\pm}10''$ to ensure that the mean weak--shear computed from the
galaxies lying exterior to $R_{\rm min}$ is insensitive to the
perturbation in $R_{\rm min}$.  

We summarize the strong-- and weak--lensing model constraints in
Table~\ref{constr}, which is the key output from \S\ref{optical}.
First it defines, on the basis of our analysis of the \emph{HST} data
and ground--based spectroscopic follow--up which of the
multiple--image constraints can be used to calibrate the absolute mass
of the clusters, and which may be used only for constraining the shape
of the cluster potentials.  Second, it lists how many faint galaxies,
from what observed regions of the clusters have been carefully
selected to provide the weak--lensing constraints.  The strength of
the weak--lensing signal is also listed as the mean tangential shear,
$\langle\tau_{\rm u}\rangle$.  In the next section we explain how we
use these constraints to model the distribution of mass in the cluster
cores.

\section{Gravitational Lens Modelling}\label{modelling}

We use the {\sc lenstool} ray--tracing code (Kneib 1993) supplemented
by additional routines to incorporate weak--lensing constraints (Smith
2002) to build detailed parametrized models of the cluster mass
distributions.  We refer the interested reader to
Appendix~\ref{method} for full details of the modelling method.  Here,
we explain the modelling process in more general terms for the
non--lensing--expert reader whom we assume would prefer not to be
distracted by the many technical details which may be found in the
Appendix.

Each model comprises a number of parametrized mass components which
account for mass distributed on both cluster-- and galaxy--scales.
The cluster--scale mass components represent mass associated
with the cluster as a whole i.e.\ DM and hot gas in the ICM.  The
galaxy--scale mass components account for perturbations to the 
cluster potential by the galaxies.

A ${\chi}^2$--estimator quantifies how well each trial lens model fits
the data, and is minimized by varying the model parameters to obtain
an acceptable ($\chi^2/{\rm dof}{\simeq}1$) fit to the observational
constraints.  This is an iterative process, which we begin by
restricting our attention to the least ambiguous model constraints
(i.e.\ the confirmed multiple--image systems) and the relevant free
parameters.  For example, in a typical cluster lens there will be one
spectroscopically--confirmed multiple--image system and a few other
candidate multiples.  The model fitting process therefore begins with
using the spectroscopic multiple to constrain the parameters of the
main cluster--scale mass component.  Once we have established an
acceptable model using the confirmed multiple--image systems, we use
this model to explore the other constraints and to search for further
counter--images.  Specifically, we test the predictive power of the
model and use this to iterate towards the final refined model.  At
each stage of this process we incorporate additional constraints
(e.g.\ faint image pairs) and the corresponding free parameters (e.g.\
the ellipticity and orientation of key mass components, or the
velocity dispersion of cluster galaxies that lie close to faint image
pairs) into the model.

\subsection{Construction of the Lens Models}\label{themodels}

This section describes how the method outlined above and described in
detail in Appendix~\ref{method} was applied to each cluster in our
sample.  The parameters and the reduced $\chi^2$ of each fiducial
best--fit model are listed in Table~\ref{pars}.  One of us (GPS)
constructed the best--fit fiducial lens models for 9 out of the 10
clusters.  The tenth cluster, A\,2218, was modelled by JPK, based on
the results described in Kneib et al.\ (1995, 1996), Ebbels et al.\
(1998) and Ellis et al.\ (2001).  The estimation of uncertainties on
relevant model parameters that is required to provide robust error
bars on the cluster mass measurements (\S\ref{mass}) was performed for
all ten clusters by GPS.

\begin{table*}
\caption{
Best--fit Parameters of the Fiducial Lens Models$^a$
\label{pars}
}
{\small
\begin{tabular}{llrrrrrrrr}
\hline
{Cluster} & {Mass$^b$}           & {\hfil$\Delta{\rm R.A.}^c$\hfil}    & {\hfil$\Delta{\rm Dec.}^c$\hfil}    & $~~a/b~~$ & {\hfil$\theta$\hfil} & {\hfil$\rc$\hfil} & {\hfil$\rt^d$\hfil}        & {\hfil$\sigo$\hfil} & {\hfil$\chi^2/{\rm dof}$\hfil}\cr
          & {Component}      & {\hfil($''$)\hfil} & {\hfil($''$)\hfil} &       & {\hfil(deg)\hfil}    & {\hfil(kpc\hfil}) & {\hfil(kpc)\hfil}        & {\hfil($\kms$)\hfil} \cr
\hline
\multispan{10}{\hfil Individually Optimized Mass Components \hfil}\cr
\hline
A\,68     & Cluster~\#1       &    ${+}0.6$ & ${-}0.7$     & $2.2$    & $37$      & $108$   & $[1000]$  & $950$  & $11.6/11$\cr
          & Cluster~\#2       & $[{-}45.8]$ & $[{+}68.4]$  & $1.0$    & $58$      &  $81$   & $[1000]$  & $707$             \cr
          & BCG             &    ${-}0.2$ & $0.0$        & $1.7$    & $37$      & $0.3$   & $83$      & $301$             \cr
\noalign{\smallskip}
A\,209    & Cluster~\#1       & $[0.0]$     & $[0.0]$      & $[1.9]$  & $[43]$    & $[50]$  & $[1000]$  & $630$  & $0.6/1$ \cr
\noalign{\smallskip}
A\,267    & Cluster~\#1       & $[0.0]$     & $[0.0]$      & $2.0$    & ${-}60$   & $115$   & $[1000]$  & $1060$ & $3.6/3$\cr
\noalign{\smallskip}
A\,383    & Cluster~\#1       & ${+}0.3$    & ${+}0.5$     & $1.13$   & $109$     & $51$    & $[1000]$  & $900$  & $12.8/16$ \cr
          & Galaxy~\#2       & $[{+}14.9]$ & $[{-}16.8]$  & $[1.13]$ & $[{-}7]$  & $2.2$   & $[40]$    & $176$ \cr
          & BCG             & ${-}0.5$    & ${+}0.1$     & $1.07$   & $126$     & $0.6$   & $110$     & $310$ \cr
\noalign{\smallskip}
A\,773    & Cluster~\#1       & $[0.0]$     & $[0.0]$      & $[1.9]$  & $[{-}38]$ & $[75]$  & $[1000]$  & $750$ & $3.6/3$\cr
          & Cluster~\#2       & $[{+}1.0]$  & $[{+}24.0]$  & $[1.8]$  & $[{-}10]$ & $[75]$  & $[1000]$  & $700$           \cr
          & Cluster~\#3       & $[{+}84.4]$ & $[{+}12.0]$  & $[1.0]$  & $...$     & $[75]$  & $[1000]$  & $550$           \cr
\noalign{\smallskip}
A\,963    & Cluster~\#1       & $[0.0]$     & $[0.0]$      & $1.7$    & $90$      & $95$    & $[1000]$  & $980$  & $1.4/2$ \cr
          & BCG             & $[0.0]$     & $[0.0]$      & $1.1$    & $[90]$    & ${<}2$  & $96$      & $320$            \cr
\noalign{\smallskip}
A\,1763   & Cluster~\#1       & $[0.0]$     & $[0.0]$      & $[1.9]$  & $[180]$   & $[70]$  & $[1000]$  & $700$  & $5.1/3$\cr
\noalign{\smallskip}
A\,1835   & Cluster~\#1       & $[0.0]$     & $[0.0]$      & $[1.5]$  & $[70]$    & $[70]$  & $[1000]$  & $1210$ & $0.7/1$\cr
\noalign{\smallskip}
A\,2218   & Cluster~\#1       & ${+}0.2$    & ${+}0.5$     & $1.2$    & $32$      & $83$    & $[1000]$  & $1070$ & $17.8/19$ \cr
          & Cluster~\#2       & $[{+}47.0]$ & $[{-}49.4]$  & $1.4$    & $53$      & $57$    & $[500]$   & $580$ \cr
          & Galaxy~\#3       & $[{+}16.1]$ & $[{-}10.4]$  & $[1.1]$  & $[70]$    & ${<}2$  & $65$      & $195$ \cr
          & Galaxy~\#4       & $[{+}4.8]$  & $[{-}20.9]$  & $[1.4]$  & $[{-}23]$ & ${<}2$  & $77$      & $145$ \cr
          & BCG             & ${+}0.3$    & ${+}0.1$     & $1.8$    & $53$      & ${<}3$  & $136$     & $270$ \cr
\noalign{\smallskip}
A\,2219   & Cluster~\#1       & ${+}0.1$    & ${+}0.2$     & $1.7$    & $35$      & $77$     & $[1000]$   & $902$ & $3.7/3$\cr
          & Cluster~\#2       & $[{+}39.2]$ & $[{-}32.0]$  & $[1.1]$  & $[8]$     & $55$     & $375$    & $515$ \cr
          & Cluster~\#3       & $[{-}22.9]$ & $[{+}4.5]$   & $[1.0]$  & $...$     & $31$     & $365$    & $395$ \cr
          & BCG             & $[0.0]$     & $[0.0]$      & $[1.6]$  & $[29]$    & ${<}3$   & $120$    & $278$ \cr
\hline
\multispan{10}{\hfil Luminosity Scaled Mass Components \hfil}\cr
\hline
\multispan2{$L_K^{\star}{\rm~galaxy}^c$\dotfill} & ...    & ...      & ...   & ...      & $0.2$   & $30$   & $180$\cr
\hline
\end{tabular}
}
\begin{tabular}{l}
\parbox[b]{6.5in}{\footnotesize\addtolength{\baselineskip}{-5pt} 
$^a$ Parameter values listed in parenthesis were not free parameters.  

}\\
\parbox[b]{6.5in}{\footnotesize\addtolength{\baselineskip}{-5pt} 
$^b$ Individually optimized mass components are numbered and
  identified as being cluster-- or galaxy--scale.

}\\
\parbox[b]{6.5in}{\footnotesize\addtolength{\baselineskip}{-5pt}
$^c$ The position of each mass component is given relative to the optical centroid of the central galaxy in each cluster (Table~\ref{sample}).

}\\
\parbox[b]{6.5in}{\footnotesize\addtolength{\baselineskip}{-5pt}
$^d$
Cluster galaxies are included in the lens models down
to the limit where the mass of additional components would be
comparable with the uncertainties in the overall cluster mass, which
equates to 
a magnitude limit of $K{\le}K^{\star}{+}2.5$.

}\\
\end{tabular}
\end{table*}


\smallskip
\noindent
{\bf A\,68} --- We first constrained the model with just the
multiply--imaged ERO at $z{=}1.6$ (C0 -- Table~\ref{mult}),
identifying nine distinct knots of likely star--formation in each
image of this galaxy.  A model containing just one cluster--scale mass
component (\#1), did not fit these data well: $\chi^2/{\rm
dof}{\gs}5$.  We therefore added a second cluster--scale mass
component (\#2) to the North--West of the central galaxy.  Despite the
strong evidence for the presence of this mass component in the
weak--shear map (Fig.~\ref{hstdata}), no single bright cluster galaxy
dominates the group of galaxies found in this region.  We therefore
adopt the brightest of this group of galaxies as the center of
Clump~\#2, for which we adopt a circular shape.  C0 places strong
constraints on the mass required in this second clump because the
spatial configuration of the images is very sensitive to the details
of the bi--modal mass structure of the cluster.  We find an acceptable
fit without optimizing the spatial parameters of Clump~\#2.  The
South--West corner of C0 straddles the $z{=}1.6$ radial caustic in
this best--fit lens model, causing an additional, radially amplified
image of this portion of the galaxy to be predicted.  We search the
\emph{HST} frame in the vicinity of the predicted radial image, and
find a faint radial feature (C19) $4''$ North--West of the central
galaxy which is consistent with the model prediction.  Further
constraining the model with C15/C16/C17, at $z{=}5.4$ confirms the
validity of the model thus far, and helps to constrain the spatial
parameters of the NW cluster--scale mass component.  This model is
also able to reproduce the observed morphology of the other candidate
multiple--image systems.

\smallskip
\noindent
{\bf A\,209} --- Given the weak constraints on this cluster from the
\emph{HST} data, we restrict our attention to a simple model in which
the velocity dispersion of the central cluster--scale mass component
is the only free parameter.

\smallskip
\noindent
{\bf A\,267} --- The important difference between this cluster and
A\,209 is that it contains a candidate multiple--image pair (E2a/b).
In addition to constraining $\sigo$ and $\rc$ for the central
cluster--scale mass component we therefore use this image pair to
constrain the shape of the cluster potential.

\smallskip
\noindent
{\bf A\,383} --- We use the many constraints available for this
cluster to determine precisely the full range of geometrical and
dynamical parameters for the cluster--scale and central galaxy mass
components.  Despite the overall relaxed appearance of this cluster,
the bright cluster elliptical South--West of the central galaxy
actually renders this a bi--modal cluster (albeit with very unequal
masses) on small scales.  We therefore also obtain a constraint on the
velocity dispersion of this galaxy (A\,383~\#2 in Table~\ref{pars}).
Sand et al.'s (2004) spectroscopic redshifts for B1a/b and B0b,
placing them both at $z{=}1.01$, i.e.\ the same redshift as B0a,
slightly modifies Smith et al.'s (2001) multiple--image interpretation
of this cluster.  However the parameter space occupied by this cluster
is consistent with that of Smith et al.'s model.

\smallskip
\noindent
{\bf A\,773} --- Although no multiple--image systems have been
identified yet in this cluster, the large number of early--type
galaxies and the strength of the weak--shear signal suggest that this
cluster is probably quite massive.  First we use the shapes of the
weakly--sheared galaxies (Table~\ref{constr}) to constrain a model
that contains a single cluster--scale mass component centered on the
BCG (A\,773~\#1 in Table~\ref{pars}).  The best--fit velocity
dispersion of Clump~\#1 in this model is ${\sim}1000\kms$.  However,
the spatial structure in the residuals reveals that this simple model
does not reproduce the strong shear signal observed to the North of
the second brightest cluster galaxy and to the East of the BCG, i.e.\
in the saddle region between the BCG and the group of cluster
ellipticals at the Eastern extreme of the WFPC2 field--of--view
(Fig.~\ref{hstdata}).  We therefore introduce two more cluster--scale
mass components: A\,773~\#2 is coincident with the second brightest
cluster elliptical and A\,773~\#3 coincides with the brightest member
of the Eastern group of galaxies.  This model faithfully reproduces
the global shear strength, and crucially it also reproduces the spatial
variation of the shear and thus provides a superior description of the
cluster potential than the initial simple model.

\smallskip
\noindent
{\bf A\,963} --- H0 provides a straightforward yet powerful constraint
on the potential of this relaxed cluster, enabling the dynamical and
spatial parameters of both the cluster--scale and BCG mass components
to be constrained.

\smallskip
\noindent
{\bf A\,1763} --- This cluster is similar to A\,209 in that there are
no confirmed multiple--image systems and the weak--shear signal is
relatively low (Table~\ref{constr}).  We therefore fit a model that
contains the velocity dispersion of the (single) cluster--scale mass
component as the only free parameter.  Overall, this simple model is
an acceptable fit to the global weak--shear signal, however it fails
to reproduce the large observed shear signal to the West of the
central galaxy (Fig.~\ref{hstdata}).  We interpret these residuals as
a signature of substructure in this cluster, indicating that the mass
distribution may be more complex than a single cluster--scale mass
plus galaxies.  Unfortunately the weak--shear signal is not strong
enough to place any further constraints on this cluster at this time.

\smallskip
\noindent
{\bf A\,1835} --- The multiple--image interpretation of Schmidt et
al.\ (2001) is ruled out by the new WFPC2 data presented in this
paper, specifically, the differences in surface brightness between K0,
K1 and K3.  The absence of multiple--image constraints therefore
results in a model similar to those of A\,209 and A\,1763, with just a
single free parameter -- the central velocity dispersion of the
central cluster--scale mass component.

\smallskip
\noindent
{\bf A\,2218} --- The model of A\,2218 builds on the models published
by Kneib et al.\ (1995; 1996) and incorporates for the first time the
spectroscopic redshifts of the M2 (Ebbels et al.\ 1998) and M3 (Ellis
et al.\ 2001) multiple--image systems.  In addition to the central
cluster--scale mass component (A\,2218~\#1), this model contains a
second cluster--scale mass component (A\,2218~\#2) centered on the
second brightest cluster galaxy which lies South--East of the BCG.  The
velocity dispersion and cut--off radius of the two bright cluster
galaxies (A\,2218~\#3 \& \#4) that lie adjacent to the M0
multiple--image system are also included as free parameters.

\smallskip
\noindent
{\bf A\,2219} --- We first attempt to find an acceptable solution that
is based on a single cluster--scale mass component centered on the
BCG, constrained just by P0.  This model succeeds in reproducing the
straight morphology of P0 (Fig.~\ref{hstzooms}), however when P3/P4/P5
are added to the constraints, the fit deteriorates substantially.  We
therefore add a second cluster--scale mass component (A\,2219~\#2) at
the position of the second brightest cluster galaxy  (South--East
of the central galaxy).  The second clump improves the fit somewhat,
but the tight constraints from these two multiple--image systems on
the saddle region between clumps \#1 and \#2 necessitate the addition
of a third cluster--scale component (A\,2219~\#3 -- see
Fig.~\ref{hstdata}).  This tri--modal model is a good fit, and readily
accommodates the additional constraints from P2a/b/c with a minimum of
further modifications.  This best--fit model is also able to reproduce
faithfully the details of the candidate multiple--image systems.

\subsection{Calibration of Weak Lensing Constraints}\label{tests}

We investigate the systematic uncertainty that may arise as a result
of confirmed multiple--image systems not being available for all of
the clusters.  First, we focus on the five clusters for which both
multiple--image and weak--shear constraints are available (A\,68,
A\,383, A\,963, A\,2218, A\,2219).  We ignore the multiple--image
constraints and construct a model of each of these clusters using just
the weak--shear information.  In common with the five lens models that
are based solely on weak--shear constraints (A\,209, A\,267, A\,773,
A\,1763, A\,1835) we find that the weak--shear signal alone can
generally only constrain one free parameter (${\sigo}$) per
cluster--scale mass component.  Individually, the velocity dispersions
of the cluster--scale mass components in the weak--shear constrained
models agree within the uncertainties with the velocity dispersions
obtained in the multiple--image constrained models.  However, when
treated as an ensemble, the mean ratio of weak--shear constrained
velocity dispersions to multiple--image constrained velocity
dispersion is $0.94{\pm}0.04$.  Based on just five clusters, it
therefore appears that $\sigo$ for the cluster--scale mass components
in the models of weak--shear only clusters may be under--estimated, on
average, by ${\sim}6\%$.  Mass scales as ${\sigo}^2$; this possible
systematic error in ${\sigo}$ therefore translates into a possible
${\sim}12\%$ under--estimate in cluster mass.

This uncertainty probably arises from contamination of the faint
background galaxy catalogues by faint cluster galaxies, which we
estimated conservatively to be ${\sim}20\%$ in \S\ref{extract}.  Our
cross--calibration of the strong-- and weak--lensing constraints
therefore suggests that the contamination is somewhat lower than
previously thought.  We choose not to correct the parameters of the
weak--shear constrained models for this effect because the
uncertainties in these models are dominated by the statistical
uncertainty which is typically ${\Delta}{\sigo}{\sim}10$--20\%.  A
global 6\% correction to the velocity dispersions of the weak--lensing
constrained cluster lens models would also neglect the dependence of
the contamination, for a given cluster, on the optical richness of
that cluster.  This uncertainty has a negligible affect on the results
that rely on absolute cluster mass measurements (\S\ref{masslx})

\section{X--ray Data and Analysis}\label{xray}

We complement the detailed view of the distribution of total mass in
the cluster cores that is now available to us from the lens models
with high--resolution X--ray observations with \emph{Chandra}.  The
purpose of including these data is to compare the underlying mass
distribution derived from lensing with the properties of the ICM.
Specifically, we wish to compare the mass and X--ray morphologies of
the clusters, and to explore how the lensing--based mass measurements
are correlated with the X--ray temperature of the clusters
(\S\ref{results}).

We therefore exploit archival \emph{Chandra} data for nine of the
clusters (Table~\ref{chobs}).  In the spectral and imaging analysis we
used only chips I0--I1--I2--I3 and chip S3 for observations in ACIS--I
and ACIS--S configurations respectively.  All of the \emph{Chandra}
observations were performed in ACIS--I configuration except A\,383
(ID: 2321), A\,963 and A\,1835 which were observed in ACIS--S
configuration.  To reduce the data we used the procedures described by
Markevitch et al.\ (2000), Vikhlinin et al.\ (2001a), Markevitch \&
Vikhlinin (2001), and Mazzotta et al.\ (2001).  We note that the three
observations of A\,383 were not all performed in the same
configuration.  The spectral response and background for each
observation was therefore generated individually before combining the
data.  The data were also cleaned for the presence of strong
background flares following the prescription of Markevitch et
al. (2000a).  The net exposure time for each observation is listed in
Table~\ref{chobs}.  Adaptively smoothed flux contours are also
over--plotted on the \emph{HST} frames in Fig.~\ref{contours}.

\begin{table}
\caption{
Summary of Archival X--ray Observations
\label{chobs}
}
\begin{tabular}{p{10mm}p{10mm}p{7mm}p{10mm}p{10mm}}
\hline
Cluster & {Obs.}   & \centering{${\texp}$}  & \centering{${\txtot}^a$}               & \centering{${\txann}^b$} \cr
        & {ID No.} & \centering{(ks)}       & \centering{(keV)}                      & \centering{(keV)}               \cr
\hline
A\,68   & 3250       & \raggedleft{$8.4$}   & \raggedleft{$9.5^{{+}0.9}_{{-}0.7}$}   & \raggedleft{$9.5^{{+}1.5}_{{-}1.0}$} \cr
A\,209  &  522       & \raggedleft{$10.0$}  & \raggedleft{$8.4^{{+}0.5}_{{-}0.5}$}   & \raggedleft{$8.7^{{+}0.6}_{{-}0.5}$} \cr
A\,267  & 1448       & \raggedleft{$6.4$}   & \raggedleft{$5.9^{{+}0.5}_{{-}0.4}$}   & \raggedleft{$6.0^{{+}0.7}_{{-}0.5}$} \cr
A\,383  &  524       & \raggedleft{$7.4$}   & \raggedleft{$4.3^{{+}0.2}_{{-}0.1}$}   & \raggedleft{$5.2^{{+}0.2}_{{-}0.2}$} \cr
        & 2320       & \raggedleft{$17.9$}  &                                        & \cr
        & 2321       & \raggedleft{$14.3$}  &                                        & \cr
A\,773  &  533       & \raggedleft{$11.3$}  & \raggedleft{$8.0^{{+}0.5}_{{-}0.4}$}   & \raggedleft{$8.2^{{+}0.5}_{{-}0.5}$} \cr
A\,963  &  903       & \raggedleft{$3.6$}   & \raggedleft{$7.3^{{+}0.3}_{{-}0.3}$}   & \raggedleft{$7.2^{{+}0.3}_{{-}0.3}$} \cr
A\,1763$^c$ & 801049 & \raggedleft{$18.0$}  & \raggedleft{$8.9^{{+}0.5}_{{-}0.4}$}   & ... \cr
A\,1835 &  496       & \raggedleft{$10.5$}  & \raggedleft{$7.7^{{+}0.3}_{{-}0.2}$}   & \raggedleft{$9.3^{{+}0.6}_{{-}0.4}$} \cr
A\,2218 & 1454       & \raggedleft{$9.7$}   & \raggedleft{$6.9^{{+}0.5}_{{-}0.5}$}   & \raggedleft{$6.8^{{+}0.5}_{{-}0.5}$} \cr 
        &  553       & \raggedleft{$5.4$}   &                                        &    \cr 
A\,2219 & 896        & \raggedleft{$42.3$}  & \raggedleft{$14.0^{{+}0.8}_{{-}0.6}$}  & \raggedleft{$13.8^{{+}0.8}_{{-}0.7}$} \cr
\hline
\end{tabular}
\begin{tabular}{l}
\parbox[b]{3.2in}{\footnotesize\addtolength{\baselineskip}{-5pt} 
$^a$ ${\txtot}$ is measured in an aperture of radius $R{\le}2{\rm
    Mpc}$\hfil}\cr 
\parbox[b]{3.2in}{\footnotesize\addtolength{\baselineskip}{-5pt} 
$^b$ ${\txann}$ is measured in an annulus defined by $0.1{\le}R{\le}2{\rm
    Mpc}$\hfil}\cr 
\parbox[b]{3.2in}{\footnotesize\addtolength{\baselineskip}{-5pt} 
$^c$ A\,1763 has not been observed by \emph{Chandra}.  The ID number
  and exposure time listed for this cluster relate to the archival
  \emph{ROSAT} data available for this cluster, and the temperature is
  from Mushotzky \& Scharf's (1997) analysis of \emph{ASCA} data.\hfil}\cr 
\end{tabular}
\end{table}

Spectral analysis was performed in the 0.8--9\,keV energy band in
PI channels, thus avoiding problems connected with the poor
calibration of the detector at energies below $0.8{\keV}$.  Spectra
were extracted using circular regions centered on the X--ray centroid
of each cluster within a radius of 2\,Mpc at the cluster redshift,
being careful to mask out all the strong point sources.  An absorbed
{\sc mekal} model was used, with the equivalent hydrogen column
density fixed to the relative Galactic value (Dickey \& Lockman 1990).
The temperature, plasma metallicity, and normalization were left as
free parameters.  Because of the hard energy band used in this
analysis, the derived plasma temperatures are not very sensitive to
the precise value of ${\nH}$.  We list the temperature of each cluster
derived from the total field of view (i.e.\
${\txtot}{\equiv}{\tx}(R{\le}2{\rm Mpc})$) in Table~\ref{chobs}.

The presence of a ``cool core'' (e.g.\ Allen, Schmidt \& Fabian 2001)
could bias low the cluster temperature measurements.  As the aim is to
obtain a reliable global measurement of the cluster temperatures, we
therefore re--measured the temperatures in an annulus
${\txann}{\equiv}{\tx} (0.1{\le}R{\le}2{\rm Mpc})$ centered on the X--ray
centroid of each cluster (Markevitch 1998).  There is a significant
difference between ${\txtot}$ and ${\txann}$ in just two clusters:
A\,383 and A\,1835 (Table~\ref{chobs}).  Both of these clusters have
previously been identified as containing an emission line BCG (Smith
et al.\ 2001; Allen et al.\ 1996), which is arguably the most reliable
indicator of central cold material in clusters (Edge et al.\ 1990).
We also note that none of the seven clusters for which, within the
uncertainties, ${\txtot}{=}{\txann}$ have previously been identified
as containing a cool core (e.g.\ White et al.\ 1997).  We list the
temperature ratios, ${\txtot}/{\txann}$, in Table~\ref{masstx}.

\section{Results}\label{results}

We begin with a brief review of where the preceding three sections of
analysis and modelling have brought us toward our goals of
characterizing the dynamical maturity of X--ray luminous clusters at
$z{\simeq}0.2$ and calibrating the high--mass end of the
mass--temperature relation.

The detailed gravitational lens models (\S\ref{themodels}) reveal the
total matter content of the clusters; in \S\ref{mass} we compute and
analyze detailed mass maps using the best--fit models.  These
measurements of total cluster mass are complemented by the X--ray
pass--band (\S\ref{xray}) which reveals the details of the hot
intracluster medium.  In \S\ref{mass}, we compare the spatial
distribution of total mass with the spatial structures in the X--ray
flux maps and temperature measurements derived from the \emph{Chandra}
observations.  We also compare the total matter and ICM with the
spatial distribution of stars in the clusters using the measurements
of the $K$--band luminosity of cluster galaxies estimated in
\S\ref{clus}.  In summary, the synthesis presented in \S\ref{mass}
aims to diagnose whether or not the clusters are dynamically mature
using independent probes of dark matter (inferred from the lensing
mass maps), hot intra--cluster gas and cluster galaxies.

In \S\ref{masslx} we adopt a different approach -- we explore
correlations between the integrated properties of the clusters.
Specifically, we use the cluster mass, X--ray luminosity and X--ray
temperature measurements to normalize the cluster scaling relations
and to investigate the scatter about these normalizations.  A key
focus of this exercise is to exploit the structural results from
\S\ref{mass} to search for structural segregation in the scaling
relations.  This is an important step toward constraining the mass
assembly and thermodynamic histories of clusters as a function of
cosmic epoch, and is also relevant to pinpointing potential
astrophysical systematic uncertainties when cluster are used in the
measurement of cosmological parameters.

\subsection{Mass and Structure of Cluster Cores}\label{mass}

\begin{table*}
\caption{Mass and Substructure Diagnostics\label{masstx}}
{\small
\begin{tabular}{lcccclccl}
\hline
{Cluster} & $N_{\rm DM}$$^a$ & ${\Mtot}$ & ${\Mcen}/{\Mtot}^{b}$ & $L_{K,{\rm BCG}}/L_{K,{\rm tot}}$ & X--ray     & ${\drpeak}^c$ & ${\txtot}/{\txann}$ & Overall \cr
          &              & ($10^{14}\Msol$)         &                       &                        & Morphology & (kpc)         &                     & Classification \cr
\hline
A\,68   & 2 & $4.4{\pm}0.1$ & $0.68{\pm}0.01$ & $0.35{\pm}0.07$ & Irregular            & $50{\pm}15$   & $1.00^{+0.18}_{-0.13}$ & Unrelaxed \cr 
A\,209  & 1 & $1.6{\pm}0.5$ & $0.87{\pm}0.06$ & $0.38{\pm}0.06$ & Irregular            & $17{\pm}4$    & $0.97^{+0.09}_{-0.08}$ & Unrelaxed \cr 
A\,267  & 1 & $2.6{\pm}0.4$ & $0.96{\pm}0.01$ & $0.76{\pm}0.02$ & Elliptical           & $88{\pm}5$   & $0.98^{+0.14}_{-0.11}$ & Unrelaxed \cr 
A\,383  & 1 & $3.6{\pm}0.1$ & $0.97{\pm}0.01$ & $0.66{\pm}0.03$ & Circular             & ${<}4$       & $0.82^{+0.06}_{-0.05}$ & Relaxed   \cr 
A\,773  & 3 & $5.1{\pm}1.2$ & $0.41{\pm}0.20$ & $0.22{\pm}0.08$ & Irregular            & $42{\pm}8$   & $0.98^{+0.09}_{-0.08}$ & Unrelaxed \cr 
A\,963  & 1 & $3.3{\pm}0.2$ & $0.97{\pm}0.01$ & $0.56{\pm}0.04$ & Elliptical           & ${<}4$       & $1.01^{+0.06}_{-0.06}$ & Relaxed   \cr 
A\,1763 & 1 & $2.1{\pm}0.8$ & $0.90{\pm}0.05$ & $0.45{\pm}0.05$ & Irregular            & $80{\pm}20$   & ...                    & Unrelaxed \cr 
A\,1835 & 1 & $5.8{\pm}1.1$ & $0.97{\pm}0.01$ & $0.50{\pm}0.05$ & Circular             & ${<}5$       & $0.83^{+0.06}_{-0.04}$ & Relaxed   \cr 
A\,2218 & 2 & $5.6{\pm}0.1$ & $0.77{\pm}0.01$ & $0.19{\pm}0.08$ & Irregular            & $38{\pm}7$   & $1.01^{+0.10}_{-0.10}$ & Unrelaxed \cr 
A\,2219 & 3 & $3.4{\pm}0.1$ & $0.85{\pm}0.01$ & $0.32{\pm}0.07$ & Irregular            & $13{\pm}4$    & $1.01^{+0.08}_{-0.07}$ & Unrelaxed \cr 
\hline
\end{tabular}
}
\begin{tabular}{l}
\parbox[t]{6.5in}{\footnotesize\addtolength{\baselineskip}{-5pt}
$^a$ $N_{\rm DM}$ is the number of cluster--scale DM haloes contained
  in each best--fit lens model.

}\\
\parbox[t]{6.5in}{\footnotesize\addtolength{\baselineskip}{-5pt}
$^b$ ${\Mcen}$ is the mass that resides in the centrally--located DM
  halo of the lens model and the BCG (\S\ref{mass}).

}\\
\parbox[t]{6.5in}{\footnotesize\addtolength{\baselineskip}{-5pt} 
$^c$ The uncertainties on ${\drpeak}$ include uncertainties on the
central coordinates of the cluster mass distribution in the relevant
lens models.

}\\
\end{tabular}
\end{table*}

We begin by using the gravitational lens models to measure the mass of
each cluster, and to quantify the spatial distribution of that mass.
All of the diagnostics discussed in this section are listed in
Table~\ref{masstx}, together with the overall diagnosis of ``relaxed''
or ``unrelaxed'' -- we define these terms in this section.

\subsubsection{Total Cluster Mass and its Spatial Distribution}\label{masslk}

\begin{figure}
\centerline{
\psfig{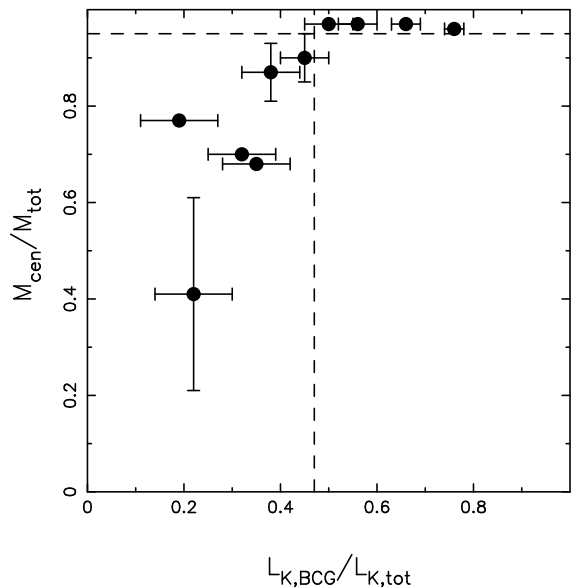} 
}
\caption
{ Central mass fraction, ${\Mcen}/{\Mtot}$ versus central $K$--band
  luminosity fraction, $L_{K,{\rm BCG}}/L_{K,{\rm tot}}$
  (\S\ref{mass}).  This plot reveals a remarkably clean separation
  between clusters with a mass distribution that is heavily dominated
  by the central mass components, and a stellar luminosity
  distribution that is dominated by the BCG.  The horizontal and
  vertical dashed lines mark the divisions between high and low
  central mass and $K$--band luminosity fractions respectively.  See
  \S\ref{masslk} for further discussion of this separation.
\label{sublk}
}
\end{figure}

We adopt a fixed projected aperture of $R{=}500{\rm kpc}$ which is
well--matched to the scales probed by the \emph{HST} data, and measure
the mass interior to that radius: ${\Mtot}{=}M(R{\le}500{\rm kpc})$.  We
also want to characterize the spatial distribution of mass in the
cluster core.  The number of cluster--scale mass components (${N_{\rm
DM}}$) in the lens models sheds some light on this question
(Table~\ref{pars}~\&~\ref{masstx}).  However, $N_{\rm DM}$ does not
contain any explicit information about mass.  We therefore complement
this quantity by measuring ${\Mcen}$, defined as the projected mass
within $R{=}500{\rm kpc}$ that is associated with the
centrally--located mass components, i.e.\ the dominant cluster--scale
mass component and the BCG.  We list the central mass fraction,
${\Mcen}{/}{\Mtot}$, in Table~\ref{masstx}.  The uncertainties in
these measurements are estimated by exploring the parameter space
occupied by each lens model, identifying the family of models that
satisfy ${\Delta}{\chi}^2{\le}1$.

The central mass fractions comprises two contributions: (i)
cluster--scale mass components in the lens models that are associated
with massive in--falling structures, and (ii) cluster galaxies that
are associated both with the central cluster--scale DM halo (and 
are presumably virialized) and with the in--falling structures.  The
central mass fraction therefore characterize the dominance of the
central concentration of mass in the overall cluster mass
distribution.  The measurements listed in Table~\ref{masstx} (see also
Fig.~\ref{sublk}) reveal that the clusters fall into two categories:
A\,267, A\,383, A\,963 and A\,1835 form a homogeneous sub--sample, all
with ${\Mcen}{/}{\Mtot}{>}0.95$, i.e.\ mass distributions heavily
dominated by the central components; the remaining six all have
${\Mcen}{/}{\Mtot}{<}0.95$ and are much more diverse than the former
sub--sample, with central mass fractions spanning
$0.4{\ls}{\Mcen}{/}{\Mtot}{\ls}0.9$.

As an independent cross--check on this sub--classification of the
clusters, we measure the distribution of stars in the clusters using
the $K$--band luminosities of cluster galaxies described in
\S\ref{clus}.  We divide the $K$--band luminosity of each BCG (i.e.\
the luminosity that is spatially coincident with the central mass
components) by the combined $K$--band luminosity of all the cluster
galaxies detected in each \emph{HST} frame.  These central $K$--band
luminosity fractions ($L_{K,{\rm BCG}}/L_{K,{\rm tot}}$) are listed in
Table~\ref{masstx} and plotted versus the central mass fractions in
Fig.~\ref{sublk}.  The central luminosity fractions span $L_{K,{\rm
    BCG}}/L_{K,{\rm tot}}{\sim}0.2$--0.8, and are not obviously more
homogeneous for low and high central fraction clusters.  Nevertheless,
there appears to be a roughly monotonic relationship between the
central mass fraction and central $K$--band luminosity fraction, thus
to first order confirming the separation of the cluster sample into
two structural classes.

This sub--classification into a homogeneous sub--sample of clusters
with ${\Mcen}{/}{\Mtot}{>}0.95$ and a diverse sub--sample with
$0.4{\ls}{\Mcen}{/}{\Mtot}{\ls}0.9$ matches the details of the cluster
lens models reasonably well.  The lens model of each of the former
clusters contains a single cluster--scale mass component.  The
situation is less clear--cut for the latter sub--sample.  Lens models
of four of the six clusters contain two or more cluster--scale mass
components, i.e.\ their mass distributions are unambiguously bi-- or
tri--modal (see also Fig.~\ref{contours}), and thus the low central
mass fractions are dominated by substructure in the cluster cores.
However the remaining two (A\,209 and A\,1763) contain a single
cluster--scale mass component.  It is therefore ambiguous whether the
moderately low central mass fractions in these clusters genuinely
reflect cluster substructure, or are simply due to the cluster galaxy
populations.  One possibility is that these two clusters are both
undergoing mergers in the plane of the sky.  This would help to
explain the absence of a confirmed strong lensing signal
(Table~\ref{mult}), the low aperture mass measurements
(Table~\ref{masstx}) and the moderately low central mass and $K$--band
luminosity fractions.  Wider--field \emph{HST} imaging would help to
resolve this uncertainty.

\subsubsection{Total Mass Versus X--ray Flux and Temperature}

\begin{figure*}
\centerline{
\psfig{file=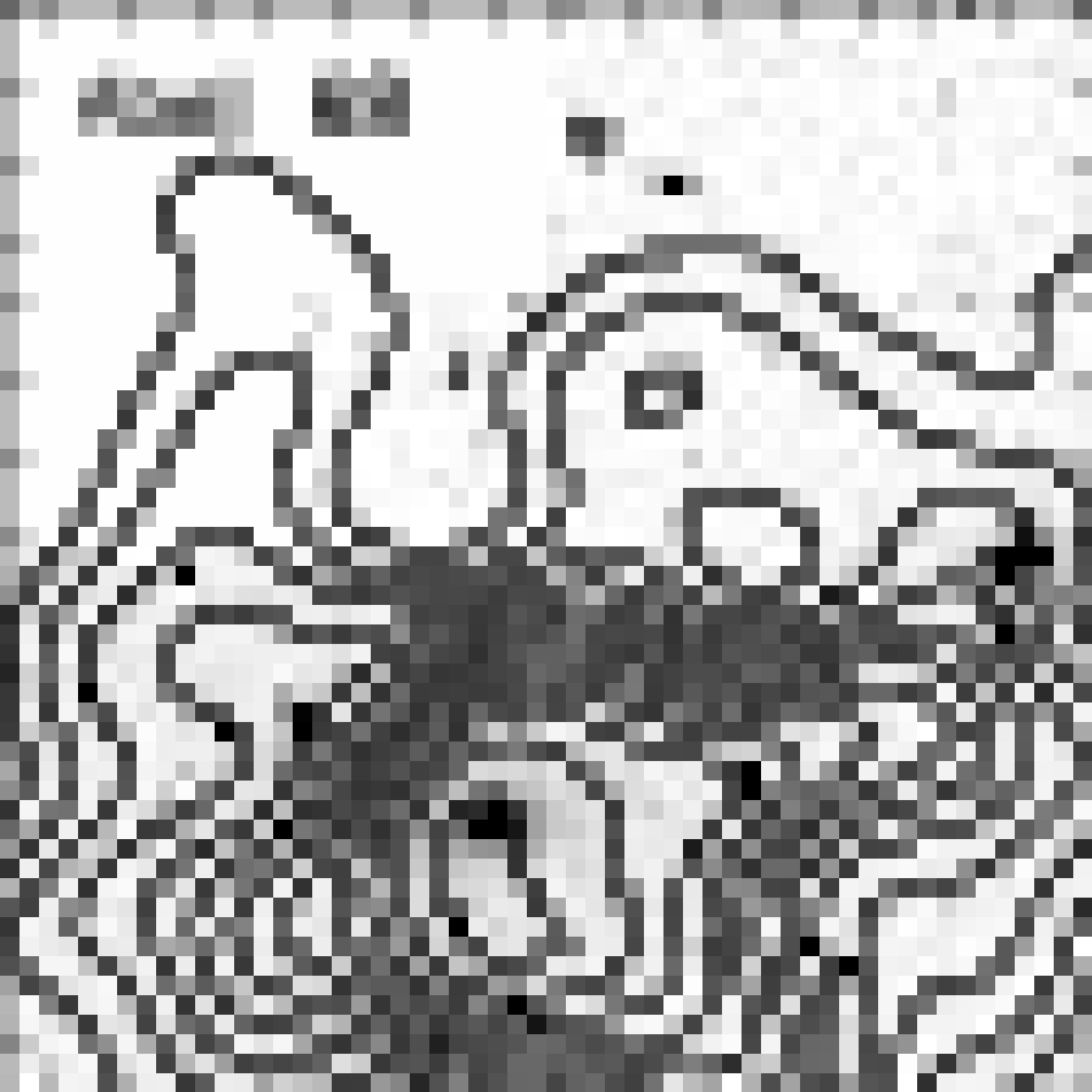,width=42mm,angle=-90}
\hspace{-1.5mm}
\psfig{file=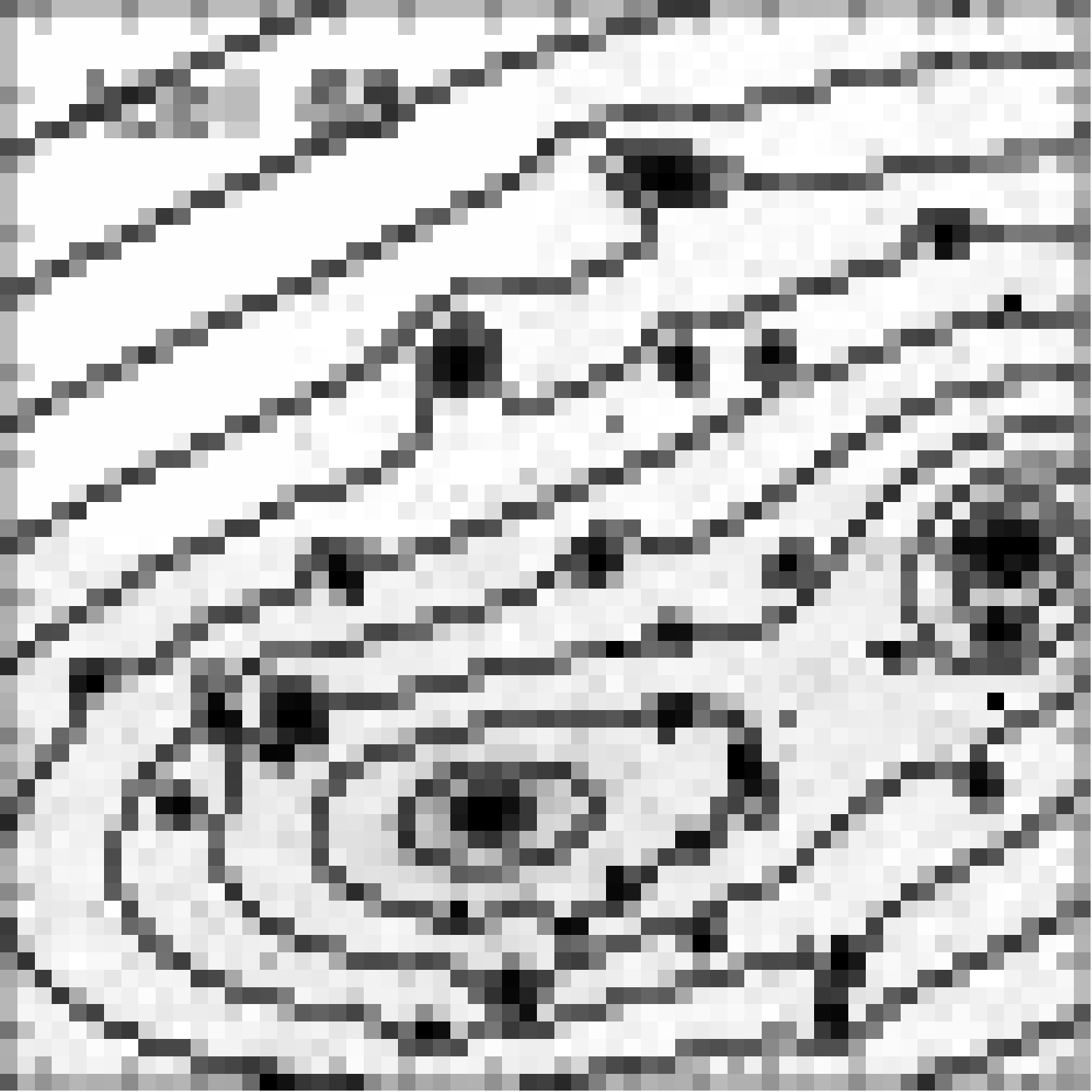,width=42mm,angle=-90}
\hspace{1mm}
\psfig{file=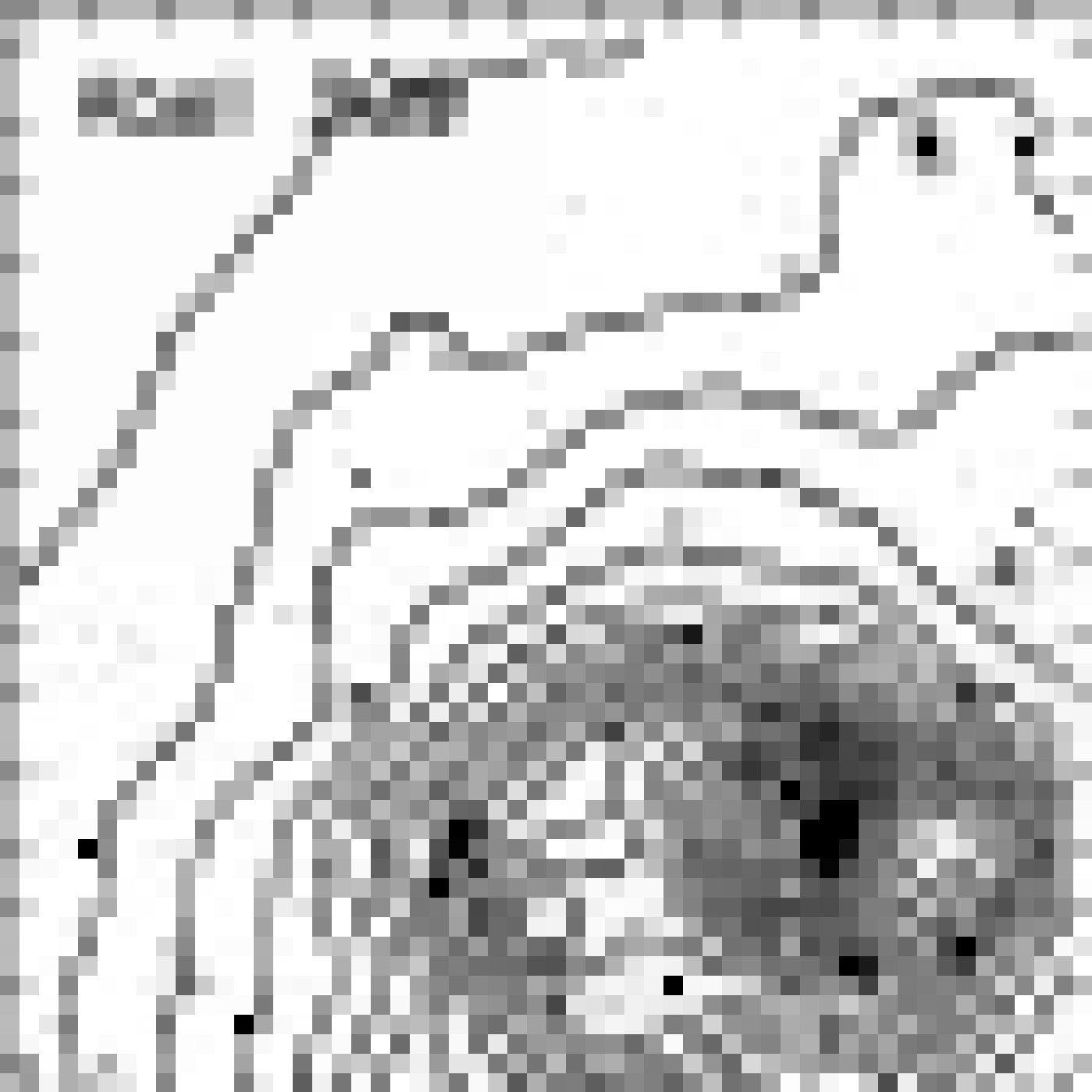,width=42mm,angle=-90}
\hspace{-1.5mm}
\psfig{file=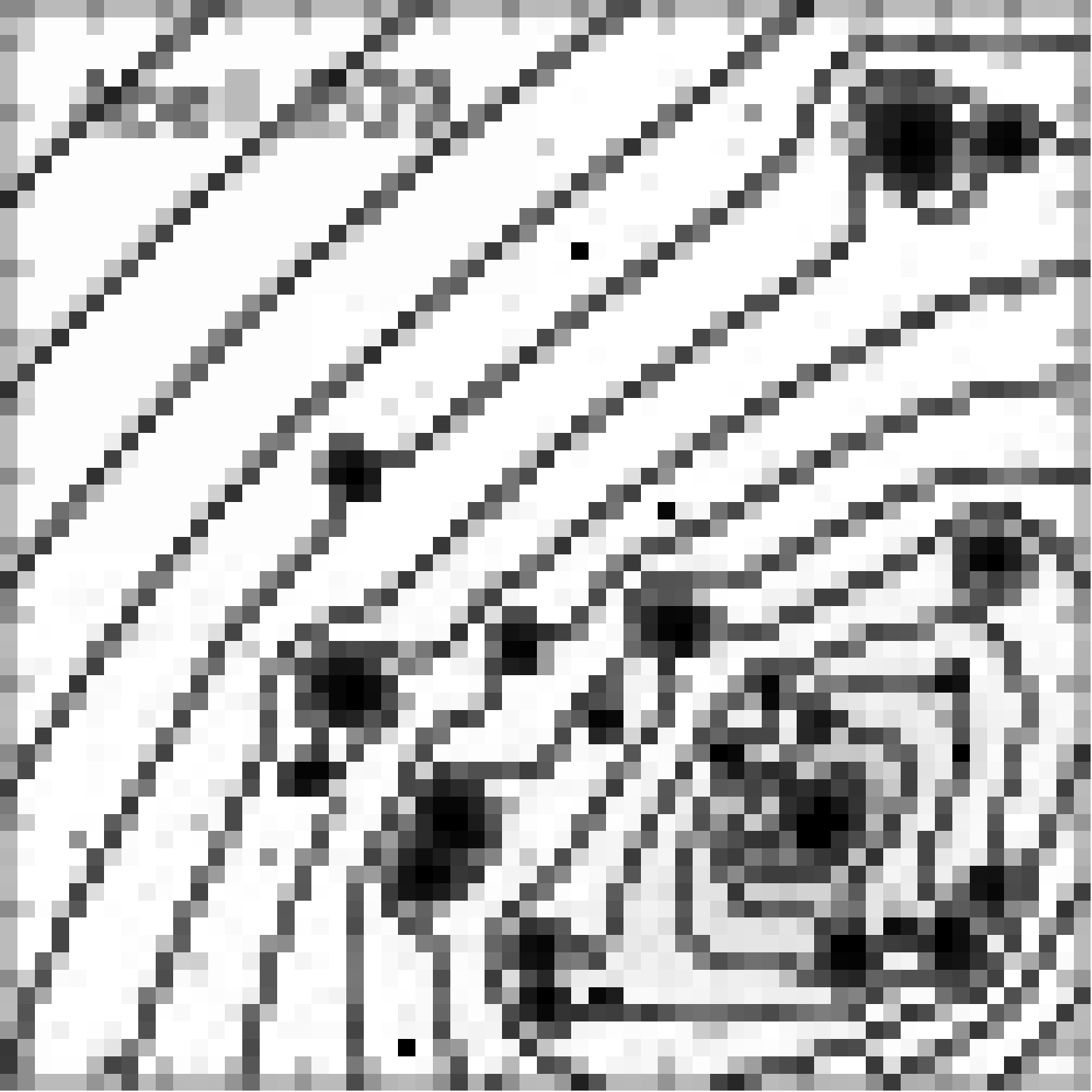,width=42mm,angle=-90}
}

\vspace{2mm}
\centerline{
\psfig{file=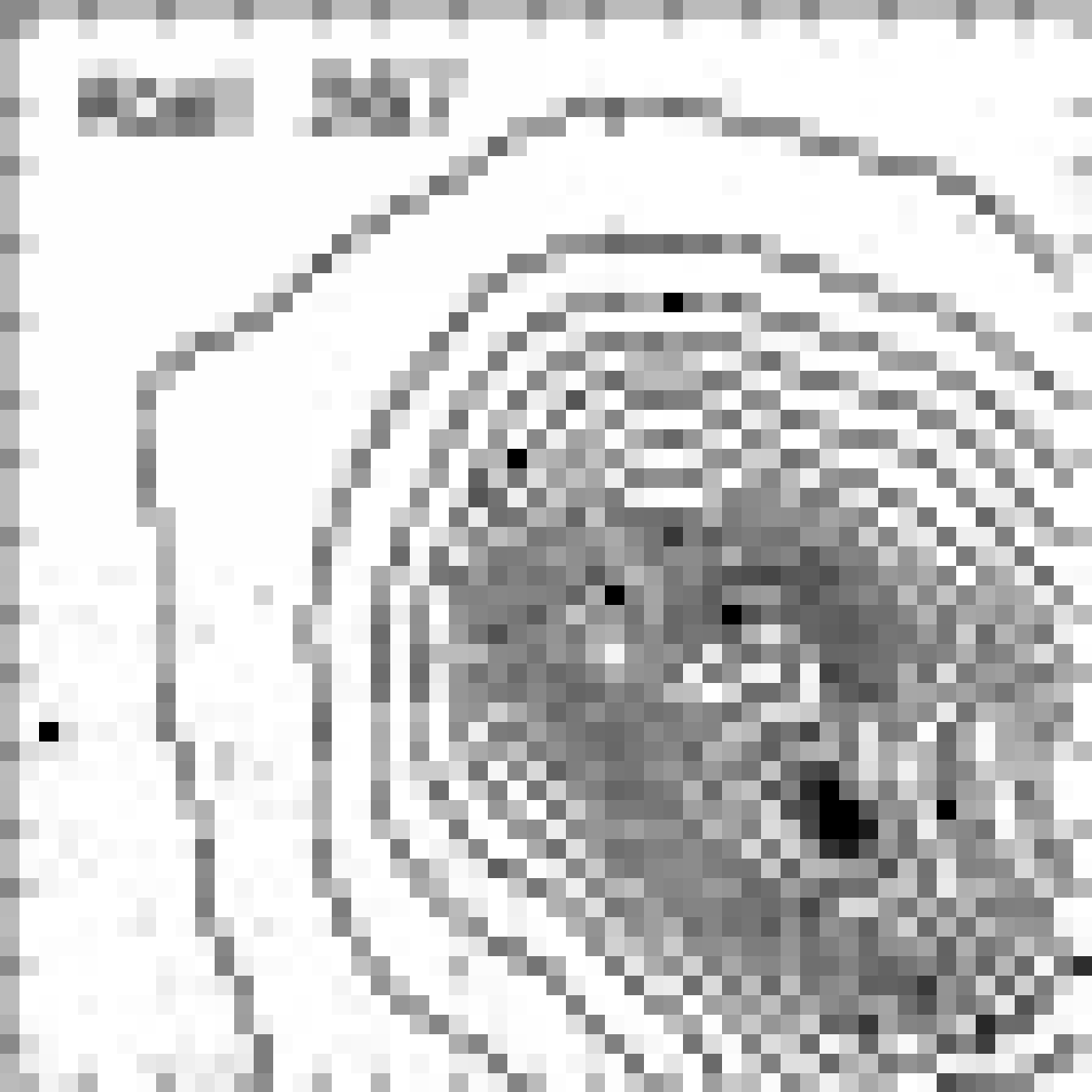,width=42mm,angle=-90}
\hspace{-1.5mm}
\psfig{file=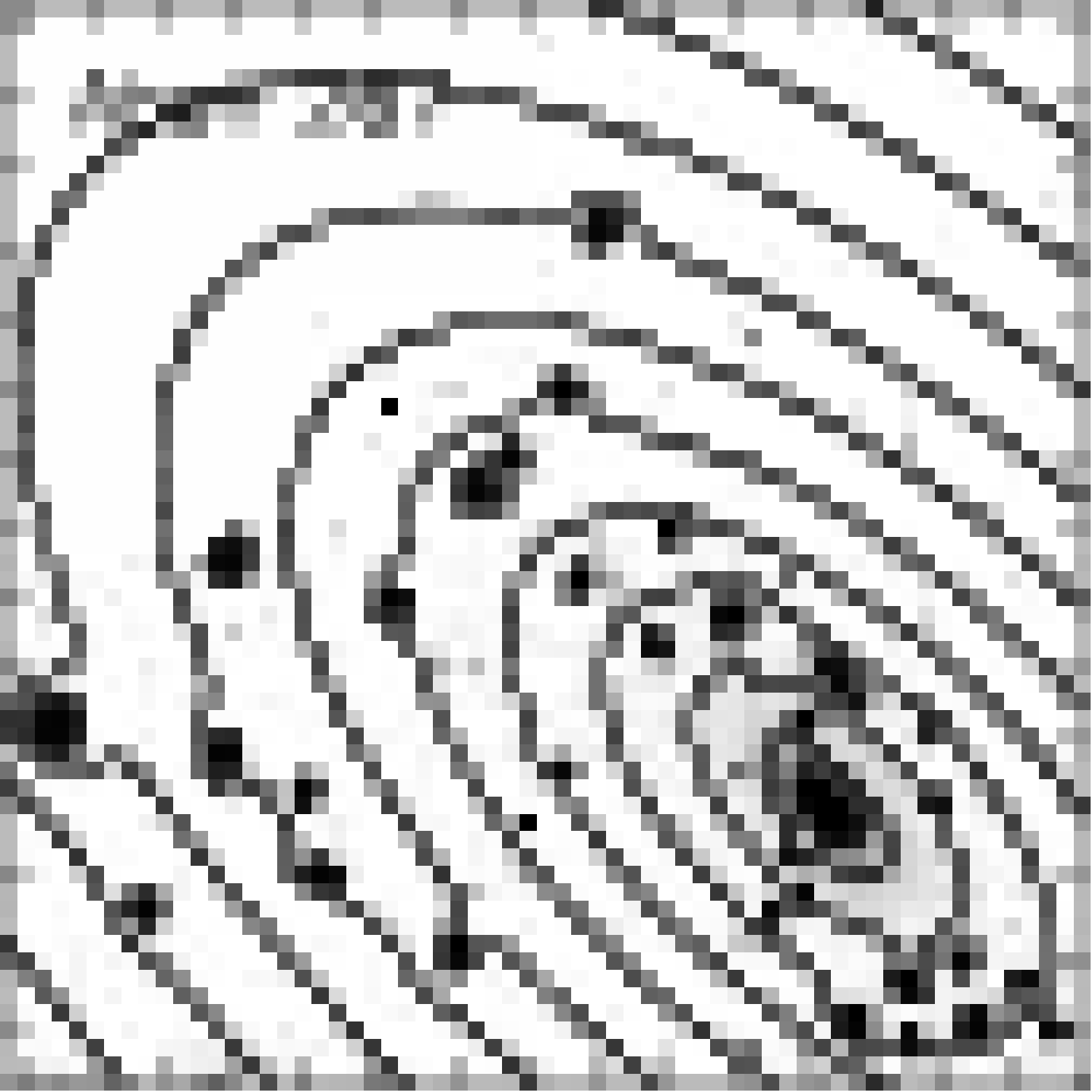,width=42mm,angle=-90}
\hspace{1mm}
\psfig{file=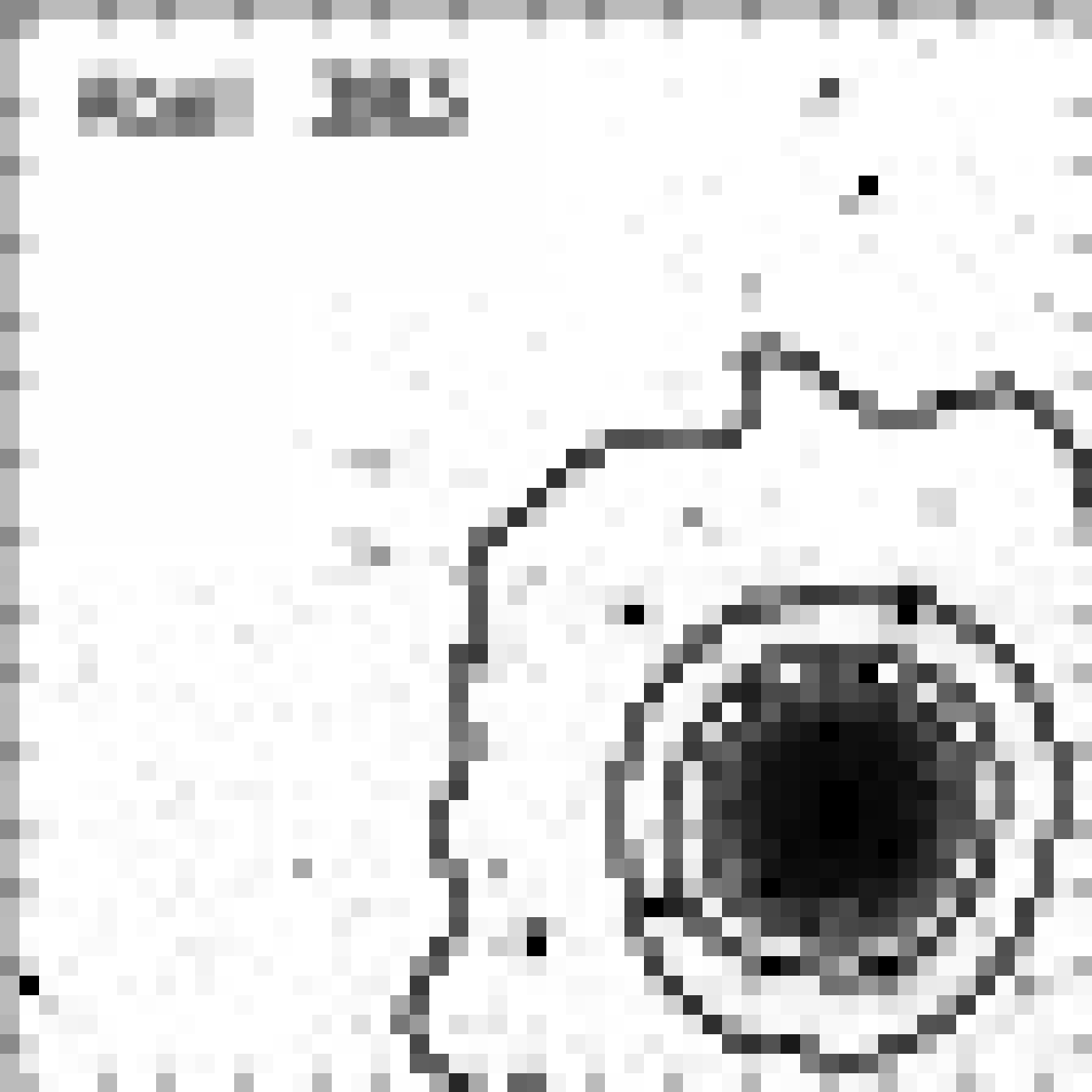,width=42mm,angle=-90}
\hspace{-1.5mm}
\psfig{file=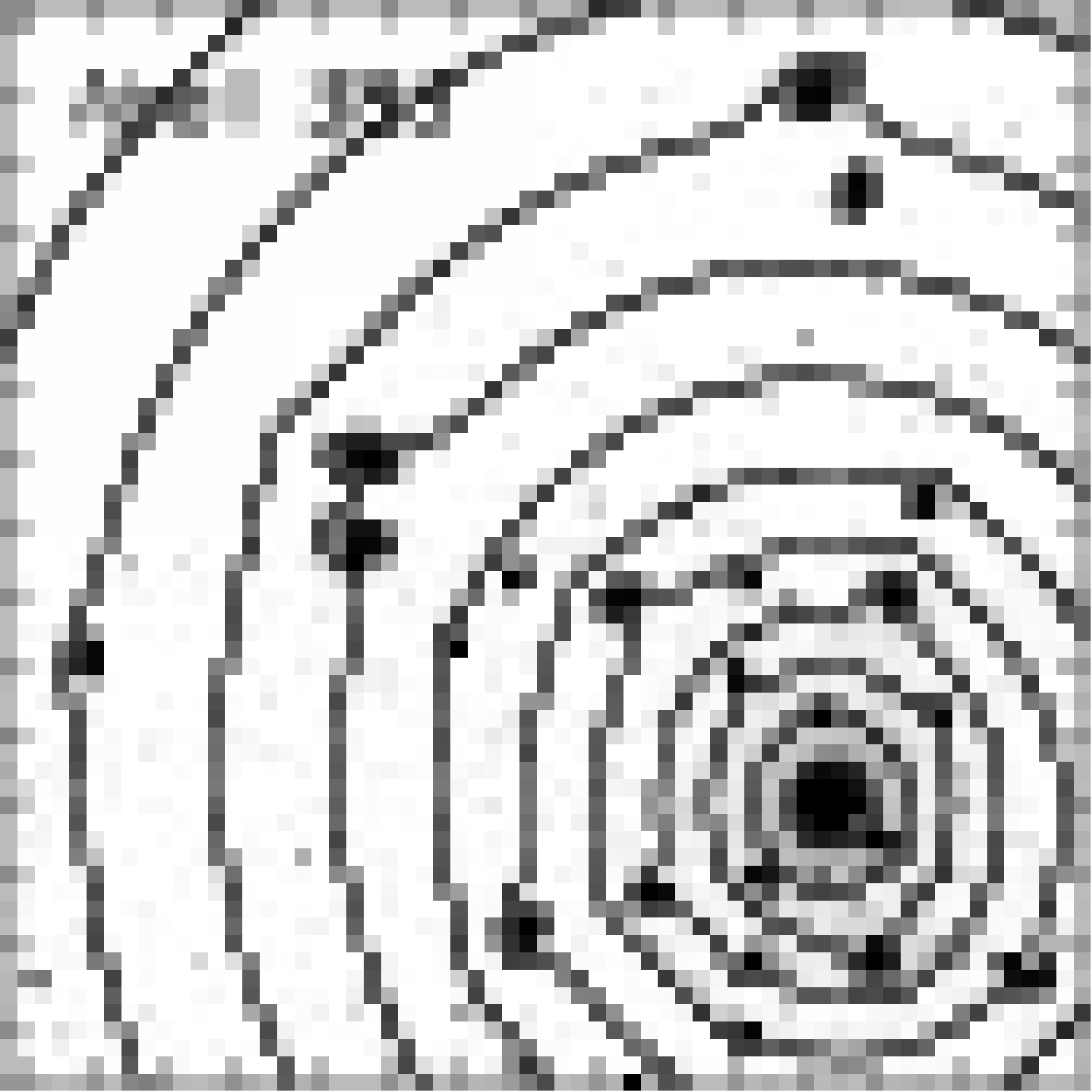,width=42mm,angle=-90}
}

\vspace{2mm}
\centerline{
\psfig{file=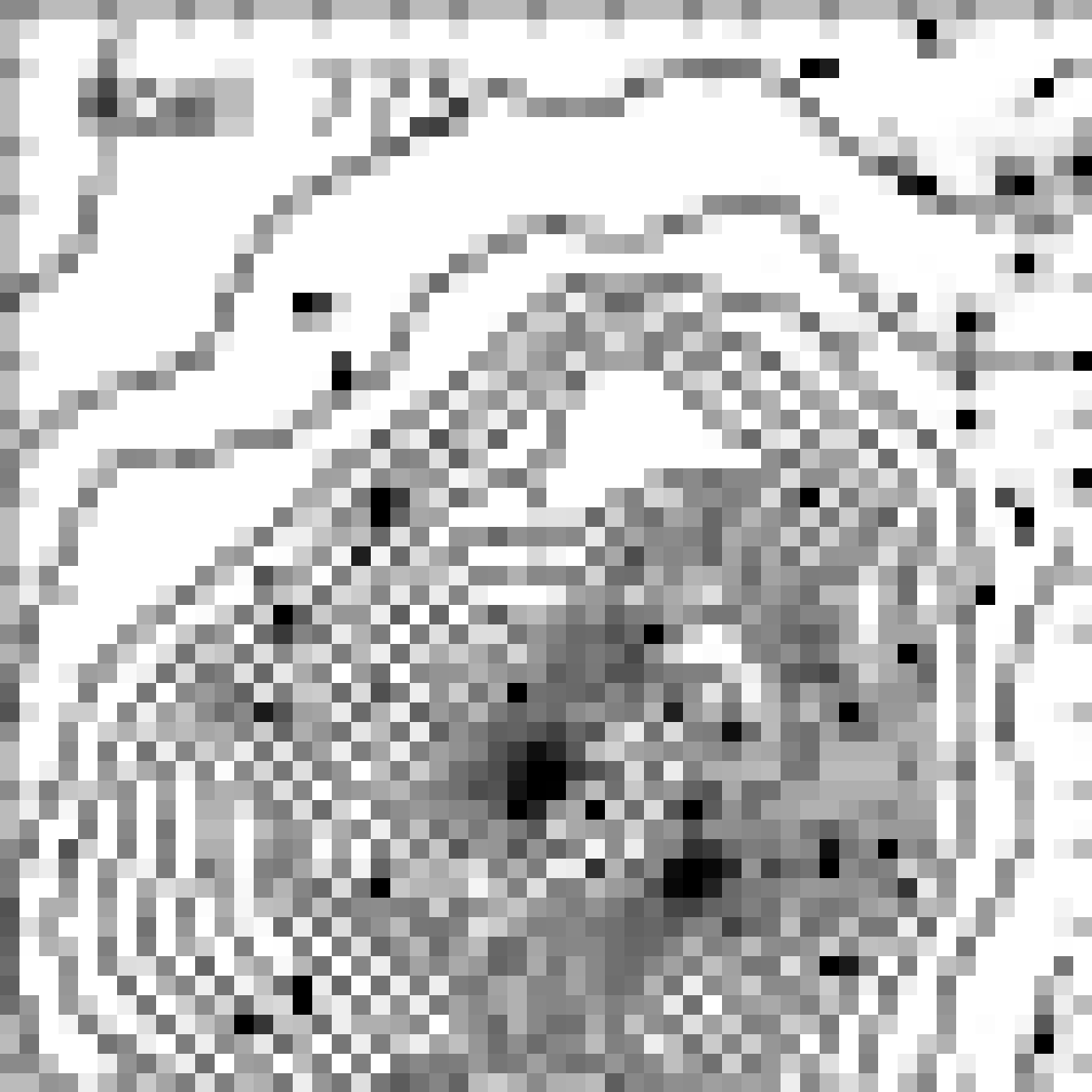,width=42mm,angle=-90}
\hspace{-1.5mm}
\psfig{file=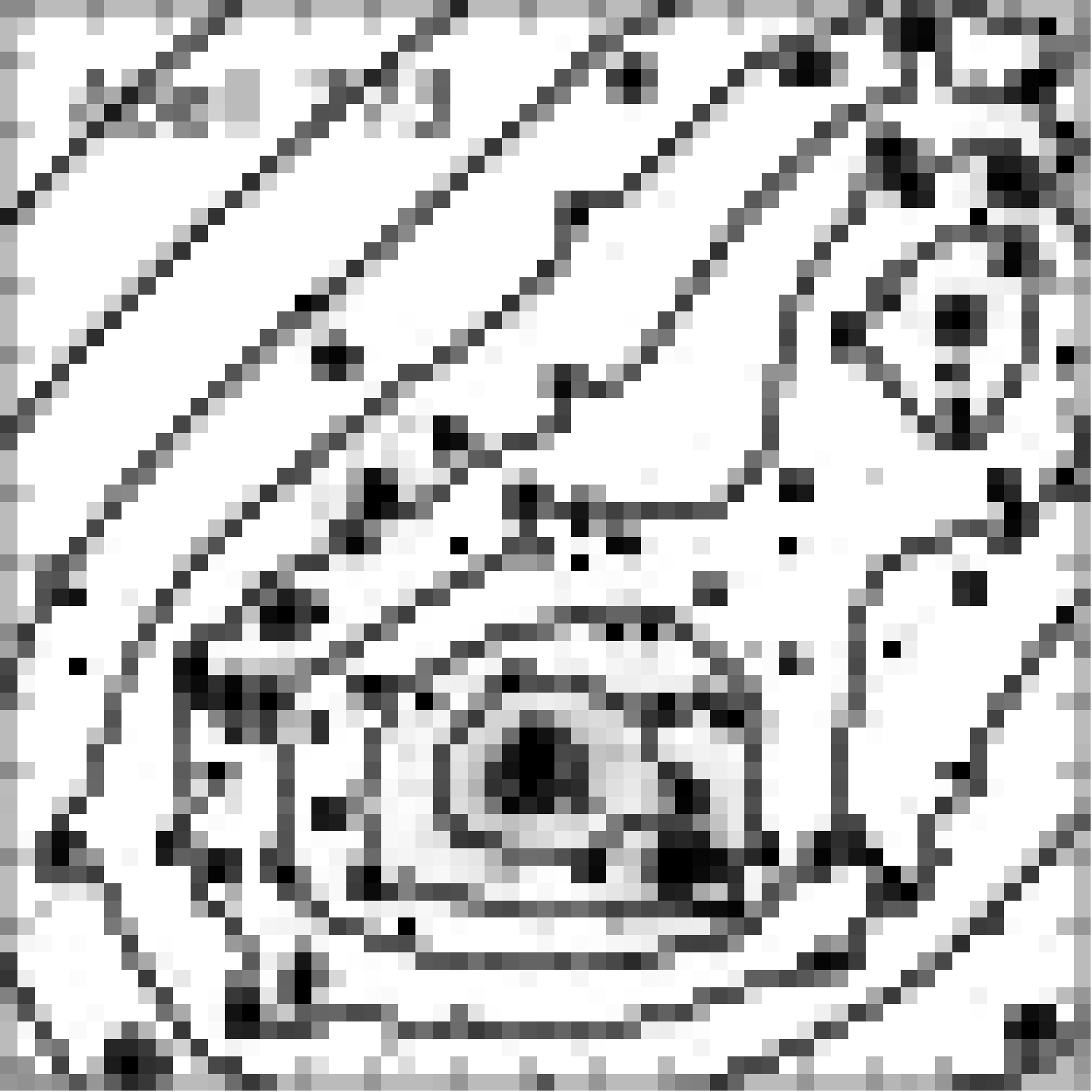,width=42mm,angle=-90}
\hspace{1mm}
\psfig{file=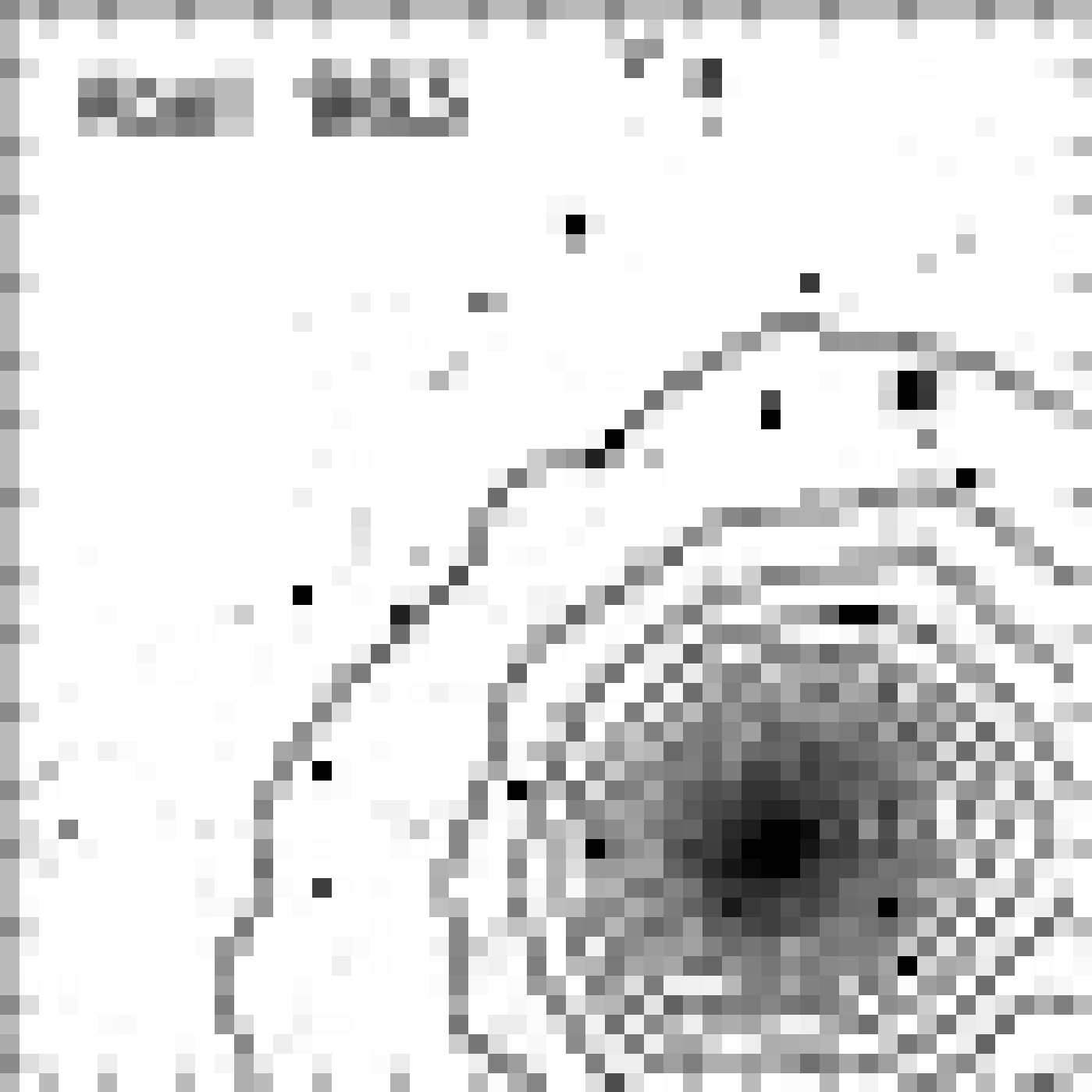,width=42mm,angle=-90}
\hspace{-1.5mm}
\psfig{file=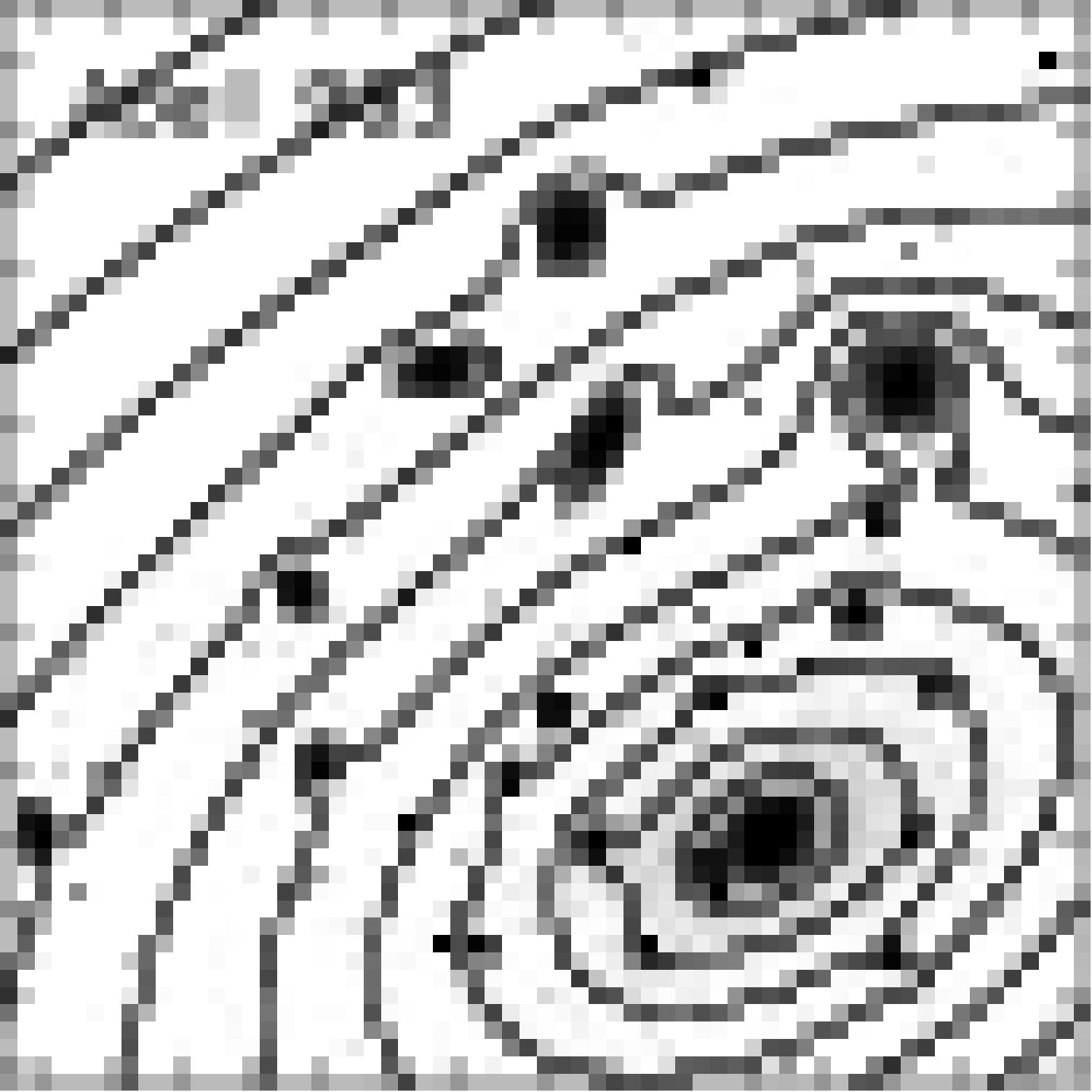,width=42mm,angle=-90}
}

\vspace{2mm}
\centerline{
\psfig{file=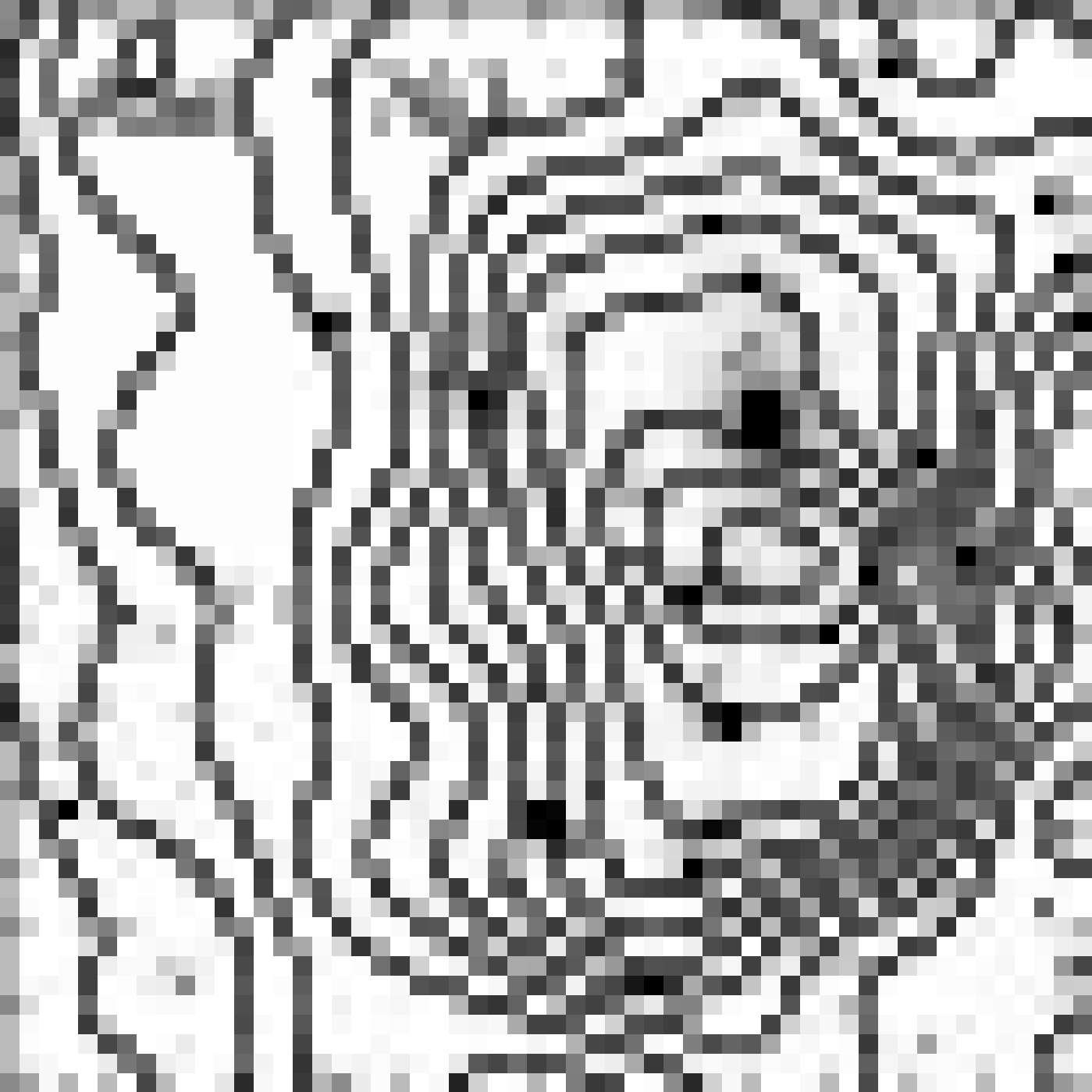,width=42mm,angle=-90}
\hspace{-1.5mm}
\psfig{file=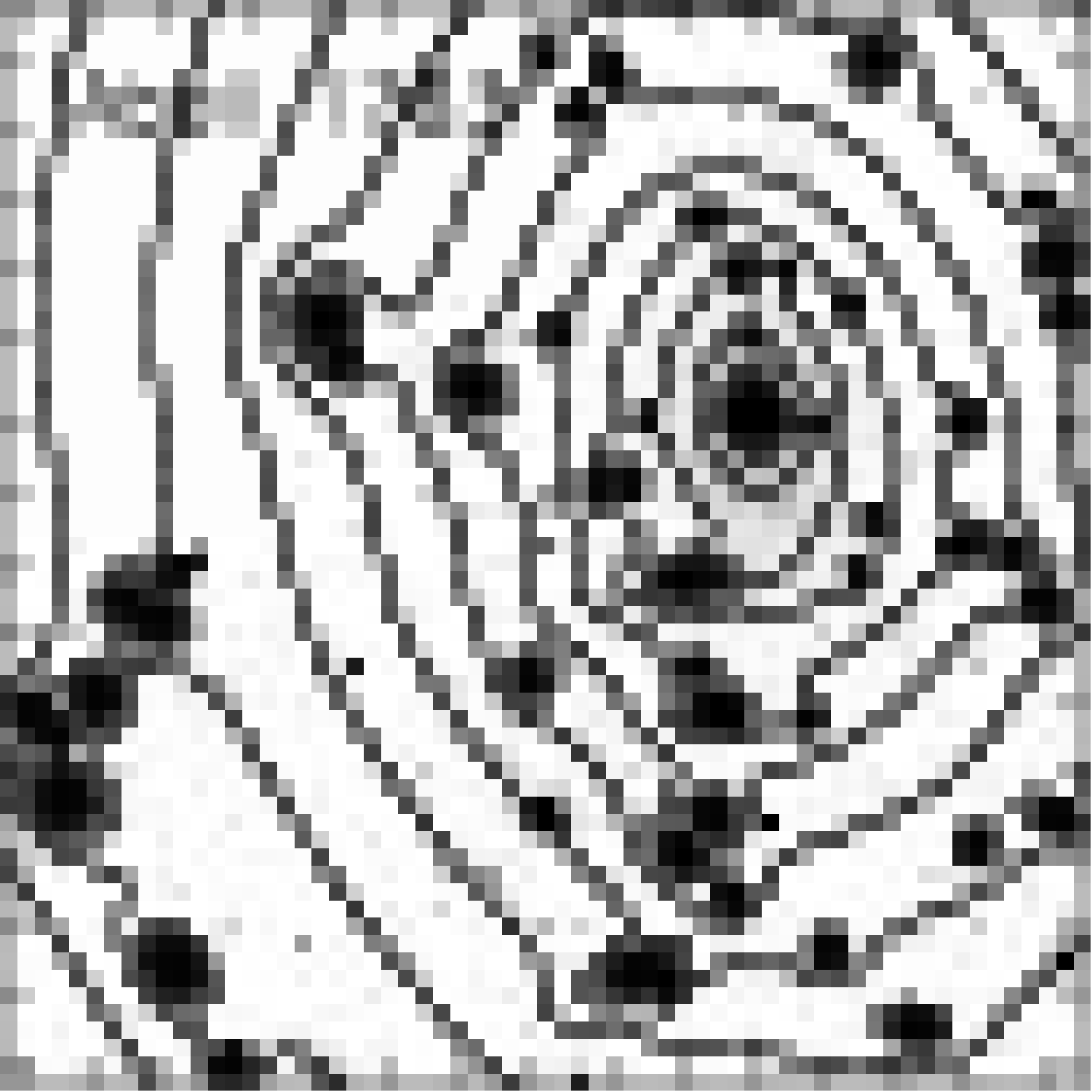,width=42mm,angle=-90}
\hspace{1mm}
\psfig{file=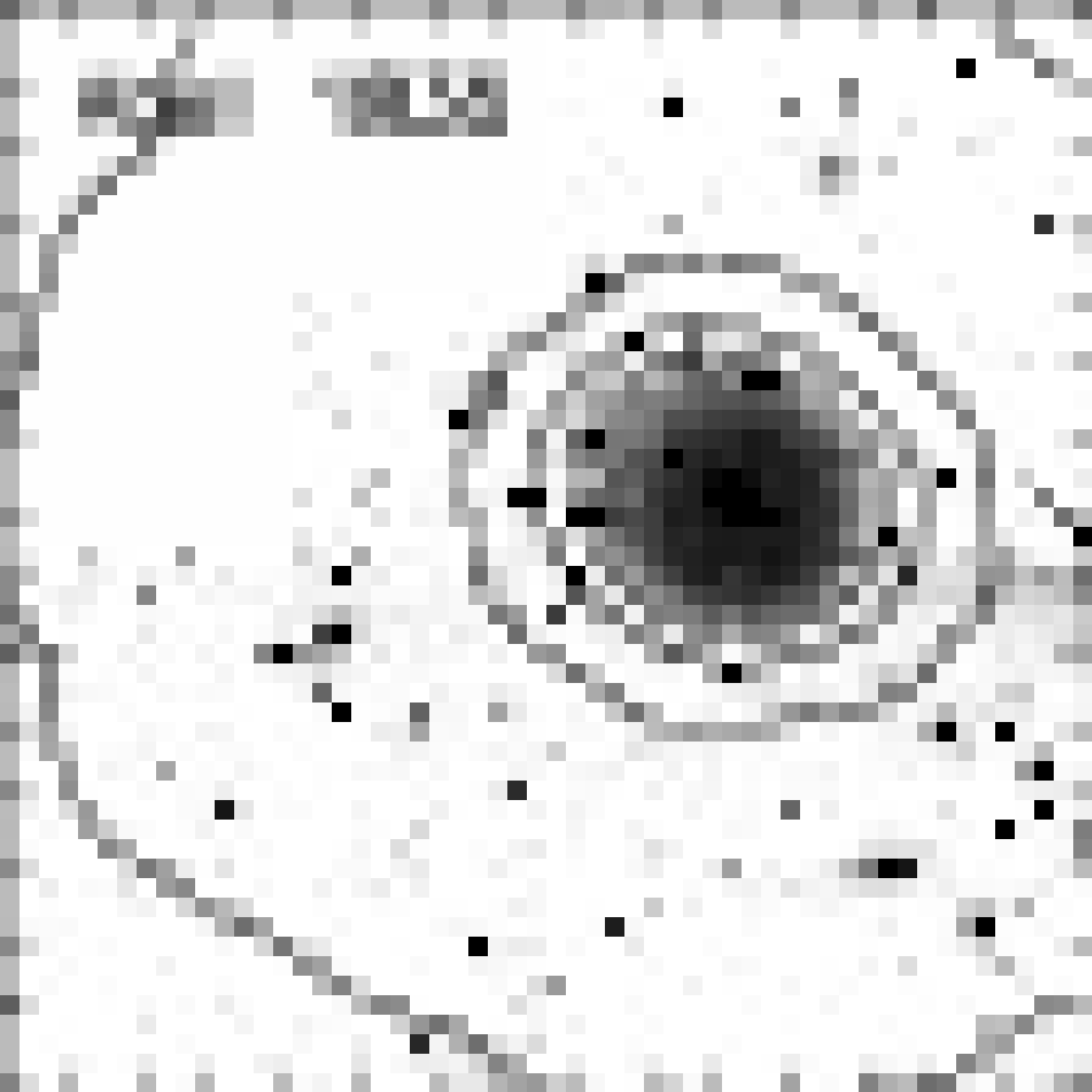,width=42mm,angle=-90}
\hspace{-1.5mm}
\psfig{file=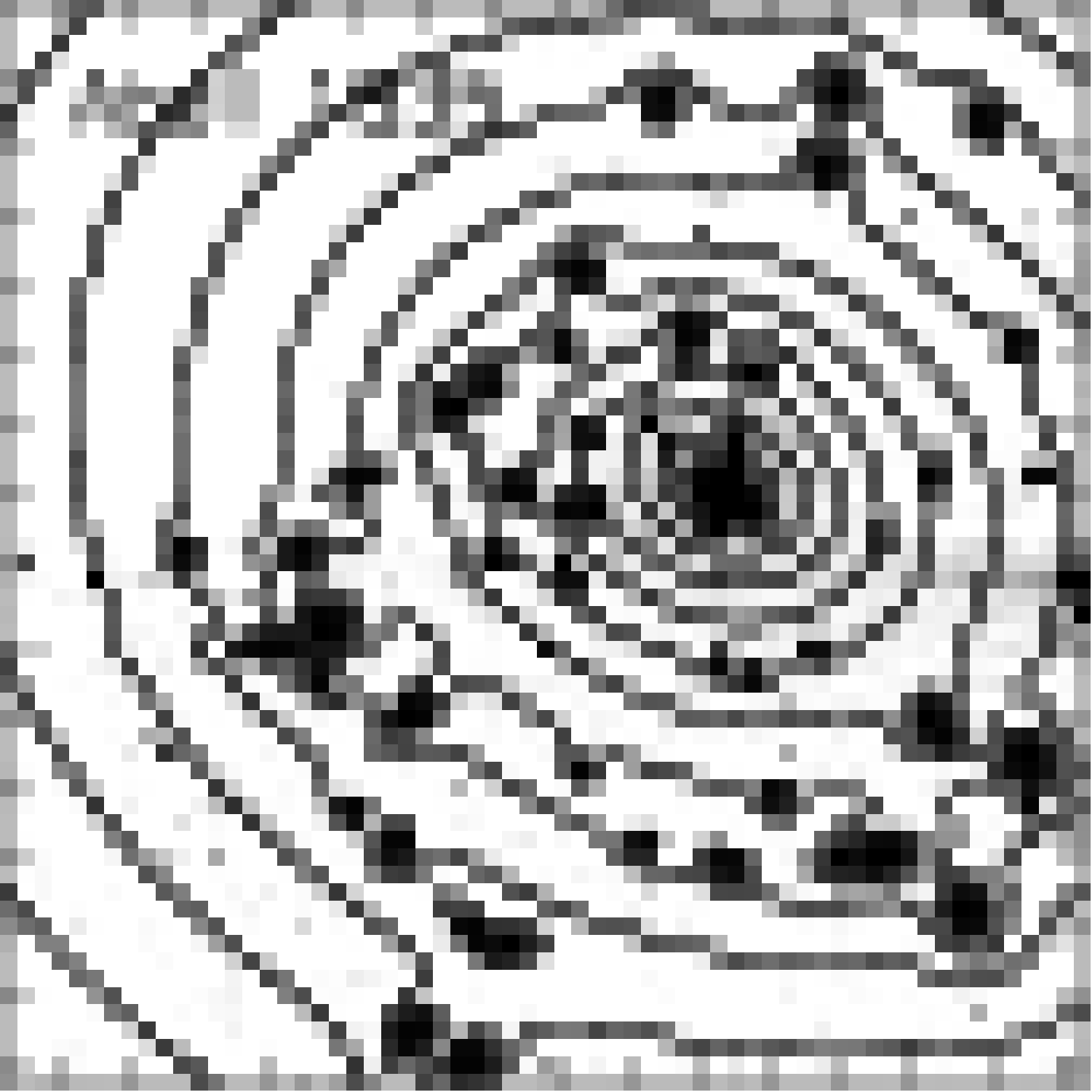,width=42mm,angle=-90}
}

\vspace{2mm}
\centerline{
\psfig{file=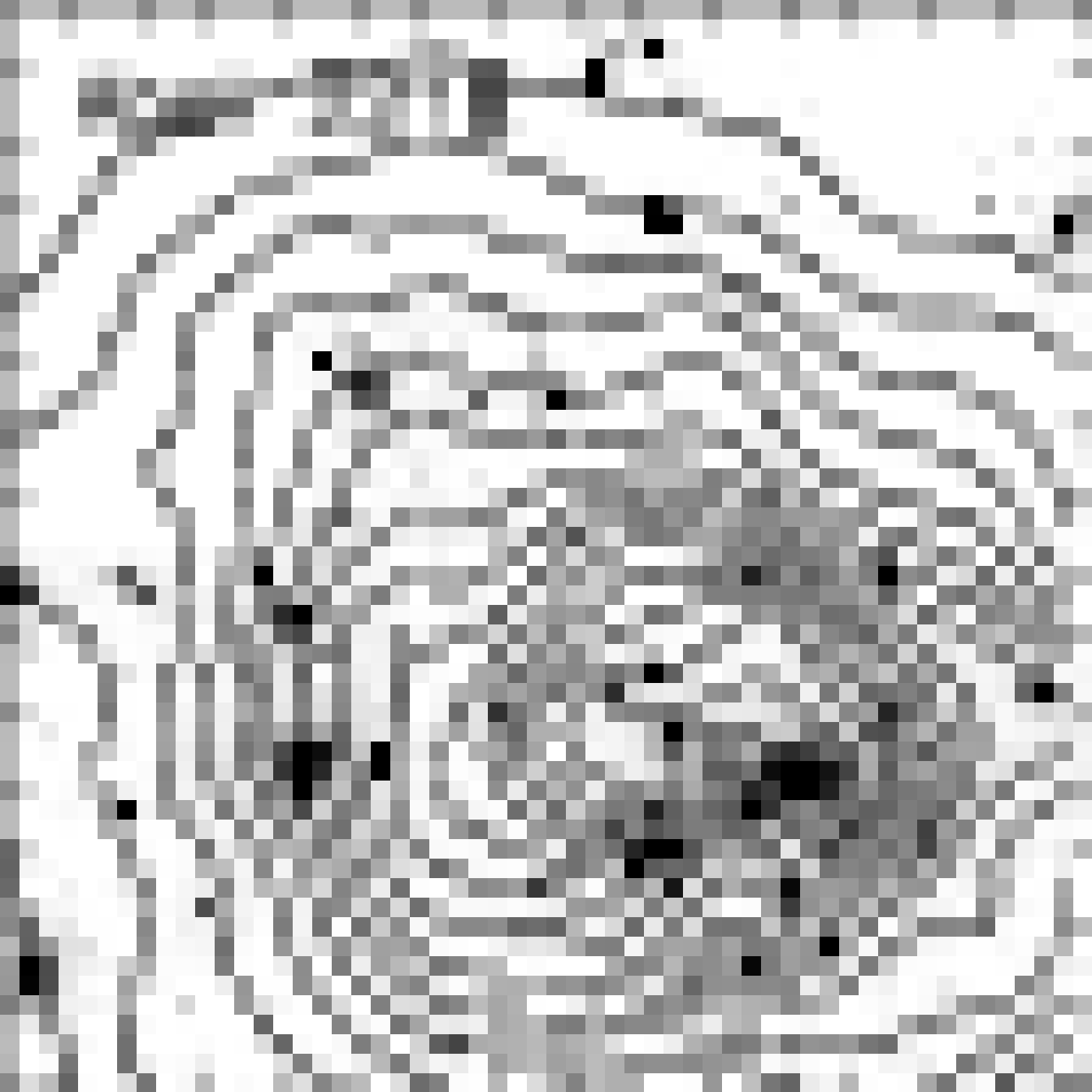,width=42mm,angle=-90}
\hspace{-1.5mm}
\psfig{file=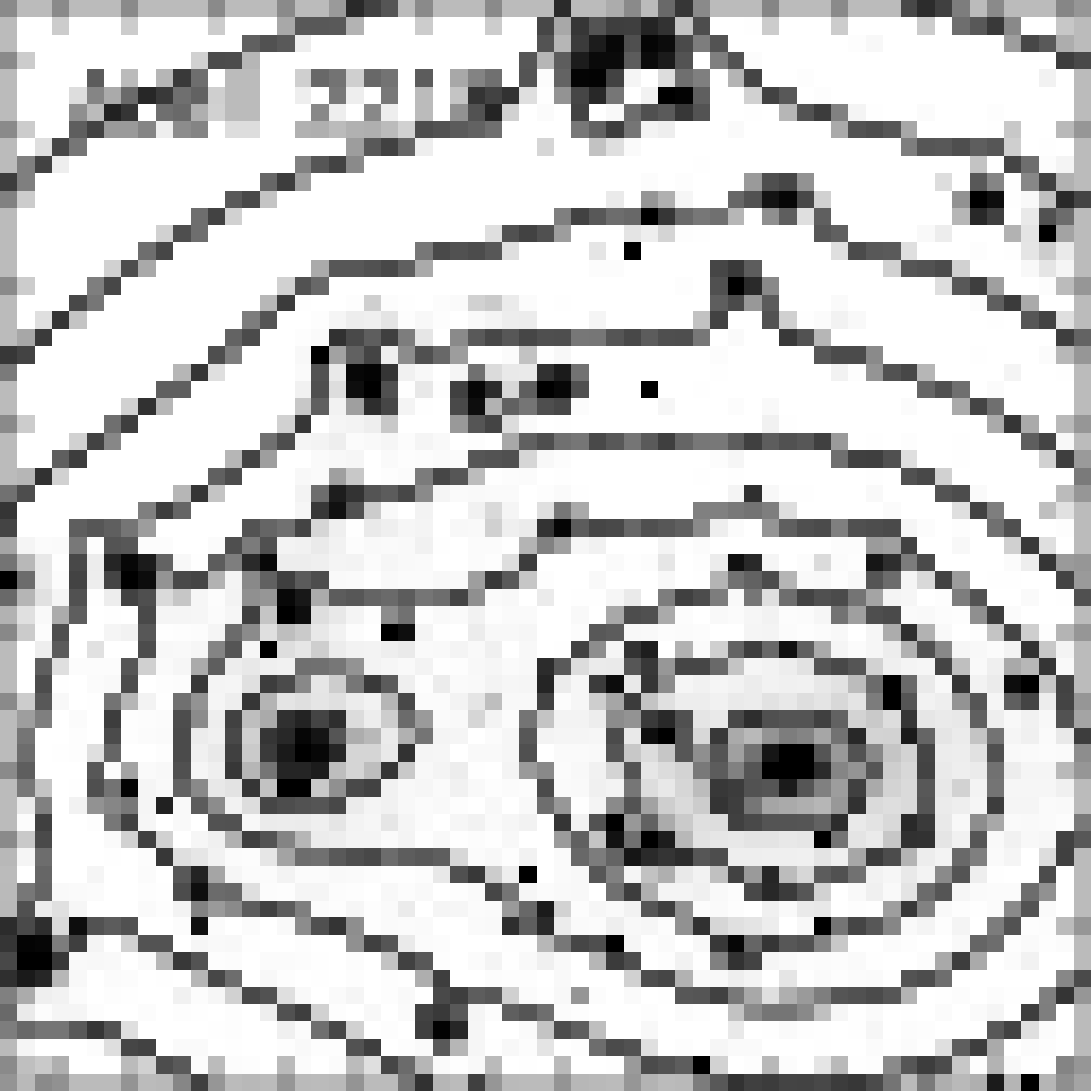,width=42mm,angle=-90}
\hspace{1mm}
\psfig{file=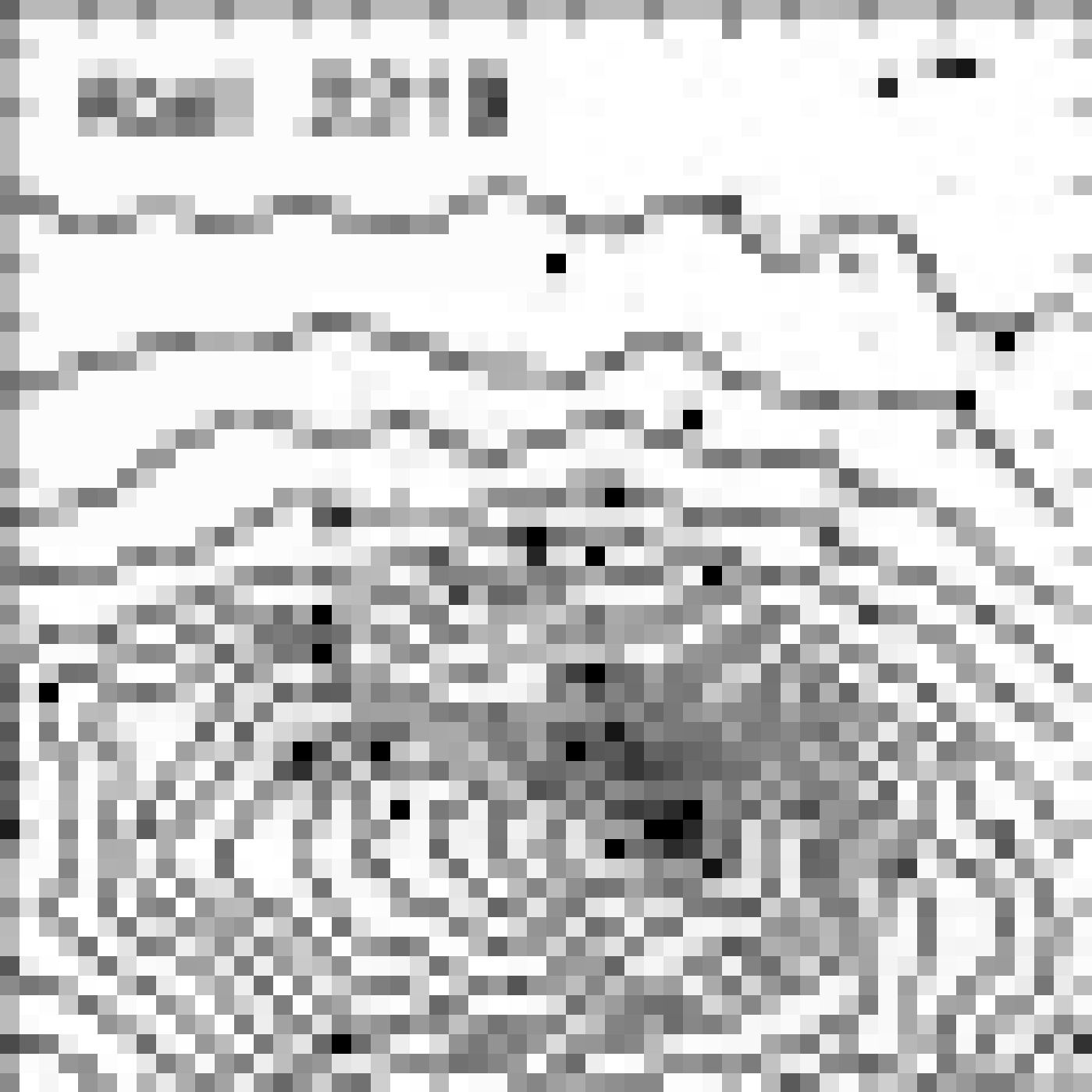,width=42mm,angle=-90}
\hspace{-1.5mm}
\psfig{file=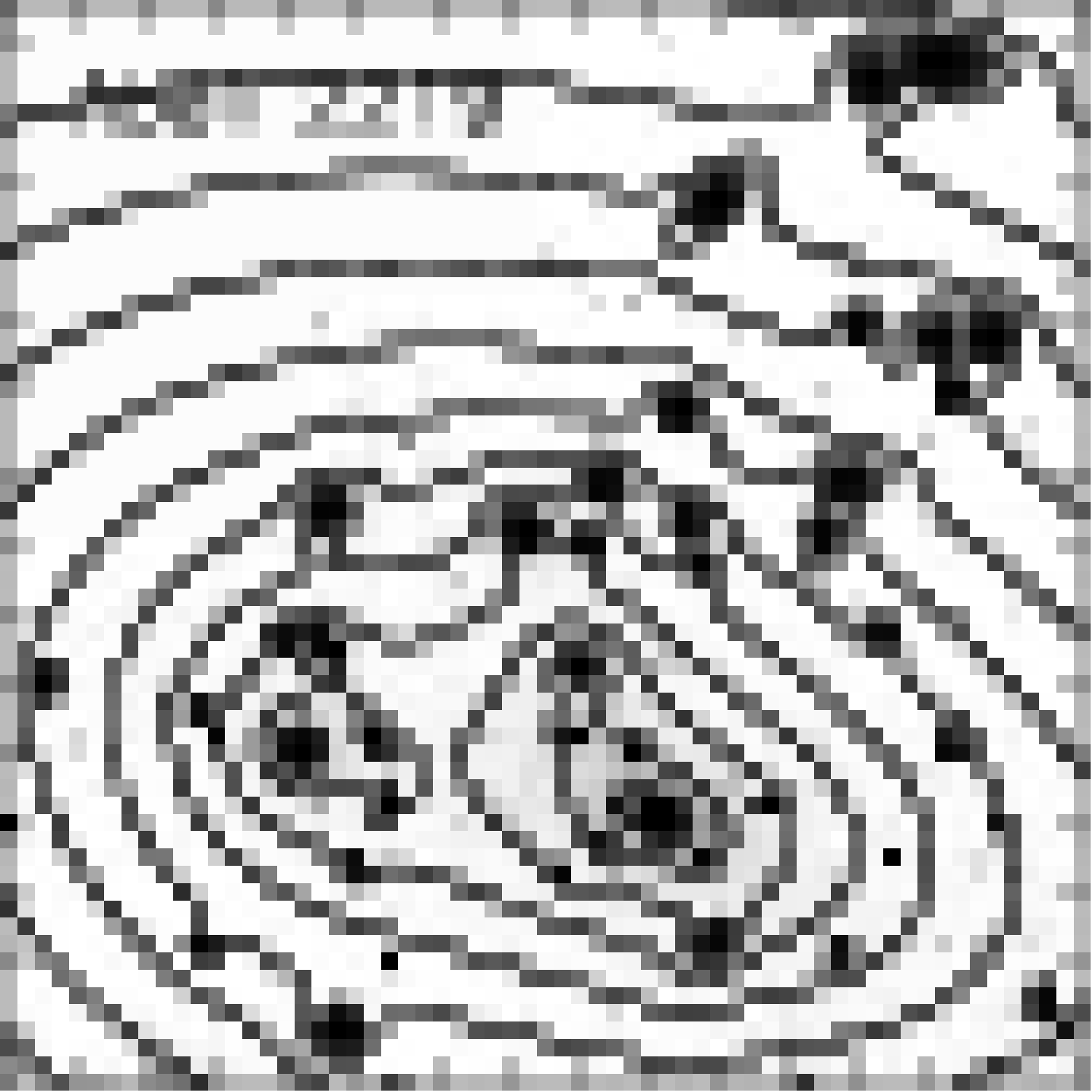,width=42mm,angle=-90}
}

\caption{
For each cluster in the sample, we show (left) the adaptively smoothed
X--ray flux contours and (right) the iso--mass contours calculated
from the best--fit lens models.  The greyscales are low--resolution
renderings of the \emph{HST}/WFPC2 frame.  All three of the clusters
classified as relaxed (A\,383, A\,963 and A\,1835; Table~\ref{masstx})
have very regular and centrally concentrated X--ray and mass
morphologies.  The seven unrelaxed clusters all have irregular X--ray
morphologies, and four of them (A\,68, A\,773, A\,2218, A\,2219) have
bi/tri--modal mass morphologies.  All contours are equally spaced in
the log.
\label{contours}}
\end{figure*}


We now compare the cluster mass distributions with the X--ray
observations to gain further leverage in diagnosing the maturity of
the full sample of 10 clusters.

Iso--mass contours computed from the best--fit lens models and
adaptively smoothed X--ray flux contours from the \emph{Chandra}
observations are overplotted on the \emph{HST} frames in
Fig.~\ref{contours}.  We also carefully check the pointing accuracy of
the \emph{Chandra} observations using $28'{\times}42'$ panoramic
CFH12k imaging of these cluster fields (Czoske 2002) to confirm that
the \emph{Chandra} astrometry matches the frame defined by the optical
data to an rms accuracy of ${\ls}2{\rm kpc}$ at the cluster redshift.
We measured the offset between the position of the X--ray peak in each
\emph{Chandra} frame and list ${\drpeak}$, the offset between this
position and the center of mass in the mass--maps in
Table~\ref{masstx}.  We compare the mass and X--ray morphologies with
the mass and luminosity fractions discussed in \S\ref{masslk}.  Three
of the four high central mass fraction clusters (A\,383, A\,963,
A\,1835) appear relaxed at X--ray wavelengths, i.e.\ circular or
mildly elliptical morphology.  A\,267, is an exception to this picture
-- its X--ray flux contours are much less regular than the other three
systems, and there appears to be a large offset between X--ray and
mass centers.  The six low central mass fraction clusters also have
irregular X--ray morphologies and misalignment between X--ray and mass
peaks.

We also list in Table~\ref{masstx} the ratio of the total to annular
temperatures (${\txtot}/{\txann}$) of each cluster measured in
\S\ref{xray} to test for the presence of cool cores.  Eight of the
clusters, comprising the six with low central mass fractions plus
A\,267 and A\,963 display no evidence of a cool core.  The absence of
evidence for a cool core in A\,267 is unsurprising given the likely
dynamical disturbance in the cluster core as indicated by its X--ray
morphology.  A temperature ratio of unity for A\,963 is also
consistent with previous work on this cluster, which has traditionally
been classified as an ``intermediate'' cluster (e.g.\ Allen 1998),
i.e.\ it appears to be quite relaxed, but has not acquired a cool core
since (presumed) previous merger activity.  This is also consistent
with the mild ellipticity in the X--ray isophotes, in contrast to the
almost circular isophotes of A\,383 and A\,1835.

\subsubsection{Summary}

Table~\ref{masstx} lists all of the diagnostics described in this
section.  Each diagnostic in isolation offers a slightly different
view of the dynamical maturity of each cluster.  We combine all of the
available information to determine a robust diagnosis of each
cluster's maturity, and to identify the remaining uncertainties.  In
making the overall classifications listed in Table~\ref{masstx}, we
define the term ``relaxed'' to mean that the cluster is dynamically
mature in all diagnostics available to us, with the exception that we
do not require it to have a cool core.  In terms of the diagnostics
listed in Table~\ref{masstx}, this means that $N_{\rm DM}{=}1$,
${\Mcen}/{\Mtot}{\ge}0.95$, $L_{K,{\rm BCG}}/L_{K,{\rm tot}}{\gs}0.5$,
${\drpeak}{<}4{\rm kpc}$, and the X--ray morphology is either
circular, or mildly elliptical.  The unrelaxed clusters do not satisfy
one or more of these criteria.

We therefore conclude that 7 out of the 10 clusters in our study,
i.e.\ $70{\pm}20\%$ of X--ray luminous cluster cores at $z{\simeq}0.2$
are dynamically immature (the error bar assumes binomial statistics --
Gehrels 1986).  Henceforth we classify A\,383, A\,963 and A\,1835 as
``relaxed'' clusters and A\,68, A\,209, A\,267, A\,773, A\,1763,
A\,2218 and A\,2219 as ``unrelaxed'' clusters (Table~\ref{masstx}).

\subsection{Cluster Scaling Relations}
\label{masslx}

We now investigate the scaling relations between cluster mass,
temperature and X--ray luminosity, focusing on the normalization of
and scatter around the relations and the impact of the dynamical
immaturity of 70\% of the sample identified in \S\ref{mass}.

\subsubsection{Mass Versus X--ray Luminosity}\label{masslum}

The sample is selected on X--ray luminosity (\S\ref{design}), we
therefore begin with the mass--luminosity relation.  First, we explore
whether we can improve on the precision of the RASS--based X--ray
luminosities (Table~\ref{sample}) using the \emph{Chandra} data.  One
of the largest uncertainties in the luminosities quoted in
Table~\ref{sample} is that \emph{ROSAT}'s large PSF limited the
efficiency with which point--sources could be excised from the cluster
data.  The sub--arcsecond PSF of the \emph{Chandra} data overcome this
problem, however we find that the corrections for point--sources are
modest and comparable with the extrapolation uncertainties that arise
from \emph{Chandra}'s field of view which is too small to embrace all
of the extended emission from clusters at $z{=}0.2$.  The
\emph{Chandra}--based luminosities are therefore no more precise than
the \emph{ROSAT} luminosities at this redshift.  We therefore adopt
the X--ray luminosities upon which the sample was selected
(Table~\ref{sample}).

We plot ${\Mtot}$ versus X--ray luminosity in Fig.~\ref{mass-sub}.
Despite selecting very X--ray luminous clusters for this study
($L_X{\ge}8{\times}10^{44}\ergs$, 0.1--2.4\,keV), these data span
sufficient dynamic range in principal to constrain both the slope and
normalization of the mass--luminosity relation (c.f.\ Finoguenov et
al.\ 2001).  We parametrize the mass--luminosity relation as follows:

\begin{equation}
({\lx}/10^{44}{\rm erg\,s^{-1}}){=}A({\Mtot}/10^{14}{\Msol})^{1/\alpha}
\end{equation}

\noindent and try to solve for $A$ and $\alpha$ following Akritas \&
Bershady (1996) to account for errors in both variables and unknown
intrinsic scatter.  Unsurprisingly, given the large scatter that is
immediately obvious upon visual inspection of Fig.~\ref{mass-sub},
this exercise fails.  We therefore fix the slope parameter and simply
measure the normalization, $A$.  This is done by computing the mean
mass and luminosity, and then solving $\log
A{=}\langle\log(L_X)\rangle{-}\langle\log(M_{\rm
tot})\rangle/{\alpha}$.  Uncertainties in both mass and luminosity are
included in the calculation by repeating it $10^4$ times, on each
occasion drawing values of $M_{\rm tot}$ and $L_X$ randomly from the
distributions defined by the error bars on X--ray luminosity and mass
listed in Tables~\ref{sample}~and~\ref{masstx} respectively.  Simple
gravitational collapse models predict that ${\alpha}{=}0.75$ (Kaiser
1986), we therefore initially measure the normalization using this
value for the slope, and also compute the intrinsic scatter around
this model.  These calculations are performed for the whole sample of
ten clusters and the relaxed and unrelaxed sub--samples, and the
results listed in Table~\ref{screl}.  Based on these calculations,
there is no evidence for segregation between relaxed and unrelaxed
clusters in the mass--luminosity plane, and the scatter is
$\sigma_{\rm M}{\simeq}0.4$.

We repeat these calculations using an empirical determination of the
slope: ${\alpha}{=}0.76^{+0.16}_{-0.13}$ (Allen et al.\ 2003), drawing
randomly from the error distribution on $\alpha$ in the same manner as
described above for the mass and luminosity data.  This has the effect
of broadening the uncertainties on the normalizations listed in
Table~\ref{screl}, but does not change the overall conclusion.  

\subsubsection{Mass Versus Temperature}

\begin{table}
\caption{Scaling Relations -- Normalizations and Scatters\label{screl}}
{\small
\begin{tabular}{llll}
\hline
Sample      & Slope$^a$      & Normalization       & Scatter\cr
\hline
\multispan4{\hfil Mass--luminosity: $L{=}A\,M^{1/\alpha}$\hfil}\cr
\hline
All         & $\alpha{=}0.75$                 & $A{=}2.64{\pm}0.36$ & $\sigma_M{=}0.41$ \cr
Relaxed     &                                 & $A{=}2.63{\pm}0.36$ & $\sigma_M{=}0.30$ \cr
\smallskip
Unrelaxed   &                                 & $A{=}2.65{\pm}0.50$ & $\sigma_M{=}0.45$ \cr
All         & $\alpha{=}0.76^{+0.18}_{-0.13}$ & $A{=}2.53{\pm}1.74$ & $\sigma_M{=}0.41$ \cr
Relaxed     &                                 & $A{=}2.50{\pm}2.05$ & $\sigma_M{=}0.30$ \cr
Unrelaxed   &                                 & $A{=}2.54{\pm}1.68$ & $\sigma_M{=}0.45$ \cr
\hline
\multispan4{\hfil Mass--temperature: $kT{=}B\,M^{1/\beta}$\hfil}\cr
\hline
All         & $\beta{=}1.5$                   & $B{=}3.59{\pm}0.24$ & $\sigma_T{=}0.42$ \cr
Relaxed     &                                 & $B{=}2.76{\pm}0.20$ & $\sigma_T{=}0.29$ \cr
\smallskip
Unrelaxed   &                                 & $B{=}4.01{\pm}0.36$ & $\sigma_T{=}0.43$ \cr
All         & $\beta{=}1.58^{+0.06}_{-0.07}$  & $B{=}3.72{\pm}0.30$ & $\sigma_T{=}0.41$ \cr
Relaxed     &                                 & $B{=}2.88{\pm}0.26$ & $\sigma_T{=}0.29$ \cr
Unrelaxed   &                                 & $B{=}4.15{\pm}0.41$ & $\sigma_T{=}0.42$ \cr
\hline
\multispan4{\hfil Luminosity--temperature: $kT{=}C\,L^{1/\gamma}$\hfil}\cr
\hline
All         & $\gamma{=}2$                    & $C{=}2.21{\pm}0.09$ & $\sigma_T{=}0.33$ \cr
Relaxed     &                                 & $C{=}1.70{\pm}0.07$ & $\sigma_T{=}0.24$ \cr
\smallskip
Unrelaxed   &                                 & $C{=}2.46{\pm}0.14$ & $\sigma_T{=}0.30$ \cr
All         & $\gamma{=}2.09^{+0.29}_{-0.29}$ & $C{=}2.22{\pm}0.71$ & $\sigma_T{=}0.33$ \cr
Relaxed     &                                 & $C{=}1.71{\pm}0.60$ & $\sigma_T{=}0.24$ \cr
Unrelaxed   &                                 & $C{=}2.48{\pm}0.76$ & $\sigma_T{=}0.30$ \cr
\hline
\end{tabular}
}
\begin{tabular}{l}
\parbox[t]{80mm}{\footnotesize\addtolength{\baselineskip}{-5pt}
$^a$ For each scaling relation, the first slope parameter listed is
  based on the self--similar collapse (e.g.\ Kaiser 1986).  The second
  value in each case is taken from recent empirical measurements:
  Allen et al.\ 2003; Finoguenov et al.\ 2001; Markevitch 1998.

}\\
\end{tabular}

\end{table}

\begin{figure*}
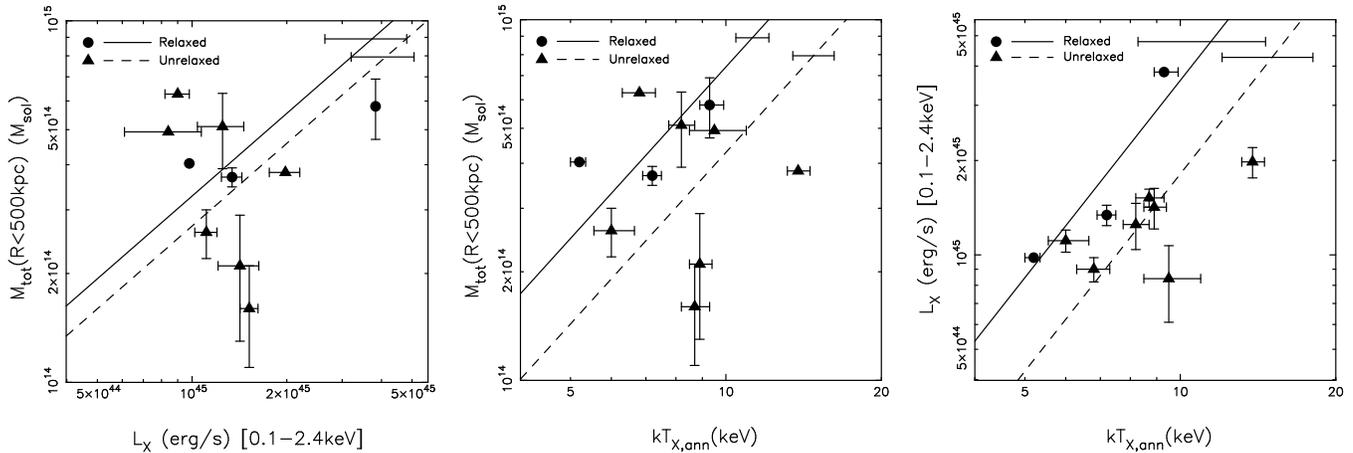

\centerline{
\psfig{file=masslx-v02.ps,width=57mm,angle=0}
\hspace{2mm}
\psfig{file=masstx-v11.ps,width=57mm,angle=0}
\hspace{2mm}
\psfig{file=lxtx-v02.ps,width=57mm,angle=0}
}
\caption{ Mass--luminosity (left), mass--temperature (center) and
luminosity--temperature (right) relations for our sample of ten
clusters.  The relaxed/unrelaxed clusters are shown by the
circular/square symbols as explained in the legend.  The solid and
dashed lines show the best--fit relations normalized by the relaxed
and unrelaxed clusters respectively; see \S\ref{masslx} and
Table~\ref{screl} for further information.  The error bars on each line
show the uncertainty on the normalization of each relation.  In
summary, these relations show that the large scatter detected in the
mass--luminosity plane appears to be symmetric; in both the
mass--temperature and luminosity--temperature planes, the scatter
appears to be asymmetric, with the unrelaxed clusters being on average
hotter than the relaxed clusters.  This segregation is statistically
insignificant in the luminosity--temperature plane, and significant at
the 2--3$\sigma$ level in the mass--temperature plane.}
\label{mass-sub}
\end{figure*}

We plot ${\Mtot}$ versus ${\txann}$ in Fig.~\ref{mass-sub}, and
parametrize the relation as:

\begin{equation}
({\txann}/1\,{\rm keV}){=}B({\Mtot}/10^{14}{\Msol})^{1/\beta}
\end{equation}

\noindent Note that we consider $\txann$, and not $\txtot$; the
results described here are therefore robust to the presence of cool
cores in relaxed clusters.

We adopt the theoretically predicted slope: ${\beta}{=}1.5$, which is
consistent with observations for the most massive clusters (e.g.\
Finoguenov et al.\ 2001; Allen et al.\ 2001).  Following
the procedures described above, we measure the normalization and
scatter, and list the results in Table~\ref{screl}.  Two significant
results emerge from this exercise.  First, the unrelaxed clusters are
on average $40\%$ hotter than the relaxed clusters at 3--${\sigma}$
significance, and second the scatter about the mass--temperature
relation for all clusters is ${\sigma_T}{\simeq}0.4$.  The statistical
significance of the temperature off--set is reduced to 2.5--${\sigma}$
if an empirically measured value of ${\beta}$ is used in place of the
theoretical value (e.g.\ ${\beta}{=}1.58^{+0.06}_{-0.07}$ --
Finoguenov et al.\ 2001).  

\subsubsection{X--ray Luminosity Versus Temperature}

Finally, we parametrize the luminosity--temperature relation as: 

\begin{equation}
({\txann}/1\,{\rm keV}){=}C({\lx}/10^{44}{\rm erg\,s^{-1}})^{1/\gamma}
\end{equation}

\noindent and repeat the analysis described above.  Adopting the
theoretical value of ${\gamma}{=}2$, the measured values of $C$
(Table~\ref{screl}) indicate that unrelaxed clusters are $30\%$ hotter
than relaxed clusters, at 2.4--$\sigma$ significance, i.e.\ less
significance than in the mass--temperature plane.  However, adopting
an empirical measurement of ${\gamma}$ in the fit
(${\gamma}{=}2.09{\pm}0.29$ -- Markevitch 1998) eliminates the
statistical significance in this difference.  Nevertheless, this hint
of structural segregation in the luminosity--temperature plane (see
also Fig.~\ref{masstx}) is significant for two reasons.  First, it
provides a lensing--independent cross--check on the results in the
mass--temperature plane, in that both luminosity and temperature
measurements are independent of the lens modelling upon which the
cluster mass measurements are based.  Second, it is consistent with
previous detections of structural segregation in the
luminosity--temperature plane (e.g.\ Fabian et al.\ 1994).  

\section{Discussion}\label{discussion}

This is the first study to combine a detailed, high--resolution
strong--lensing analysis of an objectively selected cluster sample
with analysis of a high--resolution X--ray spectro--imaging dataset.
As such, it affords the first opportunity to combine high--quality
optical and X--ray probes of cluster mass, structure and
thermodynamics.  The results presented in the previous section may be
summarized as follows:

\setcounter{fred}{0}
\begin{list}{(\roman{fred})}{\usecounter{fred}\setlength{\itemindent}{0mm}\setlength{\labelwidth}{15mm}\setlength{\labelsep}{2mm}\setlength{\leftmargin}{7mm}}
\setlength{\itemsep}{1.0mm}
\item $70{\pm}20\%$ of X--ray luminous cluster cores at $z{=}0.2$
  are dynamically immature;
\item scaling relations between cluster mass, luminosity and
  temperature display scatter of ${\sigma}{\sim}0.3$--0.6;
\item the normalization of the mass--temperature relation for unrelaxed
  (dynamically immature) clusters is $40\%$ hotter than for
  relaxed clusters. 
\end{list}

\noindent We discuss these results in \S\ref{dyn} and \S\ref{clsr},
and close by considering the implications of the results for the use
of massive clusters as cosmological probes (\S\ref{cosmology}).

\subsection{Dynamical Immaturity of Cluster Cores}\label{dyn}

Galaxy clusters grow by accreting DM, gas and galaxies from their
surroundings, including the filamentary structure (\S\ref{intro}).
The observed structure of clusters therefore probes a combination of
both the in--fall history and the relaxation processes that govern the
timescales on which clusters regain equilibrium following
cluster--cluster mergers.

The clusters can be grouped into three categories on the basis of the
mass and X--ray flux maps in Fig.~\ref{contours}.  The least ambiguous
category is the relaxed clusters (A\,383, A\,963 and A\,1835 -- see
\S\ref{mass}), all of which display a similar degree of relaxation in
both mass and X--ray.  The other two categories are sub--divisions of
the unrelaxed clusters.  First, we consider the four unrelaxed
clusters with obviously bi/tri--modal mass distributions (A\,68,
A\,773, A\,2218 and A\,2219).  Whilst the X--ray flux contours of
these clusters are both irregular and elongated in the same directions
as the mass distributions, there is no obvious evidence of
bi--modality in the X--ray flux.  On radial scales greater than a few
tens of kpc from the center of the cluster BCGs, the mass maps should
trace the distribution of DM.  The ICM in these clusters therefore
appears to be more relaxed than the DM distribution.  DM is generally
believed to be collisionless (Davis et al.\ 1985; however see also
Spergel \& Steinhardt 2000); in contrast, the ICM is baryonic and
therefore collisional.  The absence of X--ray bi--modality in clusters
with mass bi--modality therefore qualitatively supports the
collisionless DM hypothesis.

The third category comprises the remaining three unrelaxed clusters
(A\,209, A\,267 and A\,1763) for which there is no compelling evidence
in the current data for bi/tri--modality in the DM
distribution.  Nevertheless, several clues as to the true DM
distributions are present in the data.  For example, the strongly
elliptical mass distribution of A\,267 may indicate that the matter
distribution is more complex than a single elongated DM halo
plus cluster galaxy population (Edge et al.\ 2003).  Deep wide--field
space--based imaging is needed to explore this possibility,
specifically to search for evidence of other mass concentrations using
weak gravitational lensing (Kneib et al.\ 2003).  Turning to A\,1763,
galaxies appear to be falling in to this cluster along a ${\sim}2{\rm
Mpc}$ long filament (Vall\'ee \& Bridle 1982; Bardeau et al.\ 2004, in
prep.).  This cluster may therefore be experiencing a merger in the
plane of the sky, which is thus poorly sampled by the small
field--of--view of the WFPC2 data used in this study.  An alternative
interpretation of these three clusters is that sufficient time has
elapsed since the presumed merger event for the DM
distribution to relax and thus not present a bi/tri--modal structure
at the epoch of observation.  However, given the strong dynamical
disturbance in the X--ray flux maps, we consider this unlikely.

Ultimately cross--correlation of mass and X--ray maps of these and
similarly selected samples of clusters at higher redshifts will help
to constrain the relevant relaxation timescales and processes.
Critical next steps towards that goal are to obtain wider--field
space--based imaging to overcome uncertainties arising from the
current tiny pencil--beam WFPC2 observations of the cluster cores,
plus enlargement of the sample to overcome the small number statistics
of comparing sub--samples of ${\sim}3$ systems.  Uniformly deep X--ray
data would also enable tight constraints on the temperature structure
of the unrelaxed clusters to be achieved.  Such temperature maps would
help to more clearly delineate the structure of the ICM in the
unrelaxed clusters.

From a theoretical perspective, numerical simulations suggest that
both gas dynamics and substructure in the DM distribution may persist
for as long as ${\sim}5\,{\rm Gyr}$ following a merger event
(Schindler \& M\"uller 1993; Nakamura, Hattori \& Mineshige 1995;
Roettiger, Loken \& Burns 1997; Tormen, Diaferio \& Syer 1998;
de~Lucia et al.\ 2004).  This suggests that despite the expected
differences between the physics of the ICM and the DM, the relaxation
timescales for these two matter components may be comparable.  The
differences in cluster X--ray and mass morphologies noted above imply
that the X--ray and mass morphological evolution of clusters may
follow different evolutionary paths, even if the overall relaxation
timescales are indeed comparable.  This idea is supported by the
numerical simulations of Ricker \& Sarazin (2001), who found that
oscillations in the gravitational potential of a merging cluster (due
to the dominant collisionless DM) sustain turbulence and thus
non--relaxation of the ICM on timescales comparable with that required
for the DM to achieve equilibrium.

\subsection{Cluster Scaling Relations}\label{clsr}

Numerous observational and theoretical studies of cluster scaling
relationships have addressed the slope, normalization and intrinsic
scatter of these relations.  Structural segregation in the scaling
relation planes has also been discussed in the context of
cooling--flow and non--cooling--flow clusters in the
luminosity--temperature plane (Fabian et al.\ 1994; Markevitch 1998;
Allen \& Fabian 1998; Arnaud \& Evrard 1999).  Adding gravitational
lensing as a direct and clean probe of cluster mass to scaling
relation studies parallels the importance of adding this diagnostic
into the structural studies discussed above.  Lensing--based mass
estimates have been used to supplement X--ray cluster studies (e.g.\
Allen et al.\ 2003), however these studies have relied on
weak--lensing data without an absolute mass normalization from
strong--lensing constraints.  The resulting large error bars therefore
significantly degrade the advantage of using lensing as a probe of
cluster mass.  Smail et al.\ (1997) attempted to include lensing in
cluster scaling relation studies, using the weak--shear signal of
optically selected clusters as a surrogate for mass to construct a
shear--$L_X$ relation. Hjorth et al.\ (1998) went a step further,
using weak--lensing mass estimates to estimate the mass of eight
clusters, and thus construct the first lensing--based
mass--temperature relation.  However, as these authors point out,
their use of weak--lensing and heterogeneous selection function
undermines the precision of their work.

The important advantage of our study is that the lensing--based mass
measurements are based on detailed strong--lensing mass models, the
normalization of which are locked down by spectroscopically confirmed
multiple--image systems and cross--calibration between strong--lensing
and weak--lensing constrained clusters.  Combining the lensing results
with high quality X--ray spectro--imaging with \emph{Chandra} places
us in a hitherto unique position to explore the mass--temperature
plane.  We therefore concentrate our discussion of cluster scaling
relations in this area.  It is also important to note that the mass
information extracted from the lensing mass maps is two--dimensional.
In contrast, the information extracted from theoretical simulations is
three--dimensional.  Reliable calibrations between 2-- and
3--dimensional cluster masses have not yet been achieved.  We
therefore concentrate on discussing the scatter in the
mass--temperature plane and the related issue of structural
segregation, i.e.\ the \emph{relative} normalization of relaxed and
unrelaxed clusters.

\subsubsection{Scatter}\label{scatter}

Evrard, Metzler \& Navarro (1996) predicted that solely a broad--beam
measurement of the temperature of ICM in an individual cluster can be
used to measure cluster masses to an rms precision of
${\sim}10$--20\%.  This is in stark contrast to the
${\sigma}_M{\simeq}0.6$ scatter in mass that we detect in
\S\ref{masslx}.  An important clue as to the origin of this difference
is that the observational scatter appears to be dominated by the
unrelaxed clusters (see Table~\ref{screl}).  However several other
factors may also contribute to both the size of the observed scatter
and the apparent discrepancy between observation and theory.

We first consider the issue of aperture size.  The observational mass
measurements sample just the central $R{\le}500{\rm kpc}$ of each
cluster; this aperture matches approximately the radius at which the
cluster density is $2500{\times}$ the critical density, i.e.\
$\delta_c{=}2500$ (Smith et al.\ 2003).  The scatter may therefore be
dominated by systematics relevant only to the very central regions of
the clusters.  We use the ground--based weak--lensing analysis of the
same clusters by Bardeau et al.\ (2004, in prep.) to make a
preliminary estimate of how the scatter may reduce if the current
analysis were extended to larger radii ($\delta_c{\simeq}500$).  The
uncertainties on Bardeau et al.'s results are largely due to the poor
spatial resolution of ground--based data relative to our \emph{HST}
data.  However comparison of the two datasets suggests that one--third
of the scatter may be due to the small aperture size employed in this
study.  This variation in scatter as a function of overdensity is
consistent with recent observational results in the X--ray pass--band
(e.g.\ Ettori, De~Grandi \& Molendi 2002).  Therefore, after taking
account of aperture size, the scatter remains a factor of ${\sim}3$
larger than the simulations.

A further potentially important issue is that of selection effects.
Ricker \& Sarazin's (2001; see also Ritchie \& Thomas 2002)
simulations suggest that cluster--cluster mergers boost cluster X--ray
luminosities on timescales of a few Gyrs.  This may lead our X--ray
luminosity selected sample to be biased toward merging systems.
However the most luminous cluster in the sample is A\,1835, a
canonical relaxed cluster with a cool core, indeed a central excess of
X--ray flux is a signature of the most relaxed clusters.  Overall, the
details of cluster selection are a complicated subject, however there
is no strong evidence that the sample is biased toward unrelaxed
systems.  Nevertheless, a careful like--for--like comparison between
synthetic and observational cluster selection functions is an
important outstanding task.

In summary, we find a factor of 3 more scatter in the
mass--temperature relation than predicted.  Further work is required
to ensure that measurement methods applied to synthetic and observed
datasets are well matched, and to extend the space--based lensing
results to wider fields--of--view.  However, even after this work has
been completed, the disagreement between observation and theory may
persist, which would be a signature of missing physics in the
theoretical descriptions of cluster assembly and relaxation.

\subsubsection{Structural Segregation}

In \S\ref{masslx} we identified that unrelaxed clusters are 40\%
hotter than relaxed clusters at 2.5--${\sigma}$ significance.
Clearly, the aperture size issue noted above also impacts on this
result, and wider--field space--based weak--lensing observations of a
statistically complete sample are critical to a thorough investigation
of this uncertainty.  Nevertheless, the comparison with Bardeau et
al.'s (2004, in prep.) ground--based weak--lensing results provides
important reassurance that a substantial fraction of the 40\%
temperature offset is a genuine physical effect.

Indeed, recent observational and theoretical work supports the idea
that unrelaxed clusters are hotter than relaxed systems.  Using
spatially resolved spectroscopy with \emph{BeppoSAX}, Ettori,
De~Grandi \& Molendi (2002) identified the normalization of the
mass--temperature relation of ``non--cooling flow'' clusters to be
hotter than that of ``cooling flow'' clusters, however the large
uncertainties limited the statistical significance of this result.  On
the theoretical side, Ricker \& Sarazin's (2001) simulations of
cluster--cluster mergers indicate that merger--induced boosts of up to
a factor of ten (the amplitude of the boost depends on the mass ratio
between the merging clusters and the impact parameter of the
collision) in temperature can occur on short timescales (${\ls}1{\rm
Gyr}$) due to shock--heating of the gas in a major merger.  This
relatively brief luminosity boost suggests that not many clusters in
our sample should have temperatures ${\gs}3{\times}$ higher than the
mean relation.  This is indeed the case, with A\,2219 being the
possible sole example of a cluster currently experiencing a
temperature boost of this magnitude.  This cluster has previously been
identified as having recently experienced a core--penetrating merger
(e.g.\ Smail et al.\ 1995; Giovannini, Tordi \& Feretti 1999)

Despite the short--lived extreme temperature boosts, merger remnants
appear to asymptote to temperatures ${\sim}10$--40\% higher than the
pre--merger configuration, the precise long--term boost again
depending on the mass ratio and impact geometry.  Assuming that
elimination of aperture size and related issues (\S\ref{scatter})
reduces the temperature offset between relaxed and unrelaxed clusters
to ${\sim}20$--30\%, these theoretical results support the idea that
the cluster mergers are responsible for the structural segregation of
clusters in the mass--temperature plane.

If empirically clusters are either relaxed hosts of cool cores, or
unrelaxed (i.e.\ merging or merger remnant) systems without cool
cores, then merger--boosts may be sufficient to explain the structural
segregation.  However deep integrations with \emph{Chandra} and
\emph{XMM--Newton} indicate that the picture may not be so clear--cut.
For example some cool core clusters appear to be undergoing merger
activity (e.g.\ Perseus -- Churazov, et al.\ 2003), and some
dynamically relaxed clusters do not host a cool core (e.g.\ A\,963 in
this work).  The significance of merging cool core clusters and
relaxed non--cool core clusters for the merger--boost interpretation
of structural segregation in the mass--temperature plane is unclear at
this time.  For example the mass ratio of the Perseus merger may be so
large (i.e.\ the mass of the in--falling system so small relative to
Perseus) as to not be relevant to the current discussion.  However it
does suggest that alternative mechanisms such as cluster--to--cluster
variations in the level of initial pre--heating may be an additional
complication when interpreting the demographics of relaxed/unrelaxed
and cool core/non--cool core clusters (Babul, McCarthy \& Poole,
2003).

\subsection{Implications for Cluster Cosmology}\label{cosmology}

Massive galaxy clusters are one of a number of complementary probes of
cosmological parameters.  For example, many studies have used
empirical determinations of cluster scaling relations to convert the
observed X--ray luminosity and/or temperature functions into mass
functions.  The most massive clusters are rare objects, and thus the
constraints on the high--mass end of the mass function inferred from
such experiments enables constraints on a combination of
${\Omega}_{\rm M}$ and ${\sigma}_8$ in principal to be achieved.
Recent cluster--based estimates of ${\sigma}_8$ have yielded
discrepant results, with most estimates of ${\sigma}_8$ falling in the
range ${\sim}$0.6--1 (e.g.\ Eke et al.\ 1996; Nevalainen, Markevitch
\& Forman 2000; Allen, Schmidt \& Fabian 2001; Borgani et al.\ 2001;
Pierpaoli, Scott \& White 2001; Reiprich \& B\"ohringer 2002; Viana et
al.\ 2002).  The critical step in these experiments is the conversion
from observable (i.e.\ X--ray luminosity or temperature) to mass.  Our
results suggest that care must be taken to incorporate sufficient
scatter in the observable--mass relationship.  The asymmetric scatter
of the mass--temperature relationship arising from structural
segregation of clusters in this plane implies that such issues may be
most acute when converting from cluster temperature to mass,
especially if the cluster selection function is poorly characterized.
Indeed, in a companion to this article, Smith et al.\ (2003) showed
that incomplete understanding of the cluster selection function can
cause ${\sim}20\%$ systematic uncertainties in ${\sigma}_8$, favouring
values to the lower end of the range found in other recent works.

In a similar vein, we note that our results will likely impinge on
cosmological results derived from SZE surveys (see Carlstrom, Holder
\& Reese 2002 for a recent review).  Again, the key issue is the
precision to which the observable--mass relationship is known; in this
case the observable is the temperature of the ICM, as derived from the
SZE signal.  The issues for SZE surveys may be aggravated because the
goal of measuring the dark energy equation--of--state parameter $w$,
relies at least in part on measuring the \emph{evolution} of the
cluster population between two redshifts straddling the epoch at which
the dark energy is thought to take over as the dominant factor in the
expansion of the universe.  This is in contrast to measuring a
combination of ${\sigma}_8$ and ${\Omega}_{\rm M}$ from studies of
local clusters, i.e.\ just one redshift slice.  Further detailed
wide--field investigations of massive clusters at both low ($z{=}0.2$)
and higher ($z{\gs}0.6$) redshifts are therefore needed to quantify
robustly the evolution of the dynamical maturity of clusters and the
impact of that on cluster scaling relations.  We suggest that, in the
light of major imminent SZE cluster surveys, this is an urgent
exercise.

\section{Summary and Conclusions}\label{conclusions}

We have undertaken a comprehensive study of the distribution of mass
in ten X--ray luminous ($L_X{\ge}8{\times}10^{44}{\rm erg
\,s^{-1}}$[0.1--2.4\,keV]) galaxy clusters at $z{=}0.21{\pm}0.04$.  The
cornerstone of our analysis is a suite of detailed gravitational lens
models that describe the distribution of total mass in the cluster
cores.  These models are constrained by the gravitational lensing
signal detected in high--resolution \emph{HST}/WFPC2 imaging of the
clusters, including numerous multiply--imaged and weakly--sheared
background galaxies.  Analysis of archival 
\emph{Chandra} observations complements the lensing analysis and
enables us to relate the total mass and structure of the clusters to
the thermodynamics of the intra--cluster medium.
We re--cap the key results.

\setcounter{fred}{0}
\begin{list}{(\roman{fred})}{\usecounter{fred}\setlength{\itemindent}{0mm}\setlength{\labelwidth}{15mm}\setlength{\labelsep}{2mm}\setlength{\leftmargin}{7mm}} 
\setlength{\itemsep}{1.0mm}

\item Five of the ten clusters contain spectroscopically confirmed
  strong gravitational lensing, i.e.\ multiply--imaged background
  galaxies.  These five clusters comprise: A\,68 (Smith et al.\ 2002b;
  \S\ref{spec}), A\,383 (Smith et al.\ 2001; Sand et al.\ 2004),
  A\,963 (Ellis et al.\ 1991), A\,2218 (Pell\'o et al.\ 1992; Ebbels
  et al.\ 1998; Ellis et al.\ 2002) and A\,2219 (\S\ref{spec}).

\item Of the remaining five clusters, two contain relatively
  unambiguous examples of strong lensing for which spectroscopic
  redshifts are not yet available (A\,267 and A\,1835).  The other
  three clusters, A\,209, A\,773 and A\,1763 do not contain any
  obviously multiply--imaged galaxies, however the optical richness
  and massive nature of A\,773 imply that this cluster may well
  contain strong lensing that has yet to be uncovered.

\item Based on our search for strong--lensing in these clusters down
  to a surface brightness limit of
  $\mu_{702}{\simeq}25$\,mag\,arcsec$^{{-}2}$, we therefore put a firm
  lower limit on the fraction of the cluster sample that have a
  central projected mass density in excess of the critical density
  required for gravitational lensing of $50\%$.  Including A\,267
  and A\,1835 increases this limit to $70\%$.

\item We use the strong-- and weak--lensing signals to constrain
  parametrized models of the cluster potential wells, and from these
  models compute maps of the total projected mass in the cluster
  cores.  Spatial analysis of these maps reveals that four of the
  clusters form a homogeneous sub--sample with very high central mass
  fractions (${\Mcen}/{\Mtot}{>}0.95$).  The remaining six are
  strongly heterogeneous, with central mass fractions in the range
  $0.4{\le}{\Mcen}/{\Mtot}{\le}0.9$.  The central mass fraction of
  ${\Mcen}/{\Mtot}{\simeq}0.95$ that divides these two populations
  corresponds to a $K$--band central luminosity fraction of $L_{\rm
  K,BCG}/L_{\rm K,tot}{\sim}0.5$.

\item All of the six low central mass fraction clusters have an
  irregular, but not obviously bi/tri--modal X--ray morphology.  
  Four of the six are constrained by the current lensing
  data to have a bi/tri--modal mass morphology (A\,68, A\,773,
  A\,2218, A\,2219).  The other two (A\,209 and A\,1763) may be
  merging in the plane of the sky and thus any multi--modality in
  their mass distributions is not well--sampled by our WFPC2
  pencil--beam survey of the cluster cores.  

\item Three of the four high central mass fraction clusters also have
  relaxed X--ray morphologies.  The remaining cluster (A\,267) has a
  disturbed X--ray morphology, with a ${\sim}90$\,kpc offset between
  its centers of X--ray emission and mass.  The distribution of mass
  in this cluster may therefore be more complex than the single dark
  matter halo (plus cluster galaxies) that the current data are able
  to constrain.

\item Combining all of the information available to us, we define
  ``relaxed'' clusters to be those which appear relaxed in all
  available diagnostics, with the exception that we do not require a
  cool core.  Quantitatively relaxed clusters therefore have a single
  cluster--scale DM halo in their lens model ($N_{\rm DM}{=}1$), a
  high central mass fraction (${\Mcen}/{\Mtot}{\ge}0.95$) and central
  $K$--band luminosity fraction ($L_{K,{\rm BCG}}/L_{K,{\rm
  tot}}{\gs}0.5$), no evidence for an offset between the X--ray
  emission and the center of mass ${\drpeak}{<}4{\rm kpc}$) and the
  X--ray morphology is either circular, or mildly elliptical.  The
  unrelaxed clusters do not meet at least one of these criteria.

\item Applying these criteria to the cluster sample, we conclude that
  seven of the ten clusters are dynamically immature, i.e.\ unrelaxed
  (A\,68, A\,209, A\,267, A\,773, A\,1763, A\,2218, A\,2219) and three
  are relaxed (A\,383, A\,963, A\,1835); thus, formally
  $70{\pm}20\%$ of X--ray luminous cluster cores at $z{=}0.2$ are
  unrelaxed.

\item We detect a factor of three more scatter in the
  mass--temperature plane than predicted by Evrard, Metzler \& Navarro
  (1996), implying that great care should be exercised when using such
  relations to convert cluster temperature functions to mass functions
  in pursuit of cosmological parameters.  We also consider a number of
  key uncertainties that may artificially inflate our estimate of the
  scatter.  This exercise suggests that approximately one third of the
  scatter detected in this study may be due to issues related to the
  small field--of--view of our WFPC2 observations.  

\item The scatter in the mass--temperature plane is asymmetric,
  presenting evidence of structural segregation.  The normalization of
  the mass--temperature relation for unrelaxed (dynamically immature)
  clusters is $40\%$ hotter than for relaxed clusters at
  2.5--${\sigma}$ significance.  This result is consistent with recent
  simulations of cluster--cluster mergers (Ricker \& Sarazin 2001;
  Randall, Sarazin \& Ricker 2002), implying that merger induced
  temperature boosts may be the dominant factor behind the hotter
  normalization of unrelaxed systems.

\end{list}

In summary, this study is the first of its kind, exploiting detailed
strong--lensing constraints on the distribution of mass in X--ray
luminous cluster cores, complemented by X--ray spectro--imaging with
\emph{Chandra}.  The high frequency of dynamical immaturity, coupled
with the structural segregation of the clusters in the
mass--temperature plane have profound implications for our
understanding of how clusters form and evolve.  Perhaps of greatest
importance at this time is the implications of these results for using
clusters to constrain the cosmological parameters, ${\Omega}_M$,
${\sigma}_8$ and the dark energy equation--of--state parameter $w$.
In a companion paper we demonstrate that inadequate knowledge of the
cluster selection function can lead to $20\%$ systematic errors in
cluster--based measurements of ${\sigma}_8$.  Turning to $w$,
forthcoming Sunyaev--Zeldovich Effect experiments designed to detect
and measure the mass of clusters out to high redshifts, using
mass--temperature scaling relations may be compromised by unidentified
and/or poorly calibrated astrophysical systematic uncertainties (see
also Majumdar \& Mohr 2003).

Our future program will build on these results in three ways.  First,
we aim to overcome the principal uncertainties in the current work:
small number statistics and tiny field--of--view.  Wide--field
space--based imaging of a statistically complete sample of clusters
would be essential to achieve this goal.  Second, detailed comparison
of selection effects and measurement techniques between theoretical
and observational studies will enable more detailed and rigorous
comparison between observational and synthetic datasets.  Finally, we are
gathering \emph{HST}/ACS imaging of an identically selected sample of
clusters at $z{\simeq}0.55$ drawn from the MACS sample (Ebeling et
al.\ 2001b).  We will combine these data with observations in the
X--ray pass--band and compare the results to those found here.  The
${\sim}3$\,Gyr difference in look--back--time between $z{=}0.2$ and
$z{=}0.55$ will enable us to search for evolutionary trends in the
most massive clusters.

\section*{Acknowledgments}

GPS thanks Alastair Edge for much encouragement and assistance during
this project.  We also thank Steve Allen, Michael Balogh, Sebastien
Bardeau, John Blakeslee, Richard Bower, Kevin Bundy, Warrick Couch,
Sarah Bridle, Richard Ellis, Gus Evrard, Andy Fabian, Masataka
Fukugita, Henk Hoekstra, Phillipe Marty, Ben Moore, Bob Nichol, Johan
Richard, David Sand, Tommaso Treu and Mark Voit for a variety of
helpful discussions, comments and assistance.  GPS acknowledges
financial support from PPARC.  JPK acknowledges support from CNRS.
IRS acknowledges support from the Royal Society and the Leverhulme
Trust.  PM acknowledge supports from the European commission contract
number HRPN--CT--2000--00126 and by CXC grant GO2--3177X.  HE
acknowledges financial support under NASA grants NAG 5--6336 and NAG
5--8253.  OC acknowledges support from the European Commission under
contract no.\,ER--BFM--BI--CT97--2471.  We also acknowledge financial
support from the UK--French ALLIANCE collaboration programme
\#00161XM.  Finally, we recognize and acknowledge the cultural role
and reverence that the summit of Mauna Kea has within the indigenous
Hawaiian community. We are most fortunate to have the opportunity to
conduct observations from this mountain.

\section*{References}

\begin{list}{}{\setlength{\labelwidth}{0mm}\setlength{\labelsep}{0mm}\setlength{\itemindent}{-5mm}\setlength{\leftmargin}{5mm}\setlength{\parsep}{0mm}\setlength{\itemsep}{0mm}}

\item Abell G.O., Corwin H.G.\ Jr., Olowin R.P., 1989, ApJS,
           70, 1
\item Adami C., Ulmer M.P., Durret F., Nichol R.C., Mazure A., Holden
           B.P., Romer A.K., Savine C., 2000, A\&A, 353, 930
\item Akritas M.G., Bershady M.A., 1996, ApJ, 470, 706
\item Allen C.W., 1973, Astrophysical Quantities, 3rd edn.\
           Athlone Press, London 
\item Allen S.W., Fabian A.C., Kneib J.-P., 1996, MNRAS, 279,
           615
\item Allen S.W., 1998, MNRAS, 296, 392
\item Allen S.W., Fabian A.C., 1998, MNRAS, 297, 57
\item Allen S.W., Schmidt R.W., Fabian A.C., 2001, MNRAS, 328, L37
\item Allen S.W., Schmidt R.W., Fabian A.C., Ebeling H., 2003,
           MNRAS, 342, 287
\item Arnaud M., Evrard A.E., 1999, MNRAS, 309, 631
\item Babul A., McCarthy I.G., Poole G.B., 2003, to appear in the
  proceedings of the ``Multiwavelength Cosmology'' Conference held in
  Mykonos, Greece, June 2003, ed. M. Plionis (Kluwer), astro--ph/0309543 
\item Balogh M.L., Smail I., Bower R.G., Ziegler B.L., Smith
           G.P., Davies R.L., Gaztelu A., Kneib J.-P., Ebeling H.,
           2002, ApJ, 566, 123 
\item Bertin E., Arnouts S., 1996, A\&A, 117, 393
\item B\'ezecourt J., Hoekstra H., Gray M.E., AbdelSalaam H.M.,
           Kuijken K., Ellis R.S., 2000, A\&A, submitted,
           astro--ph/0001513 
\item Bond J.R., Kofman L., Pogosyan D., 1996, Nature, 380, 603
\item Borgani S., et al., 2001, ApJ, 561, 13
\item Brainerd T.G., Blandford R.D., Smail I., 1996, ApJ, 446, 623
\item Buote, D.A., Tsai, J.C.,  1996 ApJ, 458, 27
\item Carlstrom J.E., Holder G.P., Reese E.D., 2002, ARA\&A, 40,
           643
\item Casertano S., Wiggs M., 2001, ``An Improved Goemetric
           Solution for WFPC2'', \emph{ISR WFPC2-2001-10}
\item Churazov E., Forman W., Jones C., B\"ohringer H., ApJ,
           590, 225
\item Cole S., Norberg P., Baugh C.M., Frenk C.S., \& the 2dFGRS team,
           2001, MNRAS, 326, 255
\item Crawford C.S., Allen S.W., Ebeling H., Edge A.C.,
           Fabian A.C., 1999, MNRAS, 306, 857
\item Czoske O., Kneib J.-P., Soucail G., Bridges T.J., Mellier Y.,
           Cuillandre J.-C., 2001, A\&A, 372, 391
\item Czoske O., Moore B., Kneib J.-P., Soucail G., 2002, A\&A,
           386, 31 
\item Czoske O., 2002, PhD Thesis, Observatoire Midi-Pyr\'en\'ees,
           Universit\'e Paul Sabatier, Toulouse, France 
\item Davis M., Efstathiou G., Frenk C.S., White S.D.M., 1985, ApJ,
           292, 371 
\item De~Grandi, S.\,et al., 1999, ApJ, 514, 148
\item De~Lapparent V., Geller M.J., Huchra J.P., 1986, ApJ, 302, 1
\item De~Lucia G., et al., 2004, MNRAS, 348, 333
\item De~Propris R., et al., 2003, ApJ, 598, 20
\item Dickey, J.M., Lockman, F.J., 1990, ARA\&A, 28, 215
\item Dressler A., Shectman S.A., 1988, AJ, 95, 985
\item Ebbels T.M.D., Ellis R.S., Kneib J.-P., Le~Borgne J.-F., Pell\'o
           R., Smail I., Sanahuja B., 1998, MNRAS, 295, 75
\item Ebbels T.M.D., 1998, PhD Thesis, University of Cambridge, UK
\item Ebeling H., et al., 1996, MNRAS, 281, 799
\item Ebeling H., et al., 1998, MNRAS, 301, 881
\item Ebeling H., Edge, A.C., Henry J.P., 2000, in Large Scale
           Structure in the X--ray Universe, Proceedings of the 20-22
           September 1999 Workshop, Santorini, Greece, eds.\ Plionis, M.,
           Georgantopoulos, I., Atlantisciences, Paris, France, p.39
\item Ebeling H., Edge A.C., Henry J.P., 2001, ApJ, 553, 668
\item Edge A.C., Stewart G.C., Fabian A.C., Arnaud K.A., 1990,
           MNRAS, 245, 559 
\item Edge A.C., Smith G.P., Sand D.J., Treu T., Ebeling H.,
           Allen S.W., van Dokkum P.G., 2003, ApJ, 599, L69 
\item Eke, V.R., Cole, S., Frenk, C.S., 1996, MNRAS, 282, 263
\item Ellis R.S.,  Allington-Smith J., Smail I., 1991, MNRAS,
           249, 184 
\item Ellis R.S., Santos M.R., Kneib J.-P., Kuijken K., 2001,
           ApJ, 560, L119 
\item Ettori S., De~Grandi S., Molendi S., 2002, A\&A, 391, 841
\item Evrard A.E., Mohr J.J., Fabricant D.G., Geller M.J., 1993,
  ApJ, 419, L9 
\item Evrard A.E., Metzler C.A., Navarro J.F., 1996, ApJ, 469, 494
\item Evrard A.E., et al., 2002, ApJ, 573, 7
\item Fabian A.C., Crawford C.S., Edge A.C., Mushotzky R.F.,
           1994, MNRAS, 267, 779
\item Fern\'andez-Soto A., Lanzetta K.M., Yahil A., 1999, ApJ, 513, 34
\item Finoguenov A., Reiprich T.H., B\"ohringer H., 2001, A\&A,
           368, 749 
\item Fruchter, A.S., Hook, R.N., 1997, in Applications of Digital
             Image Processing, Proc.\,SPIE, 3164, ed.\,Tescher, A., p120
\item Geller M.J., Beers T.C., 1982, PASP, 94, 421
\item Gehrels N., 1986, ApJ, 303, 336
\item Gilmozzi R., Ewald S., Kinney E., 1995, ``The Geometric
  Distortion Correction for the WFPC Cameras'', \emph{ISR
  WFPC2-95-02}
\item Gioia I.M., et al., 1990, ApJ, 356, L35
\item Giovannini G., Tordi M., Feretti L., 1999, NewA, 4, 141
\item Goto T., Okamura S., McKay T.A., Annis J., Bahcall N.A.,
           Bernardi M., Brinkmann J., Gomez P.L., Hansen S., Kim R.S.J.,
           Sekiguchi M., Sheth R., 2002, PASJ, accepted,
           astro-ph/0205413 
\item Gunn J.E., Gott J.R.III, 1972, ApJ, 176, 1
\item Hjorth J., Oukbir J., van Kampen E., 1998, MNRAS, 298, L1
\item Hoekstra H., Franx M., Kuijken K., Squires G., 1998, ApJ,
  504, 636 
\item Hogg D.W., Pahre M.A., McCarthy J.K., Cohen J.G., Blandford
           R., Smail I., Soifer B.T., 1997, MNRAS, 288, 404
\item Holtzman, J.A., Burrows, C.J., Casertano, S., Hester, J.J.,
             Trauger, J.T., Watson, A.M., Worthey, G., 1995, PASP,
             107, 1065 
\item Johnson H.L., 1966, ARA\&A, 4, 193
\item Jones C., Forman W., 1984, ApJ, 276, 38
\item Kaiser N., 1986, MNRAS, 222,323
\item Kashikawa N., et al., 2002, PASJ, 54, 819
\item Kassiola A., Kovner I., 1993, ApJ, 417, 450
\item King C.R., Ellis R.S., 1985, ApJ, 288, 456
\item Kneib J.-P., 1993, PhD Thesis, Universit\'e Paul Sabatier,
  Toulouse, France 
\item Kneib J.-P., Mellier Y., Fort B., Mathez G., 1993, A\&A, 273, 367
\item Kneib J.-P., Mellier, Y., Fort, B., Soucail, G., Longaretti,
             P.Y., 1994, A\&A, 286, 701
\item Kneib J.-P., Mellier Y., Pell\'o R., Miralda-Escud\'e J.,
 Le Borgne J.-F., B\"ohringer H., Picat J.-P., 1995, A\&A, 303, 27 
\item Kneib, J.-P., Ellis, R.S., Smail, I., Couch, W.J., Sharples,
             R.M., 1996, ApJ, 471, 643
\item Kneib J.-P., Hudelot P., Ellis R.S., Treu T., Smith G.P.,
  Marshall P., Czoske O., Smail I., Natarajan P., 2003, ApJ, 598,
  804 
\item Luppino G.A., Gioia I.M., Hammer F., Le~F\`evre O., Annis J.A.,
1999, A\&A, 136, 117 
\item Majumdar S., Mohr J.J., 2003, ApJ, 585, 603
\item Mannucci F., Basile F., Cimatti A., Daddi E., Poggianti
           B.M., Pozzetti L., Vanzi L., 2001, MNRAS, 326, 745 
\item Markevitch M., 1998, ApJ, 504, 27
\item Markevich, M. et al. 2000a, CXC memo
             (http://asc.harvard.edu/cal ``ACIS'', ``ACIS Background'')
\item Markevitch, M., et al.\ 2000, ApJ, 541, 542
\item Markevitch, M., \&  Vikhlinin, A. 2001, ApJ, 563, 95
\item Markevitch, M., 2002, astro-ph/0205333
\item Marty P.B., Kneib J.-P., Sadat R., Bernard J.-P., Czoske
  O., Ebeling H., Smail I., Smith G.P., 2004, A\& A, submitted 
\item Mazzotta P., Markevitch M., Vikhlinin A., Forman W.R.,
             David L.P., Van~Speybroeck L., 2001, ApJ, 555, 205
\item Mellier, Y., Fort, B., Kneib J.-P., 1993, ApJ, 407, 33
\item Miralda--Escud\'e J., Babul A., 1995, ApJ, 449, 18
\item Miyazaki S., et al., 2002, ApJ, 580, L97
\item Nakamura F.E., Hattori M., Mineshige S., 1995, A\&A,
           302, 6009
\item Nevalainen J., Markevitch M., Forman W., 2000, ApJ, 532, 694
\item Oke, J.B., et al., 1995, PASP, 107, 375
\item Peacock J.A., et al., 2001, Nature, 410, 169
\item Peebles P.J.E., 1980, \emph{``Large Scale Structure of the
Universe''}, Princeton University Press, Princeton
\item Pell\'o R., Le~Borgne J.-F., Sanahuja B., Mathez G., Fort B.,
           1992, A\&A, 266, 6 
\item Peres, C.B., Fabian, A.C., Edge, A.C., Allen, S.W., Johnstone,
             R.M., White, D.A., 1998, MNRAS, 298, 416
\item Pierpaoli E., Scott D., White M., 2001, MNRAS, 325, 77
\item Randall S.W., Sarazin C.L., Ricker P.M., ApJ,
           577, 579
\item Reiprich T.H., B\"ohringer H., 2002, ApJ, 567, 716
\item Rhodes J.D., Refregier A., Groth E.J., 2000, ApJ, 536, 79
\item Richstone D., Loeb A., Turner E.L., 1992, ApJ, 393, 477
\item Ricker P.M., Sarazin, C.L., 2001, ApJ, 561, 621
\item Ritchie B.W., Thomas P.A., 2002, MNRAS, 329, 675
\item Roettiger, K., Loken C., Burns J.O., 1997, ApJS,
           109, 307
\item Sand D.J., Treu T., Ellis R.S., 2002, ApJ, 574, L129
\item Sand D.J., Treu T., Smith G.P., Ellis R.S., 2004, ApJ, 604, 88
\item Schindler S., Mueller E., 1993, A\&A, 272, 137
\item Schmidt R.W., Allen S.W., Fabian A.C., 2001, MNRAS, 327, 1057
\item Schuecker P., B\"ohringer H., Reiprich T.H., Feretti L.,
  2001, A\& A, 378, 408
\item Shectman S.A., et al., 1996, ApJ, 470, 172
\item Smail, I., Couch, W.J., Ellis, R.S.,Sharples, R.M., 1995, ApJ,
           440, 501
\item Smail I., Hogg D.W., Blandford R., Cohen J.G., Edge A.C.,
           Djorgovski S.G., 1995b, MNRAS, 277, 1 
\item Smail, I., et al., 1996, ApJ, 469, 508
\item Smail, I., Ellis, R.S., Dressler, A., Couch, W.J., Oemler,
             A., Butcher, H., Sharples, R.M., 1997, ApJ, 470, 70
\item Smith G.P., Kneib J.-P., Ebeling H., Csozke O., Smail I.,
           2001, ApJ, 552, 493 
\item Smith G.P., Smail I., Kneib J.--P., Czoske O., Ebeling H.,
           Edge A.C., Pello R., Ivison R.J., Packham C., Le Borgne
           J.--F., 2002, MNRAS, 330, 1 
\item Smith G.P., Smail I., Kneib J.-P., Davis C.J., Takamiya M.,
           Ebeling H., Czoske O., 2002, MNRAS, 333, L16 
\item Smith G.P., 2002, PhD Thesis, University of Durham, UK,
           available upon request from gps@astro.caltech.edu
\item Smith G.P., Edge A.C., Eke V.R., Nichol R.C., Smail I.,
           Kneib J.-P., 2003, ApJ, 590, L79 
\item Spergel D.N., Steinhardt P.J., 2000, Phys.\ Rev., 84,
           L3760 
\item Squires G., Neumann D.M., Kaiser N., Arnaud M., Babul A.,
           B\"ohringer H., Fahlman G., Woods D., 1997, ApJ, 482, 648
\item Squires G., Kaiser N., Fahlman G., Babul A., Woods D.,
           1996, ApJ, 469, 73
\item Swinbank A.M., Smith J., Bower R.G., Bunker A., Smail I.,
           Ellis R.S., Smith G.P., Kneib J.-P., Sullivan M.,
           Allington-Smith J.R., 2003, ApJ, 598, 162 
\item Tormen G., Diaferio A., Syer D., 1998, MNRAS, 229, 728
\item Trauger J.T., Vaughan A.H., Evans R.W., Moody D.C., 1995,
           ``Geometry of the WFPC2 Focal Plane'', in \emph{Calibrating
           HST: Post Service Mission}, eds.\ A.\ Koratkar \& C.\
           Leitherer  
\item Tyson J.A., Kochanski G.P., dell'Antonio I.P., 1998, ApJ,
           498, L107 
\item Vall\'ee J.P., Bridle A.H., 1982, ApJ, 253, 479
\item Vettolani G., et al., 1997, A\&A, 325, 954
\item Viana, P.T.P., Liddle, A.R., 1996, MNRAS, 281, 323
\item Viana P.T.P., Nichol R., Liddle A.R., 2002, ApJ, 569, L75
\item Vikhlinin, A., Markevitch, M., Murray, S.S. 2001a, ApJ,
             551, 160
\item West M.J., Bothun G.D., 1990, ApJ, 350, 36
\item White D.M., Jones C., Forman W., 1997, MNRAS, 292, 419
\item Wittman D., et al., 2003, ApJ, 597, 218
\item Wu X., Fang L., 1997, ApJ, 483, 62
\item Wu X.; Chiueh T., Fang L., Xue Y., 1998, MNRAS, 301,
  861
\item Wu X., 2000, MNRAS, 316, 299
\item Yoshida N., et al., 2001, MNRAS, 325, 803
\item Zehavi I., et al., 2002, ApJ, 571, 172
\end{list}

\onecolumn

\begin{appendix}

\section{Gravitational Lens Modelling -- Method}\label{method}

This appendix describes relevant details of how the {\sc lenstool}
ray--tracing code is used to construct robust models of the
distribution of mass in galaxy clusters using both strong-- and
weak--lensing constraints.  

\subsection{Mathematical Overview}
\label{math}

Consider a single source at $\zs$ that appears to an observer under
the action of a gravitational lens at $\zl$ as $N$ distinct images at
positions $\vec{u}_i\,(1{\le} i{\le} N)$.  We describe the source with
$\nu$ free parameters, $\Pi_j\,(1{\le} j{\le}\nu)$, for example: the
position of the center of the source, the ellipticity, the
orientation, the surface brightness.  We write the transformation
equations in the following form:

\begin{equation}
\Pi_j^S{=}f_j(\Pi_{ji}^I, \varphi(\vec{u}_i))~~~~~(1{\le} i{\le} N)~~(1{\le} j{\le}\nu)
\end{equation}

\noindent
where $f_j$ are functions that depend on the parameters that describe
the observed images and the gravitational potential of the lens.  The
source parameters , $\Pi_j^S$, and the lens potential,
$\varphi(\vec{u}_i)$, are the unknowns in these equations.  We use the
image parameters , $\Pi_{ji}^I$ (i.e.\ the observables), to constrain
both the source parameters and the lens potential.  If we are able to
recover $\nu$ parameters for each image, then we have $\nu(N{-}1)$
constraints on our lens model.  Generalizing this to $n$ sets of
multiple images of sources at redshifts $\zs_i$, each multiple-image
system being characterized by $(\nu_i, N_i)$, then the total number of
constraints $n_c$ is given by:

\begin{equation}
n_c{=}\sum_{i}^{n}[\nu_i(N_i{-}1){-}\epsilon_i]
\label{theory:nc}
\end{equation}

\noindent
where $\epsilon_i{=}0$ if $\zs_i$ is known and $\epsilon_i{=}1$ if $\zs_i$
is not known (Kneib et al.\ 1993).  Strictly, Equation~\ref{theory:nc} only
applies in the idealized case of all multiple-images being resolved,
and none of the images being merging pairs.  Clearly higher resolution
imaging will increase $n_c$.

We describe each observed gravitational image with the following
parameters: $\Pi^I{=}(\vec{u}^I, S^I, \vec{\tau}^I)$, where $\vec{u}^I$
is the position of the image, $S^I$ is the observed flux and
$\vec{\tau}^I{=}\tau^Ie^{2i\theta^I}$ is the complex deformation of the
image, describing its ellipticity ($\tau$) and orientation ($\theta$).
We use these quantities and their counterparts in the source-plane 
to write down the transformation equations:

\begin{equation}
\begin{array}{ll}
{\rm Position:} & \vec{u}^S{=}\vec{u}^I{-}\vec{\nabla}\varphi(\vec{u}^I) \cr
{\rm Flux:}     & S^S{=}|{\rm det}\,J|S^I \cr
{\rm Shape:}    & {\rm sgn}({\rm det}\,J)\tv^S{=}\tv^I{-}\tv_{\rm pot}[\delta^I{-}\tau^I\Re(\gv^I\gv^*_{\rm pot})] \cr
\end{array}
\label{theory:posfshape}
\end{equation}

\noindent
where the first equation is simply the lens equation, $J$ is the
Jacobian matrix of the lensing transformation,
$\delta{=}(1{+}\tau^2)^{1/2}$, $\gv{=}\gamv/(1{-}\kappa)$, the subscript
``pot'' denotes quantities applicable to a circular source (see Kneib
et al.\ (1996) and references therein for a detailed derivation of the
shape transformation equation), $\Re$ takes the real part of
dot-product between $\gv^I$ and $\gv^*_{\rm pot}$, and $\gv^*$ is the
complex conjugate of $\gv$.

\subsection{Parametrization of the Lens Plane Mass Distribution}

Each mass component is parametrized as a smoothly truncated
pseudo-isothermal elliptical mass distribution (PIEMD -- Kassiola \&
Kovner 1993; Kneib et al.\ 1996).  This functional form is physically
well motivated (it avoids the central singularity and infinite spatial
extent of singular isothermal models) and can describe mass
distributions of arbitrarily large ellipticities.  Each PIEMD mass
component is parametrized by its position ($\xc$, $\yc$), ellipticity
($\epsilon$), orientation ($\theta$), core radius ($\rc$), cut-off
radius ($\rt$) and central velocity dispersion ($\sigo$), and the
projected mass density, $\Sigma$ is given by:

\begin{equation}
\Sigma(x,y){=}\frac{\sigo^2}{2G}\,\frac{\rt}{\rt{-}\rc}\left(\frac{1}{(\rc^2{+}\rho^2)^{1/2}}{-}\frac{1}{(\rt^2{+}\rho^2)^{1/2}}\right)
\end{equation}

\noindent
where $\rho^2{=}[(x{-}\xc)/(1{+}\epsilon)]^2{+}[(y{-}\yc)/(1{-}\epsilon)]^2$ and
the ellipticity of the lens is defined as $\epsilon{=}(a{-}b)/(a{+}b)$.  The
geometrical parameters ($\xc$, $\yc$, $\epsilon$, $\theta$) of each
mass component are matched to the observed light distribution of the
related cluster galaxy.  The dynamical parameters ($\rc$, $\rt$,
$\sigo$) of the ``major mass components'' (i.e. cluster-scale mass
components and selected bright cluster galaxies, including each
central galaxy) are kept as free parameters.  To minimize the number
of model parameters, the dynamical parameters of the remaining mass
components are scaled with the luminosity of their associated galaxy
following Brainerd, Blandford \& Smail (1996):

\begin{equation}
\rc{=}\rc^{\star}(L/L^{\star})^{1/2};\hspace{5mm}
\rt{=}\rt^{\star}(L/L^{\star})^{1/2};\hspace{5mm}
\sigo{=}\sigo^{\star}(L/L^{\star})^{1/4}
\end{equation}

\noindent We also scale the mass of individual galaxies with their 
luminosity, using: 

\begin{equation}
M{=}(\pi/G)(\sigo^{\star})^2\rt^{\star}(L/L^{\star})  
\end{equation}

\noindent
These scaling laws are physically well motivated and conserve the
mass-to-light ratio of the galaxies in a manner analogous to the
observed Faber-Jackson and Tully-Fisher scaling relations for spiral
and elliptical galaxies respectively.

\subsection{Model Optimization}

We construct a $\chi^2$--estimator to quantify how well each trial lens
model fits the observational data:

\begin{equation}
\chi^2{=}\chi^2_{\rm pos}{+}\chi^2_{\rm shape}{+}\chi^2_{\rm flux}{+}\chi^2_{\rm crit}{+}\chi^2_{\rm weak}
\label{theory:chi2}
\end{equation}

\noindent
The first three terms compare the source-plane properties computed for 
each observed image ($x_i^S,~y_i^S,~\tau_i^S,~\theta_i^S,~S_i^S$ --
as defined in equation \ref{theory:posfshape}).  We 
define $\chi^2_{\rm pos}$, $\chi^2_{\rm shape}$ and $\chi^2_{\rm flux}$ 
as follows:

\begin{equation}
\chi^2_{\rm pos}{=}\sum_{i{=}1}^{N{-}1}\frac{(x_i^S{-}x_{i{-}1}^S)^2{+}(y_i^S{-}y_{i{-}1}^S)^2}{\sigma_{\rm pos}^2}
\label{theory:chipos}
\end{equation}

\begin{equation}
\chi^2_{\rm shape}{=}\sum_{i{=}1}^{N{-}1}\frac{(\tau_i^Scos(2\theta_i^S){-}\tau_{i{-}1}^Scos(2\theta_{i{-}1}^S))^2{+}(\tau_i^Ssin(2\theta_i^S){-}\tau_{i{-}1}^Ssin(2\theta_{i{-}1}^S))^2}{\sigma_{\rm shape}^2}
\label{theory:chishape}
\end{equation}

\begin{equation}
\chi^2_{\rm flux}{=}\sum_{i{=}1}^{N{-}1}\frac{(S_i^S{-}S_{i{-}1}^S)^2}{\sigma_{\rm flux}^2}
\label{theory:chiflux}
\end{equation}

\noindent
where $\sigma_{\rm pos}^2$, $\sigma_{\rm shape}^2$ and $\sigma_{\rm flux}^2$ 
are the accuracies with which we can measure the position, shape and
flux of galaxies in our \emph{HST} data.

The fourth term in equation~\ref{theory:chi2} compares how well the
symmetry breaks in the observed gravitational images (i.e.\ locations
of critical lines) are reproduced by the lens model.  We define
($x_{\rm ct}^{\rm obs}$, $y_{\rm ct}^{\rm obs}$) and ($x_{\rm ct}^{\rm
mod}$, $y_{\rm ct}^{\rm mod}$) as the observed and model critical line
positions respectively and construct $\chi^2_{\rm crit}$, where
$\Delta x_{\rm crit}$ and $\Delta y_{\rm crit}$ are the uncertainties
in the position of the symmetry break.

\begin{equation}
\chi^2_{\rm crit}{=}\frac{(x_{\rm ct}^{\rm obs}{-}x_{\rm ct}^{\rm mod})^2{+}(y_{\rm ct}^{\rm obs}{-}y_{\rm ct}^{\rm mod})^2}{\Delta x_{\rm crit}^2{+}\Delta y_{\rm crit}^2}
\end{equation}

Finally, we construct $\chi^2_{\rm weak}$ in a similar manner to
$\chi^2_{\rm shape}$; the differences being that we sum over the $F$
faint galaxy images detected in the cluster field, $\sigma_{\rm weak}$
is the width of the distribution of galaxy shapes from surveys of
field galaxies (e.g.\ Ebbels 1998) and we compare the image-plane
galaxy shapes with that induced by the trial mass distribution at the
faint galaxy image on a circular source.

\begin{equation}
\chi^2_{\rm weak}{=}\sum_{i{=}1}^{F{-}1}\frac{(\tau_i^Icos(2\theta_i^I){-}\tau_{\rm pot}^Icos(2\theta_{\rm pot}^I))^2{+}(\tau_i^Isin(2\theta_i^I){-}\tau_{\rm pot}^Isin(2\theta_{\rm pot}^I))^2}{\sigma_{\rm weak}^2}
\label{theory:chiweak}
\end{equation}

The $\chi^2$ estimator is minimized by varying the model parameters to
obtain an acceptable ($\chi^2{\sim}1$) fit to the observational
constraints.  This is an iterative process, which we begin by
restricting our attention to the least ambiguous model constraints
(i.e.\ the confirmed multiple-image systems) and the relevant free
parameters.  For example, in a typical cluster lens there will be one
spectroscopically-confirmed multiple-image system and a few other
candidate multiples.  The model fitting process therefore begins with
using the spectroscopic multiple to constrain the dynamical parameters
of the main cluster-scale mass component.  

\end{appendix}

\end{document}